\documentclass[twocolumn]{aastex631}
\usepackage{amsmath,amsfonts,amssymb,graphicx,chngcntr,multirow, float,booktabs, afterpage}
\usepackage[]{hyperref}
\hypersetup{colorlinks=true}

\received{}
\revised{}
\accepted{}

\usepackage{CJK}

\newcommand{\athenak}{\texttt{AthenaK}}
\newcommand{\athenapp}{\texttt{Athena++}}

\newcommand{\revision}[1]{{#1}} 

\usepackage{tikz}
\usetikzlibrary{shapes}
\newcommand*\oiint{%
	\tikz \node[draw, ellipse, inner xsep=-3pt, inner ysep=-9pt] {$\displaystyle\iint$};%
}
\newcommand*\oiiint{%
	\tikz \node[draw, ellipse, inner xsep=-4pt, inner ysep=-9pt] {$\displaystyle\iiint$};%
}

\shorttitle{Long-term evolution of a black hole torus: GRMHD vs thin disk theory}

\begin{document}
\begin{CJK*}{UTF8}{gbsn}

\title{\revision{The long-term evolution of a black hole accretion torus: a comparison between GRMHD simulations and relativistic thin disk theory}}
\author{Andrew Mummery} 
\affiliation{School of Natural Sciences, Institute for Advanced Study, 1 Einstein Drive, Princeton, NJ 08540, USA}
\email{amummery@ias.edu}
\author{Minghao Guo (郭明浩)}
\affiliation{Department of Astrophysical Sciences, Princeton University, Princeton, NJ 08544, USA}
\author{Eliot Quataert}
\affiliation{Department of Astrophysical Sciences, Princeton University, Princeton, NJ 08544, USA}
\author{James M. Stone}
\affiliation{School of Natural Sciences, Institute for Advanced Study, 1 Einstein Drive, Princeton, NJ 08540, USA}

\date{\today}


\begin{abstract}
    In this paper we analyse the long-term evolution of a black hole accretion torus using novel GRMHD simulations which, for the first time, have been run out to extremely large evolutionary times $t_{\rm max} = 10^7 GM_\bullet/c^3$. This allows for detailed comparisons between the gross evolution of this GRMHD simulation and existing ``spreading disk'' analytical theories. We show that the long-term features of this evolution are \revision{well described by General Relativistic Thin Disk Theory (GRTDT), which reproduces various aspects of the GRMHD evolution to within a factor $\sim 2$}, despite the disk in our \revision{simulation} not being remotely thin, showing dynamically dominant (MAD) magnetic fields, and launching \revision{a powerful} wind. \revision{Thin disk theory} functions as an accurate and cheap effective theory for \revision{the evolution in radius and time of the bulk of the disk material, and for the time dependence of the horizon accretion rate}. We argue that this effectiveness originates from \revision{the} global conservation laws of mass and angular momentum\revision{, together with a turbulent stress which we find to be local and independent of density, lending support to the ad-hoc local-stress-tensor phenomenology employed in analytical models}. We also \revision{find that the turbulent accretion rate variability in the inner disk is log-normally distributed, consistent with} a key assumption in the modeling of observed variability.
\end{abstract}

\section{Introduction}
Accretion onto black holes powers some of the most luminous objects in the Universe, and as a physical process it is still not understood in all of its complexity. The study of black hole accretion has progressed along three parallel tracks, with analytical theories \citep[e.g.][]{SS73, LBP74, NovikovThorne73, PageThorne74} and observational studies dating back to the 1970's, while full numerical solutions of the fundamental General Relativistic Magnetohydrodynamic (GRMHD) equations followed in subsequent decades \cite[e.g.,][among many others]{HawleyKrolik01, Gammie2003, Shafee08,  Noble10, Penna10, Schnittman16, Lancova19, Dhang25, Rule25}.  

One particularly important element of the theoretical accretion problem which has remained out of reach of numerical solvers until very recently is an analysis of the very late time behavior of black hole accretion flows (i.e., the behavior at times $t \gg 10^5 GM_\bullet/c^3$ far exceeding the orbital time of the inner disk), as until very recently reaching such late times was beyond even the most powerful numerical solvers. The recent development of 
``multi-zone'' \citep{Cho2023,Cho2024} and
``cyclic zoom'' \citep{Guo2025} methods greatly accelerates the required calculations, and aids significantly in overcoming this multiscale challenge. This approach allowed \citet{Guo2025} to study long-term evolution of slowly evolving GRMHD flows  in detail, without having to resort to the use of free scaling parameters that dominated uncertainties in previous studies. 

This is an extremely important development, as the long term evolution of these full GRMHD simulations of the black hole accretion problem will have broad implications for both simple theoretical models of the accretion process and observations of X-ray binaries, active galactic nuclei, compact object mergers, and tidal disruption events (among others).  Indeed, observing baselines of even super massive black hole systems all exceed times $\Delta t \gg10^5GM_\bullet/c^3 \sim 5\, {\rm d}\, (M_\bullet/10^6M_\odot)$ where (until now) only analytical theories can be hoped to apply. 

While the properties of analytical time-dependent disk theories are well understood (the accretion rate drops typically as a power-law with time, while the bulk of the material spreads to larger radii to soak up the angular momentum of the accreting material), the lack of long-term GRMHD simulations means that in effect these analytical models (which are routinely fit to data) are yet to be carefully calibrated against more detailed theories over long temporal and radial baselines. 

\revision{The purpose of this paper is to test whether the analytical theory most commonly used to interpret evolving accretion flows (time-dependent thin disk theory, here in its general relativistic form) correctly describes the long-term evolution of an accretion flow computed from first principles. We focus on the quantities which such models are used to predict, and which are therefore most directly tied to observations: the accretion rate onto the black hole, and the radial distribution of disk mass (and temperature in thin disk theories) as it evolves in time. In this sense the present work is a first step towards benchmarking the analytical theories used to bridge between detailed simulations and observations.}

\revision{It is worth being explicit at the outset all the ways in which disk theory might fail in this comparison. Global conservation of mass and angular momentum constrains the evolution of any accretion flow with a finite mass budget, and we shall show (section \ref{sec:analysis}) that these constraints alone fix some of the scaling relationships of the problem. But conservation is equally consistent with evolution which is entirely unlike that of a thin disk. There are several physically well motivated ways in which a thick, non-radiative, magnetically arrested torus could depart from thin disk theory. First, angular momentum could be removed from the bound gas non-locally, by magnetized outflows or large scale magnetic torques, rather than being transported radially through the disk; a disk evolving in this way accretes without spreading \citep[e.g.,][]{BlandfordPayne82, BaiStone13, Tabone22}, and thick non-radiative flows have also been argued to transport angular momentum in quite different ways to thin disks \citep[e.g.,][]{Narayan00, Quataert00}. Second, and specific to the magnetically arrested state, the large-scale magnetic field could itself transport mass and angular momentum non-locally: through the episodic eruption of magnetic flux bubbles from the horizon, through torques exerted by ordered field anchored at other radii, and through magnetic pressure and tension forces comparable to gravity \citep[e.g.,][]{Tchekhovskoy11, Chatterjee22}, all of which violate the assumption that the evolution is set by a local stress acting on gas in near-Keplerian orbits. Third, even if transport is mediated by local stresses, this stress could carry an intrinsic scale, or have no simple description at all: a complicated dependence on position and time, a dependence on the history of the flow, or a key role of intermittent events would all prevent any simple parameterization of the stress from capturing the evolution \citep[see, e.g., the discussion of][]{BalbPap99}. Fourth, the turbulence itself could decay as the disk spreads to radii far beyond those which initially contained magnetic field, in which case the evolution would stall \citep[see, e.g.,][on the dependence of the turbulent stress on the net magnetic flux]{Salvesen16}. Fifth, outflows which ultimately unbind an ${\cal O}(1)$ fraction of the disk mass could reshape the distribution of the remaining bound gas, rather than simply reducing its normalization \citep[e.g.,][]{BlandfordBegelman99, Yuan12}. Finally, a disk with $H/r \sim 1$ is not obviously described by the kinematics of thin disks at all: pressure support makes the rotation sub-Keplerian \citep[e.g.,][]{NarayanYi95}, and vertical structure such as meridional circulation \citep[e.g.,][]{PhilippovRafikov17} is absent by construction from any height-averaged theory. Each of these could leave a distinct imprint on the evolution of the accretion rate and the mass distribution, and we return to each in turn in section \ref{sec:discussion}.}

\revision{We find that, despite all of these possibilities, the long-term features of the evolution of the bound gas are well described, to within a factor $\sim 2$, by General Relativistic Thin Disk Theory (GRTDT), with a single stress amplitude measured directly from the simulation. This agreement holds even though the disk in our simulations is not thin, launches a wind, and has dynamically dominant (MAD) mean magnetic fields in its inner regions. We argue that this is because the bulk of the evolution is set by the global constraints of mass and angular momentum conservation, together with a turbulent stress which is, empirically, local, independent of density and only weakly dependent on radius.}

The layout of the paper is the following. In section \ref{sec:summarystats} we introduce the main ``summary'' parameters which describe the bulk evolution of an accretion flow, with which we shall perform the majority of the analysis in this paper. In section \ref{sec:grmhd} we introduce the GRMHD simulations of long term evolution of a torus presented by~\citet{Guo2025}, and discuss the features displayed by the long time evolution of the system. In section \ref{sec:analysis} we discuss how this evolution can be interpreted on a qualitative and quantitative level. We first introduce global conservation laws of mass and angular momentum, which appear to drive much of the bulk of the disk evolution, before introducing a much more detailed (but still much simplified) theory of time evolving thin relativistic disks. In section \ref{sec:comparison} we show that these thin disk theories reproduce many of the features of the far more complicated GRMHD simulations, before discussing the failures, successes and observational implications of these results in section \ref{sec:discussion}. We conclude with an outlook to the future in section \ref{sec:conclusions}, and include some technical results pertinent to the discussion in \revision{Appendices \ref{sec:global_app}, \ref{app:sol}, \ref{app:plots} and \ref{app:cyclic}. Appendix \ref{app:cyclic} presents a direct comparison of the standard and cyclic zoom simulations over the period in which both are available.} 


\section{Quantities of interest}\label{sec:summarystats}
In this paper we are seeking to understand the long-time evolution of a black hole accretion torus. We shall ultimately analyze such systems using two  distinct but (hopefully) complementary approaches: full numerical solutions of the equations of (ideal) General Relativistic Magnetohydrodynamics (GRMHD), and analytical solutions of a much simpler theory, namely time-dependent General Relativistic Thin Disk Theory (GRTDT). Before we introduce either approach, or the results of our simulations,  we provide an overview here of the main quantities of interest for describing an accretion flow (in both theories), and how they will be computed. 

The accretion disks in our simulations will be characterized by a finite initial mass supply $M_0$, and a corresponding finite angular momentum budget $J_0$. As they evolve, mass and angular momentum will be (turbulently) redistributed within the flow, some material (and angular momentum) will cross over the black hole's event horizon and some will be launched into outflows (winds, or in principle into jets). 

On the global level an accretion disk can be well characterized by two quantities, which we shall denote $M'$ and $\dot M$.  The first quantity of interest is the ``mass distribution'' of the disk at a given time, namely 
\begin{equation}
    M'(r, t) \equiv {{\rm d}  M_{\rm disk}\over {{\rm d} \log r}} .
\end{equation}
We choose a logarithmic radial interval here as the accretion flows in our simulations will become significantly radially extended at large times\revision{ (in all figures $M'$ is shown per decade in radius)}. In effect $M'$ quantifies {\it where} the bulk of the material in an accretion flow is at a given time. 

The second quantity of interest is the ``mass accretion rate'', or 
\begin{equation}
    \dot M(t) \equiv \oiint{\left.\left(\sqrt{|g| } \rho U^r \right)\right|_{r=r_+,\,\, t=t} \, {\rm d}\theta \, {\rm d}\phi}, 
\end{equation}
where $U^r$ is the radial velocity of the flow, $|g|$ the (absolute value of) the metric determinant and $\rho$ the density of the gas. 
This quantifies how much material is flowing onto the black hole (with horizon radius $r_+$) at a given time $t$, and in the absence of any outflows would equal the rate of change of the disk mass ${\rm d}M_{\rm disk}/{\rm d}t$\revision{ (see section \ref{sec:massloss}; we plot $|\dot M|$ throughout)}. As the luminosity of black hole accretion flows (although not explicitly captured in ideal GRMHD) is dominated by the innermost regions, this is a reasonable proxy for the luminosity emitted from the disk at a given time (and therefore has obvious observational significance). 

While $\dot M$ is quantified by a single number at each time, $M'$ is a more complicated variable as it has both radial and temporal structure. For ease of comparison we shall also define three ``summary statistics'' describing the variable $M'$, namely 
\begin{align}
    M'_{\rm max}(t) &\equiv {\rm the \, maximum\, value\, of }\, M'(r, t), \\
    r_{\rm max}(t) &\equiv {\rm the\, radius\, at\, which\, this\, maximum\, occurs}, \\
    \sigma_{\rm eff}(t) &\equiv {\rm the\, radial\, extent\, of\, one} \nonumber \\ &\quad\quad {\rm\, ``standard\, deviation"\, of \, }M'(r, t), 
\end{align}
or to be more explicit $2\sigma_{\rm eff}(t) \equiv (r_2 - r_1)$, where $r_1$ and $r_2$ are the two radii at which the mass distribution falls to $M'(r, t) = M'_{\rm max}/\sqrt{e}$, and $r_2 > r_1$. In other words  $r_{\rm max}$ and $\sigma_{\rm eff}$ \revision{are Gaussian-equivalent effective quantities: they are the mean and standard deviation of a Gaussian with the same peak ($M'_{\rm max}$) and the same ``one standard deviation width'' as the mass distribution, although in reality the distributions themselves are not Gaussian}. 

These four summary statistics $\dot M, M'_{\rm max}, r_{\rm max}$ and $\sigma_{\rm eff}$ will form a major part of our analysis of the long time evolution of black hole accretion flows. 


\section{Long term disk evolution in GRMHD}\label{sec:grmhd}
In this section we briefly overview the governing equations of GRMHD and our implementation of numerical solvers, before discussing the gross features of the long time evolution seen in our simulations. 

\subsection{The fundamental equations of ideal GRMHD}

In GRMHD one solves the coupled equations of mass conservation, energy-momentum conservation, and the Maxwell equations in a spacetime described by a given metric $g_{\mu\nu}$. Explicitly, mass conservation can be written 
\begin{equation}
    \nabla_\mu (\rho U^\mu) = 0, 
\end{equation}
where $\rho$ is the rest mass density of the disk, $U^\mu$ is the fluid 4-velocity, and $\nabla_\mu$ is a covariant derivative with respect to coordinate $x^\mu$. Energy-momentum conservation is described by 
\begin{equation}
    \nabla_\mu T^{\mu\nu} = 0, 
\end{equation}
where the stress energy tensor of ideal GRMHD, takes the following form
\begin{multline}
    T^{\mu \nu} = \left(\rho + {P+e + b^2/4\pi\over c^2} \right) U^\mu U^\nu + \left(P + {b^2 \over 8\pi}\right) g^{\mu \nu} \\ - {b^\mu b^\nu \over 4\pi } ,
\end{multline}
which introduces the thermal pressure $P$, internal energy density $e$, and the magnetic field 4-vector $b^\mu$ (with $b^2 \equiv b^\mu b_\mu$). The evolution equation for the magnetic field 4-vector is 
\begin{equation}
    \nabla_\mu \widetilde F^{\mu\nu} = 0,
\end{equation}
where $\widetilde F^{\mu \nu}$ is the dual of the Faraday tensor. In ideal MHD (where we make the approximation that the electric field vanishes in an appropriate frame, or explicitly $U_\mu F^{\mu\nu} = 0$), this dual tensor can be expressed as
\begin{equation}
    \widetilde F^{ \mu \nu} = b^\mu U^\nu - b^\nu U^\mu .
\end{equation}
These equations are of course implicitly dependent on the structure of the underlying spacetime, which in our work will be the Kerr metric. Many different coordinate representations of the Kerr metric exist, but in Boyer-Lindquist coordinates we have 
\begin{multline}\label{metric}
    {\rm d}s^2 = g_{\mu\nu}\, {\rm d}x^\mu {\rm d}x^\nu = - \left(1 - {2r_\mathrm{g} r \over r^2 + a^2 \cos^2\theta }\right) c^2 {\rm d}t^2 \\ - {4 r_\mathrm{g} r a c \sin^2 \theta \over r^2 + a^2 \cos^2\theta} \, {\rm d}t \, {\rm d}\phi \\ + {r^2 - 2r_\mathrm{g}r + a^2 \over r^2 + a^2 \cos^2\theta }\, {\rm d}r^2 + (r^2 + a^2 \cos^2 \theta ) \, {\rm d}\theta^2  \\ + \left(r^2 + a^2 + {2 r_\mathrm{g} a^2 r \sin^2 \theta \over r^2 + a^2 \cos^2 \theta } \right)\sin^2 \theta \, {\rm d}\phi^2 ,
\end{multline}
where $r_\mathrm{g} \equiv GM_\bullet / c^2$ is the gravitational radius, $M_\bullet$ is the black hole mass, and $a$ is the angular momentum constant (with dimensions of length) of the black hole $a = J_\bullet/M_\bullet c$, where $J_\bullet$ is the total angular momentum of the black hole. The coordinates are $t$, the time as measured at infinity, and the three spatial coordinates $(r, \theta, \phi)$ which have their usual quasi-spherical interpretations. As ideal GRMHD is a scale invariant theory we shall in the following write all distances in units of $r_\mathrm{g} = GM_\bullet /c^2$ and all times in units of $t_\mathrm{g} = GM_\bullet/c^3$. 

\subsection{Implementation and numerical solutions}
The full equations of GRMHD are far too complex to be solved analytically, and one must work instead with numerical solvers.  We work within the ideal GRMHD framework, solving the fluid equations using the code \athenak{} \citep{Stone26}, which is a performance portable version of a previous code  \athenapp{} \citep{Stone20, White16}. \athenak{} uses Cartesian Kerr-Schild coordinates, which are different from the metric introduced above. These coordinates have the advantage of behaving well at the horizon and the poles, unlike Boyer-Lindquist coordinates. We shall, however, present simulation output in Boyer-Lindquist $(t, r, \theta, \phi)$ coordinates for comparison between models, as they are most readily physically interpretable. In our simulations, we adopt the piecewise parabolic (PPM4) spatial reconstruction method, the HLLE Riemann solver, and the RK2 time integrator to solve the GRMHD equations. The adaptive mesh refinement (AMR) in \athenak{} allows us to flexibly achieve a high resolution and good performance over an extremely large dynamic range. We apply the first-order flux correction algorithm \citep[described by][]{Lemaster2009} in the rare cases that the higher-order fluxes would lead to negative density or pressure.

While for some analysis we use ``standard'' GRMHD (described above), in others, to speed up the simulation, we apply the cyclic zoom method, which was presented and tested in detail in \citet{Guo2025}. In brief, we cyclically zoom out (de-refine) and zoom in (refine) the simulation domain.
The spatial resolution is correspondingly decreased (increased) when we zoom out (in) using AMR to alleviate the timestep constraints while keeping the same relative resolution $\Delta x/x$ for the region of interest. We use a mask region in the center containing a careful treatment of the coarsened fluid variables to preserve as much of the small-scale physics as possible. 
In the Cartesian coordinates we use, the mask region is a region of cells within a certain radius $r_{\mathrm{z},i}\equiv 2^i\,r_\mathrm{g}$ for level $i$, where $i\in\{0,1,...,(n-1)\}$.  In particular, the method evolves the magnetic field within the de-refined region to preserve the divergence-free constraint using inductive electric fields that also preserve information about the outgoing feedback power (Poynting flux) from small scales. The method behaves like a ``$\Lambda$-cycle'' in the space-time diagram (or ``V-cycle'', depending on what we choose as the starting point of the simulation). A complete simulation typically consists of tens to hundreds of $\Lambda$-cycles.

Following \citet{Narayan2022}, we perform simulations of magnetically arrested disks (MAD) around a non-spinning black hole. We initialize the simulations with the \citet{Fishbone76} torus solution, which is determined by four parameters: the inner edge of the torus $r_\mathrm{in}$, the location of the pressure maximum $r_\mathrm{peak}$, the adiabatic index of the fluid $\gamma_\mathrm{ad}$, and the maximum density $\rho_\mathrm{max}$. The location of the outer edge of the torus is a function of these parameters. We set $r_\mathrm{in}=20\,r_\mathrm{g}$ and adjust $r_\mathrm{peak}=34.49\,r_\mathrm{g}$ so that $r_\mathrm{out}\approx100\,r_\mathrm{g}$. The initial torus mass is concentrated around $\sim 50\, r_\mathrm{g}$. Such a radially concentrated torus produces significant secular evolution on resolvable timescales (studying this evolution is naturally the purpose of this work). We adopt an ideal gas equation of state with an adiabatic index of $\gamma_\mathrm{ad}=13/9$.

The magnetic field is initialized with a weak single large magnetic field loop defined by the poloidal vector potential $A_\phi$:
\begin{equation}
\begin{aligned}
    A_\phi &= \max(0,q),\\ 
    q &= \left[\left(\frac{\rho}{\rho_\mathrm{max}}\right)\left(\frac{r}{r_\mathrm{in}}\right)^3e^{-r/r_\mathrm{mag}}\sin^3\theta\right]-A_{\phi,\mathrm{cut}},
    \label{eq:torus_Aphi}
\end{aligned}
\end{equation}
where we set $r_\mathrm{mag}=400\,r_\mathrm{g}$ and $A_{\phi,\mathrm{cut}}=0.2$. After determining the field components via the vector potential of \ref{eq:torus_Aphi}, we normalize the initial magnetic field strength in the disk following \citet{Porth19} such that the maximum gas pressure and the maximum magnetic pressure in the torus (which do not necessarily occur at the same location) satisfy $\beta_\mathrm{max}\equiv(p_\mathrm{gas})_\mathrm{max}/(p_\mathrm{mag})_\mathrm{max}=100$. The internal energy is perturbed by white noise with a maximum amplitude of $2\%$ to break symmetry and excite the magnetorotational instability (MRI) inside the torus.

To study the secular evolution, we perform two runs: a standard run and a cyclic zoom run. We run the standard simulations for a very long time of $10^6\, t_\mathrm{g}$.
For the cyclic zoom run, we set $n=8$ and the runtime
\begin{equation}
    t_{\mathrm{run},i}=
    \begin{cases}
      500\,t_\mathrm{g} & \text{if $i=0$},\\
      0 & \text{if $i=1,2$},\\
      100\, t_{\mathrm{z},i} & \text{otherwise},
    \end{cases}
\end{equation}
where $t_{\mathrm{z},i}=r_{\mathrm{z},i}/c = 2^i r_g/c$.
After $10^6 \,t_\mathrm{g}$, as the torus expands to a larger scale, we increase the number of levels from $n=6$ to $n=8$ to further speed up the simulation. Then, after $3\times10^6\, t_\mathrm{g}$, we again increase $n$ to $n=10$ and evolve it to $10^7\, t_\mathrm{g}$. \revision{In the analysis below, we use the data from the standard run for $t\leq 10^6\,t_\mathrm{g}$, which provides a higher output cadence, and the data from the cyclic zoom run for $t>10^6\,t_\mathrm{g}$, which captures the longer-term evolution. The cyclic zoom method was validated against a standard simulation of this torus configuration over $10^6\,t_\mathrm{g}$ in \citet[][their section 3.5]{Guo2025}. To make the present paper self-contained, we present in Appendix \ref{app:cyclic} a direct comparison of the two runs analyzed here over the period in which both are available. The bound mass distributions of the two runs agree to within $\sim 10$--$20\%$ in amplitude, total mass and width, and the horizon accretion rates agree to within a factor $\sim 1.5$, comparable to the level of stochastic variability in either run. The cyclic zoom run therefore reproduces the standard run to an accuracy well within the level of agreement between either run and the analytical theory developed below. The one exception is short-timescale variability: because the horizon is resolved only intermittently in the cyclic zoom run, all of our analysis of accretion rate variability (section \ref{sec:variability}) uses the standard run alone. Beyond $10^6\,t_\mathrm{g}$ no standard run is available for comparison, and we discuss the limitations of the method relevant to this regime in Appendix \ref{app:cyclic}.}




\subsection{Long time behavior of the solutions}

\begin{figure}
    \centering
    \includegraphics[width=\linewidth]{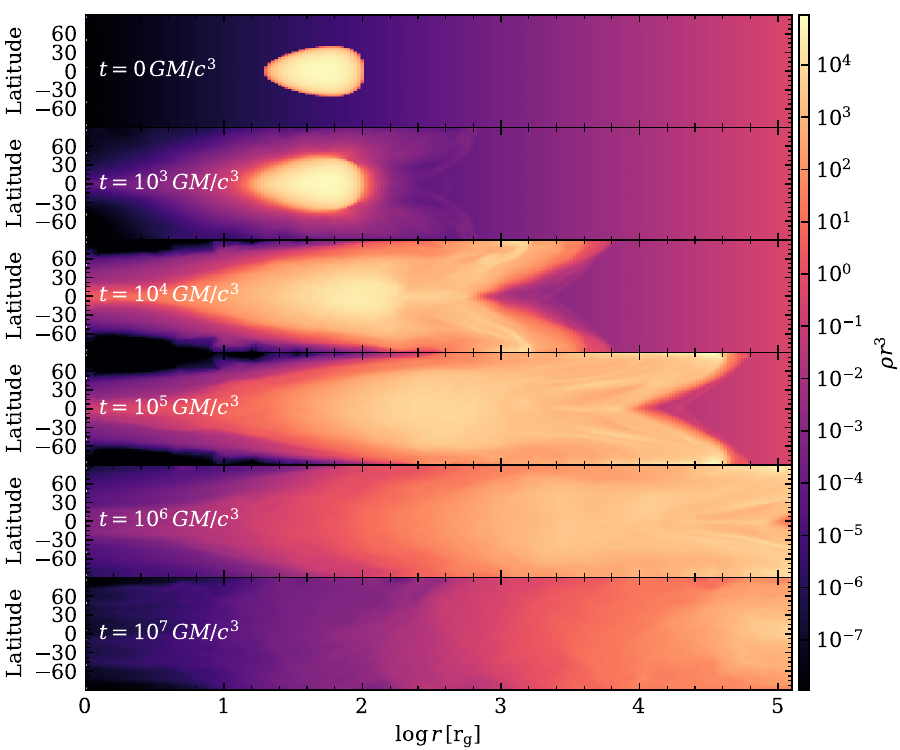}
    \caption{Azimuthally averaged distribution of mass, i.e., $M' \propto \rho r^3$, at different times. The initial, compact, Fishbone \& Moncrief torus is visible in the upper panel, followed by the subsequent evolution of the flow (later times shown on progressively lower panels). The horizontal axis of this plot shows the logarithm of the disk radius, while the vertical axis shows the latitude (defined as $L=180^\circ(\pi/2-\theta)/\pi$). 
    The solutions show a gradual decay in the magnitude of the density with time, and a spreading of the outer edge of the disk to larger radii. The aspect ratio of the flow remains broadly constant with time, and the flow (which is decidedly not thin) remains restricted to a latitude range of roughly $-60^\circ < L < 60^\circ$. At the very latest times $t\sim 10^7 t_\mathrm{g}$ (lowest panel), the disk is less obvious.}
    \label{fig:dens}
\end{figure}

\begin{figure}
    \centering
    \includegraphics[width=\linewidth]{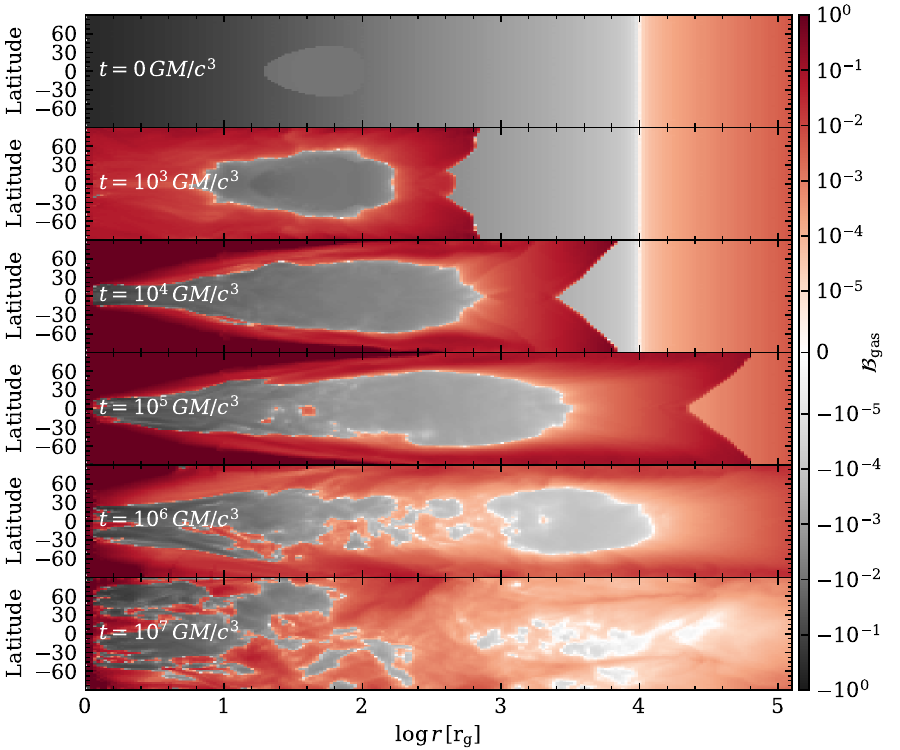}
    \caption{Azimuthally averaged hydrodynamic Bernoulli parameter, $\mathcal{B}_\mathrm{gas}$, at different times in the disk evolution (see text for precise definition). The times shown here correspond to the same evolutionary stages as in the density plot (Figure \ref{fig:dens}). The Bernoulli parameter allows ``disk'' regions to be clearly identified, while the outflowing gas can be clearly seen by the propagation of unbound material above and below the disk torus. The spreading of the disk outer edge can be clearly seen in this Figure. At the latest times the whole gas structure is at most tenuously bound, indicating an end of a normal ``disk'' phase of evolution.   }
    \label{fig:bern}
\end{figure}

\begin{figure*}
    \centering
    \includegraphics[width=0.49\linewidth]{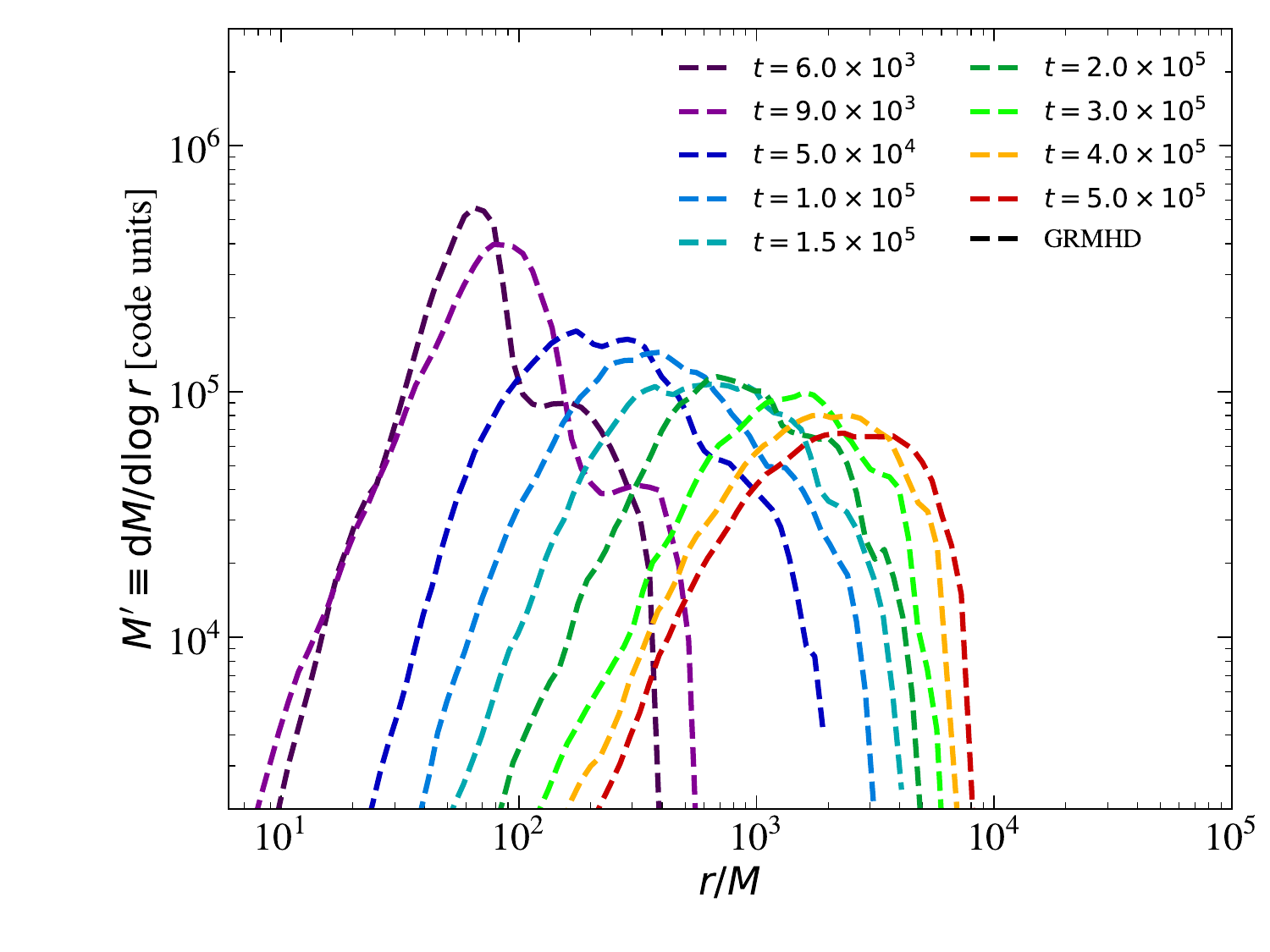}
    \includegraphics[width=0.49\linewidth]{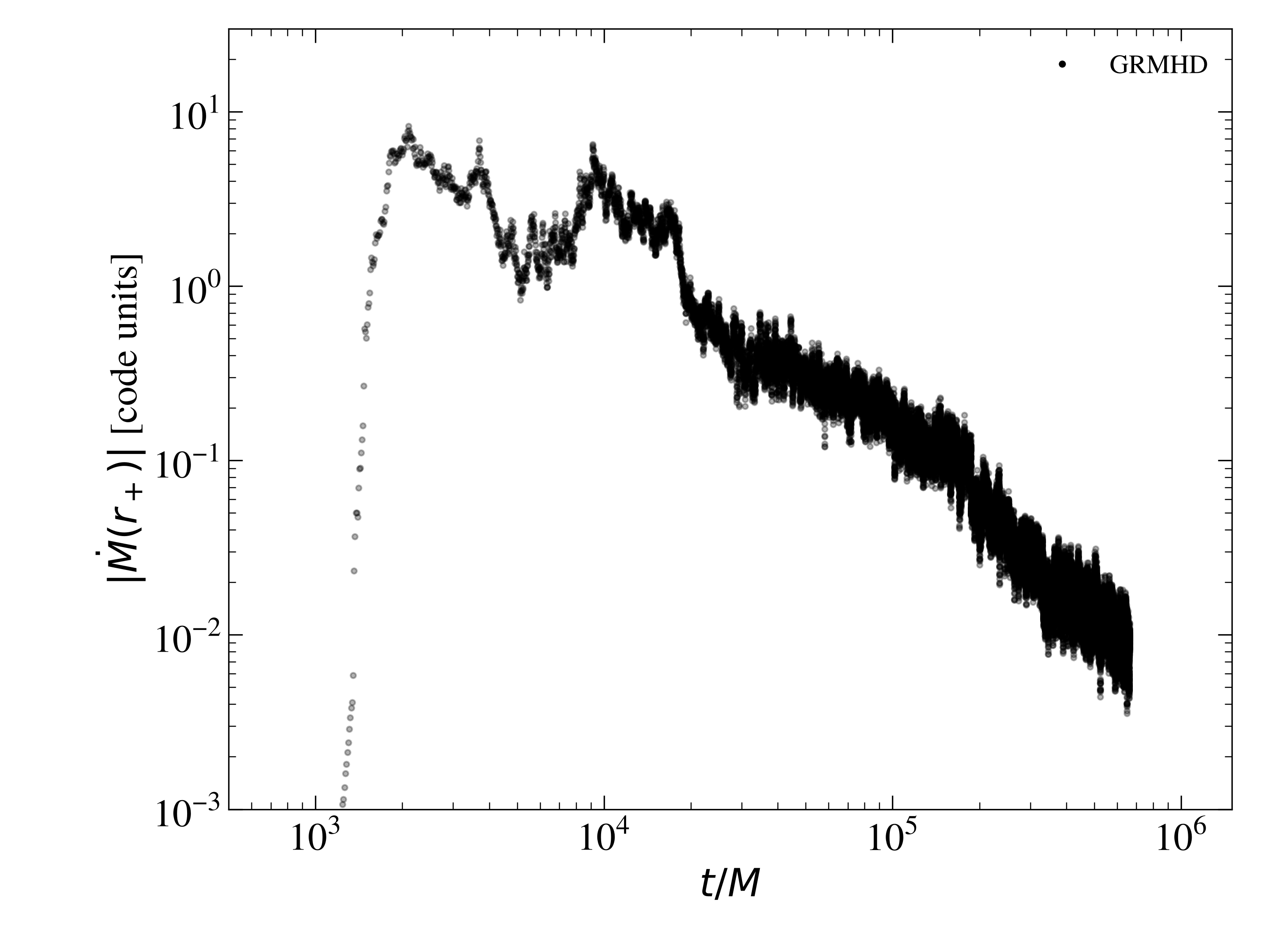}
    \includegraphics[width=0.49\linewidth]{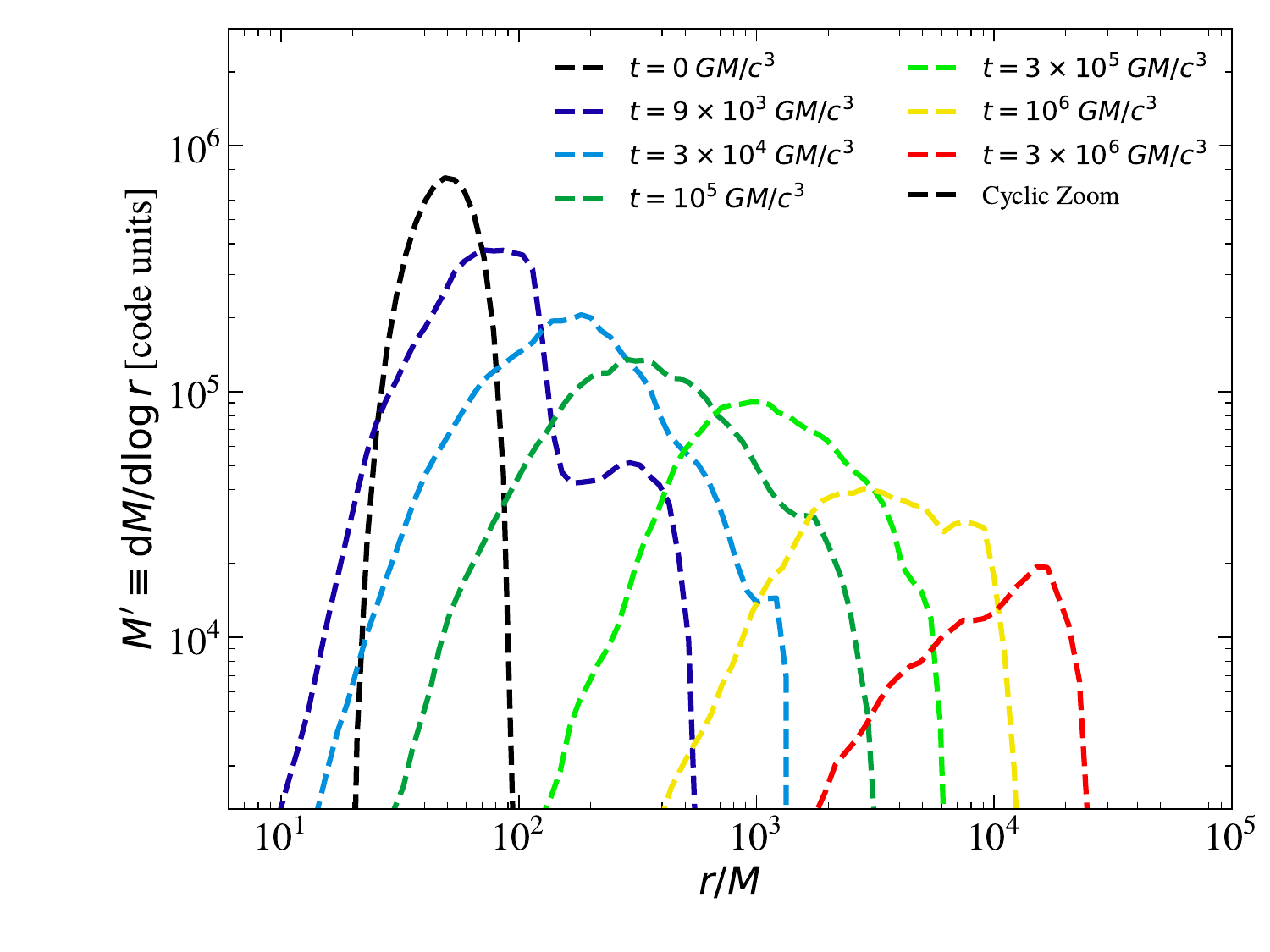}
    \includegraphics[width=0.49\linewidth]{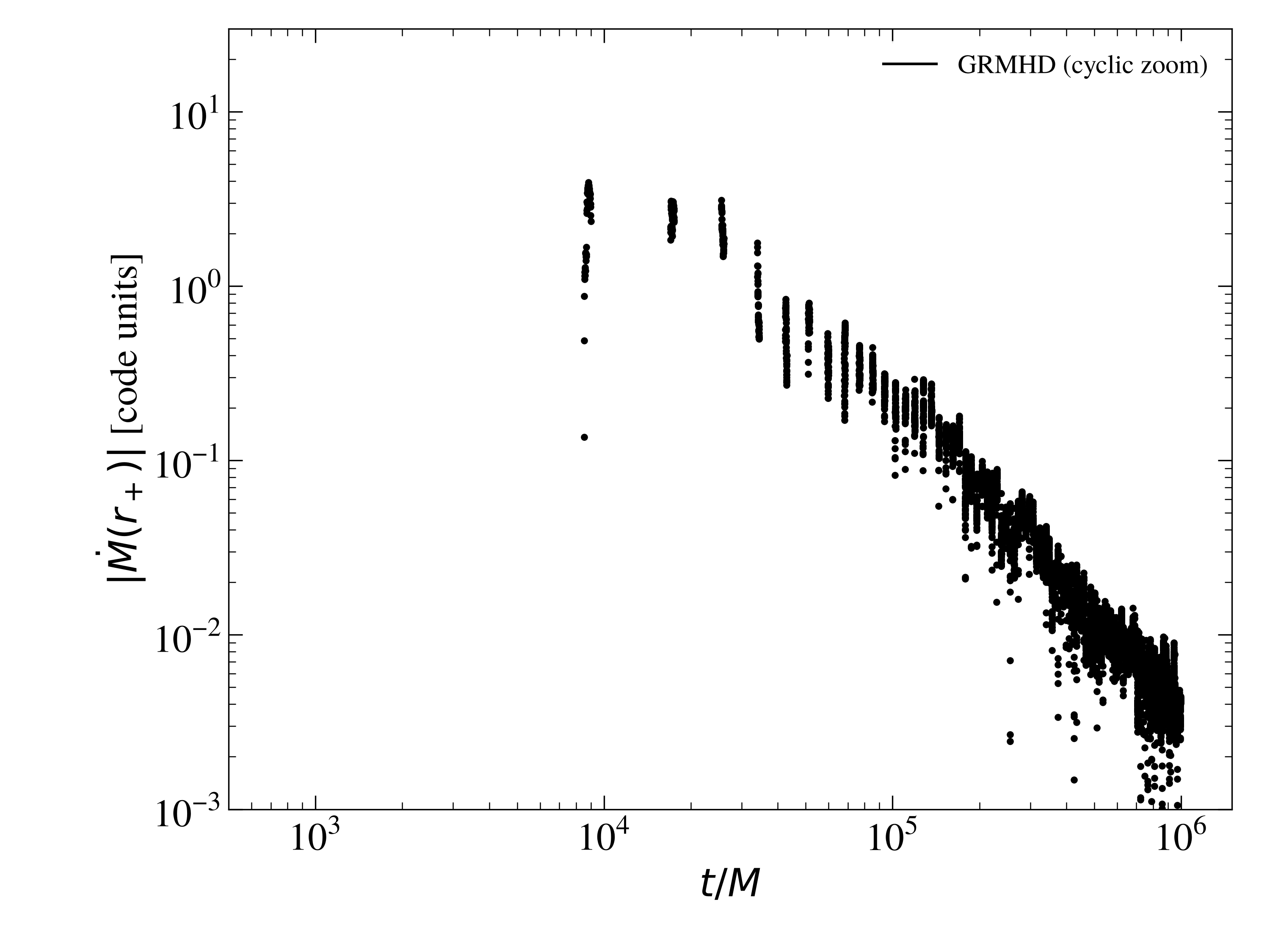}
    \caption{\revision{The evolution of the accretion flow in our GRMHD simulations. {\bf Left column:} the evolution of the bound disk mass distribution $M'$ as a function of time, for the standard GRMHD simulation (upper) and the cyclic zoom simulation (lower). Only bound mass is used in computing this distribution. In both runs the amplitude of $M'$ decays, while its peak moves outwards and its width grows. Times are labeled in each panel; the standard run is shown from $t = 6\times10^3\,t_\mathrm{g}$ to $t = 5\times10^5\,t_\mathrm{g}$. {\bf Right column:} the accretion rate onto the black hole in both simulations. In the cyclic zoom run the horizon is resolved only during the finest-level phase of each cycle, and the accretion rate is shown only for those phases (hence the clustering of points). In both simulations the accretion rate at times $t \gtrsim 10^4\,t_\mathrm{g}$ is well approximated by a power-law decay with time. A direct overlay of the two runs is shown in Figure \ref{fig:cyclic_compare}.}}
    \label{fig:long_time_grmhd}
\end{figure*}

In Figure \ref{fig:dens} we show the evolving density (multiplied by $r^3$ to give a variable proportional to $M'$) for the GRMHD run, as a function of (the logarithm of) the radius plotted against vertical angular coordinate (this is a $2\pi$ azimuthal average at each time step). We see clear global features of this evolution. Over the first $\sim 10^3\,t_\mathrm{g}$ there is minimal change to the density structure, as the magnetorotational instability \citep{BalbusHawley91} takes a few orbital timescales\footnote{Which is roughly equal to $t_{\rm orb} = 2\pi /\Omega = 2\pi (r/r_{\rm g})^{3/2} t_{\rm g}\sim 10^3\,t_\mathrm{g} $ at the relevant radius of the initial density maximum.} to initiate a turbulent evolution and redistribute material within the flow. On timescales $t\gg t_{\rm orb}$ the flow stretches down to the event horizon (the left hand edge of our plot), with an outer edge that spreads with time. We separate the mass into bound ($\mathcal{B}_\mathrm{gas}<0$) and unbound ($\mathcal{B}_\mathrm{gas}>0$) using the hydrodynamic Bernoulli parameter:
\begin{equation}
    \mathcal{B}_\mathrm{gas}=-U_t(\rho+e+P)/\rho-1.
\end{equation}

Naturally, accompanying this accretion onto the black hole is a gradual decay in the magnitude of the density with time (as material is lost into the black hole and to outflows), and a gradual spreading of the outer edge of the disk to larger and larger radii (to conserve the angular momentum of the accreted material). The aspect ratio of the flow remains broadly constant with time, and the flow (which is decidedly not thin) remains restricted to a poloidal range with a latitude of roughly $-60^\circ \lesssim L \lesssim 60^\circ$ (defined as $L=180^\circ(\pi/2-\theta)/\pi$). Accompanying the disk evolution is the gradual launching of an outflow (in this case a disk wind not a jet – possibly because the spin of the black hole is zero), this unbinds a noticeable fraction of the material (see Figure \ref{fig:bern} for a plot of the hydrodynamic Bernoulli parameter of the gas), with at late times an order $\sim {\cal O}(1)$ fraction of the mass unbound~\citep{Guo2025}. These lobe-like features are roughly symmetric with respect to the equatorial plane, and appear to be somewhat decoupled from the evolution of the rest of the disk\revision{, although they remove a substantial fraction of its mass (section \ref{sec:massloss})}. 

For the remainder of our analysis we shall restrict our attention to the mass which is {\it bound to the black hole}, defining any gas with $\mathcal{B}_\mathrm{gas} < 0$ to be part of ``the disk''. This therefore prevents our analysis of the evolving disk density structure from being contaminated by material which has been launched in an outflow, which inevitably is described by distinct physical processes from the bound, turbulently evolving, disk material. A discussion of attempts to analytically model these outflows is postponed for later sections. 

One very interesting result shown in both Figures \ref{fig:dens} and \ref{fig:bern} is that at the very latest times $(t \sim 10^7 t_\mathrm{g})$ the ``disk'' itself is no longer well defined, with the remaining gas having an extremely low density (Figure \ref{fig:dens}) and no longer remaining bound to the black hole (Figure \ref{fig:bern}). It is not clear what drove this evolutionary change, \revision{and we cannot exclude a contribution from limitations of the cyclic zoom method in this regime (Appendix \ref{app:cyclic}),} 
but this final drop in density of factor $\sim 10^3$ will clearly not be reproduced by conventional disk models. We restrict future comparisons between GRMHD and analytical theory to times prior to this dramatic transition. 

Figure \ref{fig:long_time_grmhd} shows the long term evolution of the two summary variables $M'$ and $\dot M$ of the accretion disks in the case of traditional GRMHD (upper) and the extended cyclic-zoom analysis (lower). We reiterate that only bound material is included in our definition of $M'$.  Visually we see a number of striking, and important, features of this long term evolution. First, both the classical GRMHD and cyclic-zoom runs have horizon accretion rates which clearly approach a power-law fall off with time for times $t \gg t_{\rm rise} \sim 10^4 t_\mathrm{g}$ (note that both axes are plotted on logarithmic scales in Figure \ref{fig:long_time_grmhd}).  

Second, there is a clear and consistent evolution of the mass distribution $M'(r, t)$: the amplitude $(M'_{\rm max})$ drops with time, while the location of this radial maximum $(r_{\rm max})$ grows (this can also be clearly seen in the density plot; Figure \ref{fig:dens}). The effective width of this mass distribution $(\sigma_{\rm eff})$ also shows substantial growth with time (growing from $\sigma_{\rm eff} \sim 10^2r_\mathrm{g}$ to $\sigma_{\rm eff} \sim 10^4r_\mathrm{g}$ over the course of the evolution). While it is not necessarily clear from this Figure, the evolution of $M'_{\rm max}, r_{\rm max}$ and $\sigma_{\rm eff}$ are themselves all approximately well described by a powerlaw temporal evolution (see Figure \revision{\ref{fig:summarystats}} for the explicit profiles). The fact that all four summary statistics are well described by powerlaw profiles suggests that their gross features can be determined from simple arguments. 

Now that such gross features of the evolution are clear, we turn to a theoretical framework through which this evolution can be understood. 

\section{Interpreting this long time evolution}\label{sec:analysis}
There is clear, and apparently readily understandable, structure to the long term evolution seen in the GRMHD simulations of the preceding section. In short, the horizon accretion rate, and the amplitude, location and width of the mass distribution all show evolution consistent with simple powerlaw profiles. It is the purpose of this section to derive, on physical grounds, the expected evolution of these properties from simplified versions of the full GRMHD equations. 

We begin with an analysis of the global conservation laws of black hole accretion flows, which will allow simple scaling relationships to be derived for some of our quantities, before turning to a more detailed theory for a comparison to the global evolution of $\dot M$ and $M'$.   {We stress at the outset that the theory developed here does not include the impacts of disk outflows on the evolution, and so provides a qualitatively good, but not perfectly quantitatively accurate, description of the thick disk GRMHD simulation in which outflows remove an ${\cal O}(1)$ fraction of the disk mass at late times.   Generalization of this theory to include outflows is in principle straightforward, as we discuss in \S \revision{\ref{sec:massloss}}, but we leave this to future work.}

\subsection{A global conservation law analysis}
Before we perform a detailed comparison to General Relativistic Thin Disk Theory (GRTDT), we begin by highlighting how far one can get in understanding the gross features seen in the above GRMHD simulations with a simple global conservation law analysis. 

One must be somewhat careful when constructing global conservation laws in general relativity, as in general the integral of a covariant quantity is not itself a covariant quantity. This can be simply accounted for however (we recap how this is done in the Appendix \ref{sec:global_app}), and in this section we simply use the main results of this analysis. 

Starting from the local expression of mass conservation in the fluid 
\begin{equation}
    \nabla_\mu (\rho U^\mu) = 0,
\end{equation}
 one can construct a global mass conservation constraint by integrating this expression over the four dimensional spacetime volume defined by ${\cal V} = [0, t] \times [r_+, \infty] \times [0, \pi] \times [0, 2\pi]$ (i.e., spherical in space from the horizon to infinity and linear in time as measured at infinity from $t=0$ to the current time $t$). If one assumes that the density vanishes sufficiently quickly at infinity (i.e., the mass budget is finite), and defines 
\begin{equation}
    M_d(t) \equiv \int_{r_+}^\infty \int_0^{\pi} \int_0^{2\pi} \left.\left(\sqrt{|g|} \rho U^t\right)\right|_{t=t}\, {\rm d}\phi \, {\rm d}\theta \, {\rm d}r ,
\end{equation}
and 
\begin{equation}
    \dot M \equiv \int_0^{\pi} \int_0^{2\pi} \left.\left(\sqrt{|g|} \rho U^r\right)\right|_{r=r_+, \, t=t}  \,  {\rm d}\phi \, {\rm d}\theta ,
\end{equation}
then global mass conservation can be simply expressed as 
\begin{equation}\label{mass_con}
    M_d(t) - M_d(0) = \int_0^t \dot M(t') \, {\rm d}t'. 
\end{equation}
A similar expression can be derived for global angular momentum conservation. The Kerr metric is independent of the coordinate $\phi$, meaning that there is a Killing vector $k^\mu = \delta^\mu_\phi$ which by definition satisfies $\nabla_\mu k_\nu + \nabla_\nu k_\mu = \partial_\phi g_{\mu\nu} = 0$ (i.e., $\nabla_\mu k_\nu$ is an anti-symmetric tensor). The angular momentum 4-vector 
\begin{equation}
    J^\mu \equiv k_\nu T^{\mu \nu} ,
\end{equation}
where $T^{\mu\nu}$ is the stress energy tensor, satisfies 
\begin{equation}
    \nabla_\mu J^\mu = T^{\mu\nu} \nabla_\mu k_\nu + k_\nu \nabla_\mu T^{\mu\nu} = 0,
\end{equation}
where the first term vanishes on symmetry grounds and the second by the conservation of energy momentum. By integrating this local conservation law over the same global spacetime volume ${\cal V}$, we find 
\begin{equation}\label{j_cons}
    J_d(t) - J_d(0) = \int_0^t \dot J(t') \, {\rm d}t'. 
\end{equation}
In this expression
\begin{multline}
    J_d(t) \equiv \int_{r_+}^\infty \int_0^{\pi} \int_0^{2\pi} \Bigg(\sqrt{|g|}  \Bigg[\rho U^t U_\phi \left(1 + {c_s^2 + v_A^2 \over c^2} \right) \\ - {b^tb_\phi\over 4\pi} \Bigg] \Bigg)\Bigg|_{t=t}  \, {\rm d}\phi \, {\rm d}\theta \, {\rm d}r ,
\end{multline}
and 
\begin{multline}
    \dot J \equiv \int_0^{\pi} \int_0^{2\pi} \Bigg(\sqrt{|g|} \Bigg[\rho U^r U_\phi  \\ \left(1 + {c_s^2 + v_A^2 \over c^2} \right) \Bigg] \Bigg) \Bigg|_{r=r_+, \, t=t} \, {\rm d}\phi \, {\rm d}\theta , 
\end{multline}
where $c_S^2 \equiv (P+e)/\rho$ is the sound speed (up to irrelevant factors of $\gamma_{\rm ad}$), and $v_A^2 \equiv b^2 /4\pi\rho$ is the Alf\'ven velocity. Note here our first deviation from classical Newtonian disk theory, which has $\dot J = 0$, as any fluid that reaches $r=0$ (the ``inner edge'' of a Newtonian disk) must have zero angular momentum in a Newtonian theory of gravity, while in relativity fluid elements can reach $r=0$ with a finite angular momentum $(\dot J \neq 0)$. This finite specific angular momentum which the fluid carries over the horizon is to leading order that of a circular orbit at the innermost stable circular orbit. 

At the moment equations \ref{mass_con} and \ref{j_cons} are exact, but we can make progress by making a number of simplifying (but well motivated) assumptions. The first is to postulate that, at times $t > t_\star$, one has 
\begin{equation}
    \dot M(t) = -{(n-1)M_d(0) \over t_\star} \, \left({t\over t_\star}\right)^{-n} , \quad t \geq t_\star, 
\end{equation}
where the normalisation amplitude is chosen so that as $t/t_\star \to \infty$ all material is eventually accreted (of course saying that only a factor $f$ is accreted, with the rest expelled in outflows, would not change any of the asymptotic scalings) and $t_\star$ is some currently arbitrary evolutionary timescale. At this stage this would simply represents a postulate (one which is clearly motivated from the simulations), but in fact this will be shown to be a prediction for the mass accretion rate in any simplified theory of accretion which has local, scale-free, transport of angular momentum. 

The second postulate one can motivate from the observed properties of the evolving mass distribution in the GRMHD simulations, namely that $M'(r, t)$ is (on a logarithmic radial scale), well approximated by a strongly peaked function around some time-dependent location $r_{\rm max}(t)$, with time-dependent amplitude $M'_{\rm max}(t)$ and width $\Delta$, which {\it on a logarithmic scale} is broadly time independent (again, $\partial_t \Delta \approx 0$ will be shown to be a prediction of a more detailed theory). Then 
\begin{equation}
    M_d(t) = \int_{\log r_+}^\infty M' \, {\rm d}\log r \approx M'_{\rm max} \Delta .
\end{equation}
Our final assumptions are to neglect corrections to the energy density $(c_S^2+v_A^2)/c^2 \ll 1$, and the angular momentum of the magnetic field $b^tb_\phi/4\pi \ll \rho U^t U_\phi$, and to assume that $U_\phi$ is, to leading order, set only by the radius $U_\phi \approx \overline U_\phi(r)$. This would be the case for fluid elements following approximately Keplerian orbits. Then 
\begin{multline}
    \dot J(t) \approx  \overline U_\phi(r_+)\int_0^{\pi} \int_0^{2\pi}\left. \left(\sqrt{|g|} \rho U^r \right)  \right|_{r=r_+} \, {\rm d}\phi \, {\rm d}\theta \\ = \overline U_{\phi}(r_+) \dot M(t) , 
\end{multline}
and 
\begin{multline}
    J_d(t) \approx \int_{\log r_+}^\infty \left. \left( \overline U_\phi  M' \right)\right|_{t=t}  \, {\rm d}\log r \\ = \overline U_{\phi}(r_{{\rm max}}) M'_{\rm max} \Delta  . 
\end{multline}
These equations can then be solved for the time evolution of $M'_{\rm max}$ and $r_{\rm max}$. Specifically, our approximations mean that mass conservation instantly becomes 
\begin{equation}
    M'_{\rm max}(t) \approx {M_d(0)\over \Delta }\left({t \over t_\star}\right)^{1-n}, \quad t\geq t_\star . 
\end{equation}
In addition, global conservation of angular momentum becomes 
\begin{multline}
    J_d(t) - J_d(0) \approx \overline U_\phi(r_{\rm max}) M'_{\rm max} \Delta - J_d(0) \\ = \int_0^t \dot J(t') \, {\rm d}t' \approx \overline U_\phi(r_+) M_d(0) \left[\left({t \over t_\star}\right)^{1-n} - 1\right] . 
\end{multline}
This can be inverted with the help of mass conservation 
\begin{equation}
    \overline U_\phi(r_{\rm max}) \approx \overline U_\phi(r_+) + \left[{J_d(0) \over M_d(0)} - \overline U_\phi(r_+) \right] \left({t \over t_\star}\right)^{n-1} . 
\end{equation}
Finally, let us assume the disk is quasi-Keplerian at all radii, then every $\overline U_\phi \approx \sqrt{GM_\bullet r}$, and we can define $r_0 \equiv (J_d(0)/\sqrt{GM_\bullet}M_d(0))^2$ and $\overline r \equiv (\overline U_\phi(r_+)/\sqrt{GM_\bullet})^2$. We finally therefore have that
\begin{equation}\label{rmax_global}
    r_{\rm max}(t) \approx \left[ \overline r^{1/2} + (r_0^{1/2} - \overline r^{1/2})\left({t\over t_\star}\right)^{n-1} \, \right]^2 , \quad t \geq t_\star . 
\end{equation}
We see that there is only one index of interest in the problem, namely $n$, which sets the evolution of each of $\dot M \sim t^{-n}, M'_{\rm max} \sim t^{1-n}$ and $r_{\rm out} \sim t^{2n-2}$. Global arguments cannot either (i) show that the mass accretion rate should fall as a powerlaw, (ii) relate $n$ to physical parameters in the problem, (iii) bound the possible values of $n$, or (iv) determine how the width of the matter distribution $(\sigma_{\rm eff})$ evolves with time. To answer these  points, and model the evolution of the disk in detail, one must turn to a more sophisticated theory. 

\subsection{General Relativistic Thin Disk Theory }
To construct a more sophisticated theory one starts with the local conservation laws of GRMHD, before simplifying these expressions to a more manageable theory. 

\subsubsection{Simplifying assumptions}\label{sec:assumptions}
We begin with the stress energy tensor of ideal GRMHD, namely 
\begin{multline}
    T^{\mu \nu} = \left(\rho + {P+e + b^2/4\pi\over c^2} \right) U^\mu U^\nu + \left(P + {b^2 \over 8\pi}\right) g^{\mu \nu} \\ - {b^\mu b^\nu \over 4\pi } .
\end{multline}
We begin by expanding the 4-velocities into a mean plus fluctuation 
\begin{equation}
    U^\mu = \overline U^\mu + \delta U^\mu,
\end{equation}
and introduce an averaging procedure $\left\langle \cdot \right\rangle$ such that $\left\langle \delta U^\mu \right\rangle = 0$, and therefore that $\left\langle U^\mu\right\rangle = \overline U^\mu$ (at this stage this is not actually an approximation, as one can always define this averaging procedure rigorously). Then one averages the stress energy tensor and finds 
\begin{multline}
    \left\langle T^{\mu \nu}\right\rangle  = \left(\rho + {P+e + b^2/4\pi\over c^2} \right) \, \overline U^\mu \overline  U^\nu + \left(P + {b^2 \over 8\pi}\right) g^{\mu \nu} \\  + \rho \left\langle \delta U^\mu \delta U^\nu \left(1 + {c_s^2 + v_A^2 \over c^2}\right) - {b^\mu b^\nu \over 4\pi \rho } \right\rangle,
\end{multline}
and defines \citep{BalbPap99}
\begin{equation}
    W^{\mu\nu} \equiv \left\langle \delta U^\mu \delta U^\nu\left(1 + {c_s^2 + v_A^2 \over c^2}\right)  - {b^\mu b^\nu \over 4\pi \rho } \right\rangle. 
\end{equation}
This tensor is the turbulent stress tensor, which has two contributions, the first being the Reynolds stress ($\delta U \delta U$) and the second a Maxwell stress $(-v_A v_A)$. 
Again, strictly, this is not yet an approximation as we could always solve a combination of the dynamical equations of motion and Maxwell's equations $(\nabla_\mu \widetilde F^{ \mu\nu} = 0)$ for $W^{\mu\nu}$. The future approximation which will be made is in closing $W^{\mu\nu}$ with some phenomenological relationship.

{\it Assumption 1:} We neglect all of the non-orbital speeds in the problem compared to the speed of light, i.e., 
\begin{equation}
    \rho\left(1 + {P+e + b^2/4\pi\over \rho c^2}\right) \equiv \rho\left(1 + {c_s^2 + v_A^2 \over c^2}\right) \to \rho .
\end{equation}

{\it Assumption 2:} We move to a cylindrical near-equator coordinate system $(t, r, \phi, z)$, and make explicit the averaging procedure 
\begin{equation}
    \left\langle \cdot \right \rangle \equiv {1\over 2\pi \Delta t} \int_t^{t+\Delta t}\int_{-\infty}^{+\infty} \int_{0}^{2\pi} (\cdot)\, \, {\rm d}\phi \, {\rm d}z\, {\rm d}t',
\end{equation} 
assuming that every quantity can be well described by height integrated and azimuthally averaged quantities (i.e., we move from a four dimensional to two dimensional theory $X(t, r, \theta, \phi) \to \left\langle X\right\rangle (t, r)$). An appropriate time interval over which this theory can be averaged $\Delta t$ is assumed to exist, and be long enough so that the average flow is well defined, but short compared to the secular (radial) evolution timescale itself. 

{\it Assumption 3:} The mean four velocity of the fluid is postulated to be given by 
\begin{equation}
    \overline U^\mu = \left[\overline U^t(r), \overline U^r(r, t), \overline U^\phi(r), 0\right],
\end{equation}
with $\overline U^t$ and $\overline U^\phi$ equal to circular test particle motion (and therefore $\overline U^r$ is {\it assumed small} in the meaningful sense $\overline U^r \ll r \overline U^\phi$, and pressure forces are assumed small), and there is zero flow in the vertical direction. This means that only $\overline U^r$ is a dynamic quantity, for which an equation of motion must be solved. 

A further implication of this assumption is that the governing evolution equations derived in GRTDT must only hold at radii exterior to the innermost stable circular orbit (ISCO), as within this radius the flow no longer satisfies this asymptotic scaling, as the radial velocity is dominated by gravitational acceleration not turbulent angular momentum redistribution. 

{\it Assumption 4:} The turbulent stress tensor (or more precisely the $r-\phi$ component thereof) is a {\it local} and {\it scale-free} quantity, and therefore in a rather general form will be given by
\begin{equation}
    W^r_\phi = \left\langle \delta U^r \delta U_\phi  - {b^r b_\phi \over 4\pi \rho } \right\rangle = w \left({r \over r_0}\right)^\mu \, \left({\rho \over \rho_0}\right)^\eta .
\end{equation}
where $r_0$ and $\rho_0$ are some (currently meaningless) radial and density amplitudes, $\mu$ and $\eta$ are general indices (we  require  $\eta>-1$ for stable evolution), and $w$ quantifies some ``amplitude'' of the turbulent transport. This assumption means that one does not need to solve Maxwell's equations, as all contributions of the magnetic field are now either neglected  or subsumed into a parameterized $W^{\mu\nu}$. 

While this undoubtedly represents an assumption that must be tested, we wish to stress that it is {\it apriori} not a bad assumption, as both the equations of magnetohydrodynamics and (Newtonian) gravity are {\it scale-free}, meaning that the introduction of a scale into $W^{\mu\nu}$ would be somewhat unexpected at large radii (which is typically where the bulk of the disk mass is during the evolution). An inherent scale in $W^r_\phi$ on near-horizon regions is certainly possible (and we shall show is realized in our simulations), but this is typically a region of low disk mass, and does not appear to impact the disk evolution appreciably.   The assumption of {\it locality} (i.e., $W^{\mu\nu}$ depends at most on the local conditions of the disk and the location of a fluid element within the gravitational field) is likely to be satisfied for most turbulent states, although it may break down in regions of extreme magnetic fields (which are capable of transporting angular momentum macroscopically)\revision{; we return to this in section \ref{sec:otherwise}}. It is unclear a priori whether a suitably defined $W^{\mu\nu}$ can capture angular momentum transported by outflows, a subject we return to later.

\begin{figure*}
    \centering
    \includegraphics[width=0.49\linewidth]{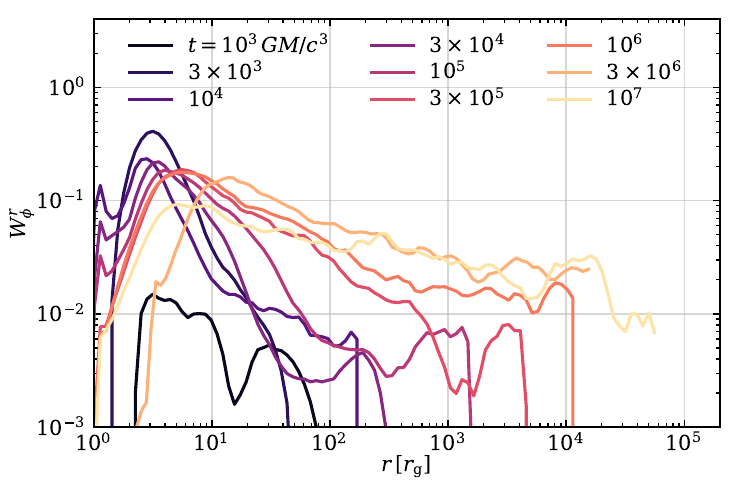}
    \includegraphics[width=0.49\linewidth]{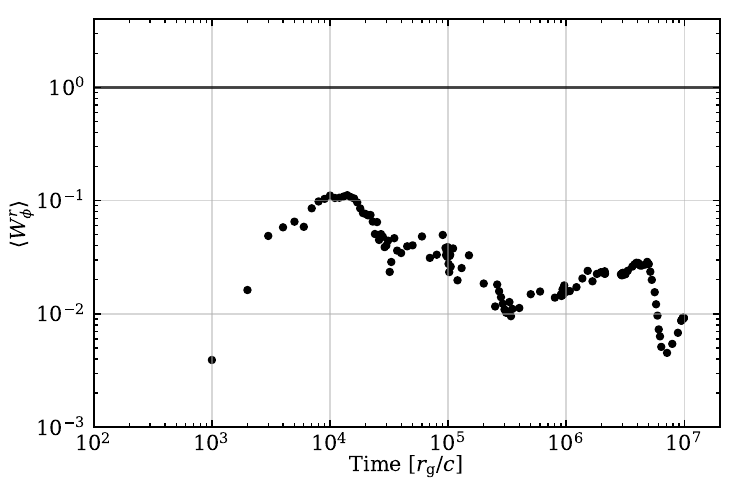}
    \caption{The $W^r_\phi$ stress extracted from simulations, at various different times. {\bf Left}: the radial dependence of the stress. The stress in our simulation is a highly stochastic variable, but otherwise shows only a very weak radial dependence and apparently no density dependence (recall that the density of the disk changes by many orders of magnitude in the inner $\sim 100 r_\mathrm{g}$ over the evolution, and there is no corresponding stress evolution). {\bf Right:} averaging over all bound mass leaves a weakly time dependent ``total stress''. 
    }
    \label{fig:stress}
\end{figure*}

Fortunately, for our purposes, we can extract $W^r_\phi$ from the GRMHD simulations themselves, by computing 
\begin{equation}
    W^r_\phi \equiv {\left\langle U^rU_\phi \right\rangle_\rho -  \left\langle U^r \right\rangle _\rho\left\langle U_\phi \right\rangle_\rho - \left\langle b^rb_\phi / 4\pi\rho \right\rangle_\rho },
\end{equation}
where $\left\langle \cdot \right\rangle_\rho$ is a density weighted spatial and temporal average (including only bound mass). This extracted stress is plotted in Figure \ref{fig:stress}, as a function of radius for a number of different times during the evolution of the disk (left panel), while the total integrated stress (over the entire disk) is shown in the right panel as a function of time. 

The results shown in Figure \ref{fig:stress} are important and interesting. We see that, to leading order, the stress is only very weakly dependent on both radius and time, with typically a weak radial decay in the bulk of the disk (recall that the bulk of the mass in the disk is at large radii at the later times shown in the left panel of Figure \ref{fig:stress}, where the stress is a weak function of radius). Recall that the density of the disk is varying by orders of magnitude over the course of this time evolution, and so it appears that $W^r_\phi$ shows little to no density dependence.  While of course the true disk stress is a somewhat chaotic variable (showing large amplitude fluctuations at a given time), it is clear that approximating the average of $W^r_\phi$ with a relatively weak radial dependence will not be egregiously in conflict with the simulations.  

\revision{We note that the weak time dependence of the total stress is not uniform over the evolution. During the first $\sim 10^4\,t_\mathrm{g}$ the stress grows as the MRI saturates, and over the period $10^4 \lesssim t \lesssim 3\times 10^5\,t_\mathrm{g}$ the bound-mass-averaged stress declines by a factor of a few, before remaining approximately constant to $t \sim 4\times 10^6\,t_\mathrm{g}$. This modest evolution should be contrasted with the change in the disk itself: between $t = 3\times 10^4$ and $3\times10^6\,t_\mathrm{g}$ the bound mass distribution at $r \sim 100\,r_\mathrm{g}$ falls by more than four orders of magnitude, while the peak of the mass distribution moves outwards by nearly two orders of magnitude in radius. It is this insensitivity of the stress to the local disk state which motivates the simple, density-independent stress parameterization used below.}

\subsubsection{Dynamical theory}
The evolution equations of general relativistic thin disk theory are then constructed from the averaged equations of mass and angular momentum conservation \citep{Balbus17}
\begin{align}
    \nabla_\mu \left(\left\langle J^{\mu}\right\rangle\right) &= 0, \\
    \nabla_\mu \left( \left\langle\rho U^\mu \right\rangle \right) &= 0.
\end{align}
Mass conservation becomes 
\begin{equation}
    \overline U^t {\partial \Sigma \over \partial t} + {1\over r}{\partial \over \partial r}(r \overline U^r \Sigma) = 0, 
\end{equation}
where we have changed notation to the surface density $\Sigma  \equiv \left\langle \rho \right\rangle$ as the averaging procedure involves a height integral. Angular momentum conservation is 
\begin{equation}
    \Sigma \overline U^\mu \partial_\mu \overline U_\phi + \nabla_\mu (\Sigma W^\mu_\phi) = 0, 
\end{equation}
or simply 
\begin{align}\label{ur}
    &\overline U^r = - {1\over r \Sigma  \overline U_\phi '} \, {\partial \over \partial r} \left[ {r \Sigma W^r_\phi } \right] ,
\end{align}
(where $\overline U_\phi'$ here denotes the radial gradient of $\overline U_\phi$). Combining the two leads to an equation for the evolution of the surface density 
\begin{equation}
    {\partial \Sigma \over \partial t} = {1 \over r \overline U^t} {\partial \over \partial r} \left({ 1\over \overline U_\phi ' } {\partial \over \partial r} \left[ {r \Sigma W^r_\phi } \right] \right) .
\end{equation}
 In this theory the mass accretion rate onto the black hole is 
\begin{equation}
    \dot M(t) = \left.\left(2\pi r \overline U^r \Sigma\right)\right|_{r = r_I} , 
\end{equation}
where $r_I$ is the innermost stable circular orbit, as the formalism breaks down within the ISCO (in effect this expression assumes that all disk material which crosses the ISCO also crosses the horizon within a short time period, this appears apriori to be perfectly reasonable). The mass distribution is 
\begin{equation}
    M'(r, t) = 2\pi r^2 \Sigma .
\end{equation}
A summary of the differences between the two theories is shown in Table \ref{tab:comparison}. 

\begin{table*}[]
    \centering
    \begin{tabular}{l|l|l}
      Quantity   & Evolution in GRMHD & Evolution in  GRTDT \\
      \hline 
        $\rho$ &  Dynamic & Dynamic \\
        $U^t$ & Dynamic & Prescribed (circular orbit) \\
        $U^r$ &  Dynamic & Dynamic (assumed small) \\
        $U^\phi$ & Dynamic & Prescribed (circular orbit) \\
        $U^\theta$ & Dynamic & Neglected \\
        $P$ & Dynamic & Neglected \\
        $e$ & Dynamic & Neglected \\
        $b^2$ & Dynamic & Neglected \\
        $b^\mu b^\nu$ & Dynamic & Absorbed into a ``stress tensor'' \\
        $W^{\mu\nu}$ & N/A & Prescribed (implicitly dynamic) \\
        \hline
    \end{tabular}
    \caption{A comparison of the various quantities that appear in both GRMHD and GRTDT, and how they are evolved within each theory. The phrase ``Dynamic'' here means that some equation of motion is solved at each time for that quantity, while prescribed means that this quantity is specified at all times. Prescribed quantities which depend on a dynamic quantity (such as a density dependent stress tensor) can be implicitly dynamic. Neglected quantities are set to zero throughout.   }
    \label{tab:comparison}
\end{table*}

\subsection{Properties of GRTDT solutions}
The turbulent stress observed in our GRMHD simulations (Figure \ref{fig:stress}) shows a weak radial dependence (in the main body of the disk), and almost no strong time or density dependence. We therefore consider turbulent stress parametrisations of the form 
\begin{equation}
    W^r_\phi = w \left({r\over r_0}\right)^\mu .
\end{equation}
This is a particularly convenient form in which to express the stress tensor, as the GRTDT equations have known \citep{Mummery23a} analytical Green's function solutions in this limit. 

These Green's functions are algebraically complex, but relatively simple to work with. In Appendix \ref{app:sol} we present these solutions in full, and derive the predicted temporal evolution of the quantities of interest $\dot M, M'$, etc. It is relatively simple to show that at large times one has
\begin{equation}
    \dot M(t) \propto t^{-n}, \quad n \equiv {4-2\mu \over 3-2\mu} ,
\end{equation}
and we note that the power law decay index $n$ can be related to the assumed microphysics of the disk stress $\mu$. The time dependence of the location of the disk outer edge can be similarly found, and scales as
\begin{equation}
    r_{\rm out}  \sim t^{2/(3-2\mu)} \sim t^{2n-2} ,
\end{equation}
precisely as intuited from a global conservation law analysis. The amplitude of $M'$ similarly evolves identically to that inferred from the global analysis
\begin{equation}
    M'_{\rm max} \sim t^{-1/(3-2\mu)} \sim t^{1-n} .
\end{equation}
What we learn from this solution, which cannot be inferred from global conservation equations, is that the equivalent width of the mass distribution inherits the same time dependence as the outer disk radius, namely 
\begin{equation}
    \sigma_{\rm eff}(t) \sim t^{2/(3-2\mu)} \sim t^{2n-2} .
\end{equation}
Note that this result justifies the earlier assumption that the fractional width of the mass distribution is independent of time $\Delta \sim (\sigma_{\rm eff}/ r_{\rm max}) \sim {\rm constant}$. 
While this final result may not appear obvious, it can be readily understood by transforming the GRTDT surface density evolution equation into a purely diffusive form. To see that GRTDT is purely diffusive, change variables to ${\rm d}h = \overline U_\phi'\, {\rm d}r$, define $\xi \equiv r \Sigma W^r_\phi$ and 
\begin{equation}
{\cal D}_h \equiv W^r_\phi \,\left( {\partial \ln (\Sigma W^r_\phi) \over \partial \ln \Sigma} \right) \, {\overline U_\phi' \over \overline U^t}, 
\end{equation}
which leaves 
\begin{equation}
    {\partial \xi \over \partial t} = {\cal D}_h {\partial^2 \xi \over \partial h^2} ,
\end{equation}
i.e., a diffusion equation with (in general) a non-linear diffusion coefficient. 

From this diffusion equation and the following scalings $h \sim r^{1/2}, U_\phi' \sim 1/h, W^r_\phi \sim r^\mu \sim h^{2\mu}$ (results that hold at large radii) one finds that the diffusion in angular momentum space goes like $(\sigma_h)^{3-2\mu} \sim t$, from which the above result follows. 
This confirms the intuition of the global analysis, that there is only one key micro-physical parameter in the problem – the index $n$.

\section{GRMHD-GRTDT comparison}\label{sec:comparison}
In this section we perform a detailed comparison between the long-timescale GRMHD simulations and the predictions of GRTDT. The GRTDT theory in principle takes as input four ``free parameters'', the stress index $\mu$, the stress amplitude $w$, the total mass of the initial disk $M_d(0)$, and how this mass is initially distributed $M'(r, 0)$. 

For comparison between GRTDT and GRMHD, it initially appears that only two of those are truly free parameters, as the disk mass and distribution are known apriori. This is generally true if one is unsure about the stress, however, in this work we can go further as we also extract the turbulent stress from our simulations (Fig. \ref{fig:stress}), \revision{and so the stress introduces {\it no free parameters}. } Indeed, we take 
\begin{equation}
    W^r_\phi = w = 0.05 \,\,  [{\rm code\,\, units}], 
\end{equation}
i.e., $\mu =0$ and a measured amplitude.

\begin{figure}
    \centering
    \includegraphics[width=0.99\linewidth]{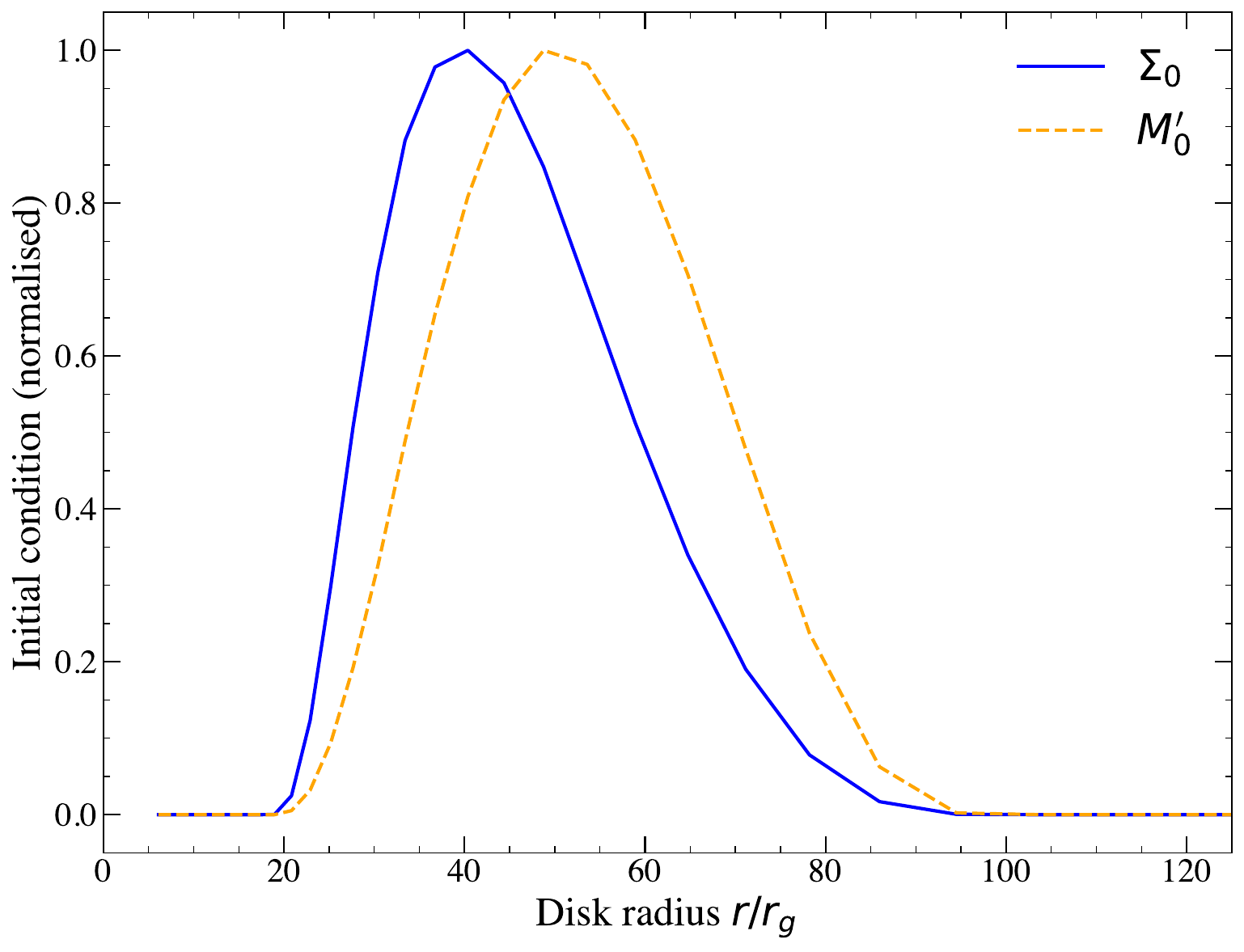}
    \caption{The initial condition of the GRMHD simulation, projected into a form appropriate for use in GRTDT. We show both the ``surface density'' (in blue), and the initial mass distribution $M'$. Both are normalised to have amplitude 1 for display purposes. }
    \label{fig:sigma0}
\end{figure}

In this case, the entire evolution of the disk in GRTDT is just given by the classic Green's function superposition  
\begin{equation}\label{green_sup}
    \Sigma(r, t) = \int_{r_I}^\infty G_\Sigma(r, t| r') \, \Sigma_0(r') \, {\rm d}r' ,
\end{equation}
with  $G_\Sigma$ the particular Green's function relevant for our choice of $W^r_\phi = 0.05$ (Appendix \ref{app:sol}), and $\Sigma_0(r')$ the initial condition of the simulation (Fig. \ref{fig:sigma0}).

\renewcommand{\arraystretch}{1.5}
\begin{table}[]
    \centering
    \begin{tabular}{c|c| c}
         Quantity & Global conservation analysis & GRTDT \\
         \hline 
         $\dot M $ & Postulated $\sim t^{-n}$ & $\sim t^{-n}$ \\
         $M'_{\rm max}$ & $\sim t^{1-n}$ & $\sim t^{1-n}$ \\ 
         $r_{\rm max}$ & $\sim t^{2n-2}$ & $\sim t^{2n-2}$ \\
         $\sigma_{\rm eff}$ & No constraints & $\sim t^{2n-2}$ \\ 
         $n$ & No constraints & ${(4-2\mu) / (3 - 2\mu)}$ \\
         \hline 
    \end{tabular}
    \caption{The scaling of the four ``summary statistics'' of the disk profile at late times, and how they do (or do not) relate to microphysical parameters of the accretion process in GRTDT (right column) and a global conservation analysis (middle column).  }
    \label{tab:my_label}
\end{table}

Implementing this explicitly, we found that GRTDT did a qualitatively excellent job at reproducing the gross evolution of the full GRMHD run (to be discussed in much more detail in the coming section), up to a (physically understandable) overall factor (plots of this implementation, and various other parameter changes, are shown in  Appendix \ref{app:plots}). In the initial (first $\sim 10^3 t_g$) evolution of the system, GRTDT simply accretes ``too quickly'' when compared to GRMHD, meaning that by the time the two different profiles come into good agreement ($t \gtrsim {\rm few}\times 10^3 t_g$) the GRTDT profiles are off by an overall normalisation factor of order $\sim 2$ (see Figure \ref{fig:m+j} for a clear illustration of this). The reason for this is simple to understand, and is a result of two effects
\begin{itemize}
    \item The GRMHD simulation evolves slowly over the first few thousand gravitational times as it takes a finite time for the MRI turbulence to reach the vigorous levels required to redistribute the disks' angular momentum. This physics is not contained in GRTDT (which has a constant turbulent stress at all times), and so they evolve at different rates initially. 
    \item GRTDT evolves quickest when there are large density gradients (see equation \ref{ur}), like those at the initial condition (see Figure \ref{fig:dens}).  This leads to further discrepancy from GRMHD, which is evolving slowly in this initial phase owing to a lack of MRI turbulence. 
\end{itemize}
As such, the naive Green's function superposition approach (equation \ref{green_sup}), deviates from the GRMHD evolution at early times. This will be discussed more quantitatively in the following section. 



We therefore allow the GRTDT theory to have {\it more initial disk mass} than is in the GRMHD simulation, to compensate for the mass accreted in the first $t \sim 5\times 10^3 t_\mathrm{g}$ when the GRMHD simulation is only very slowly evolving. We found that this mass compensation had to be at the factor $\sim 2$ level, as the accretion rate at early times in GRTDT is typically much larger than seen in the GRMHD simulations. 


\begin{figure*}
    \centering
    \includegraphics[width=0.49\linewidth]{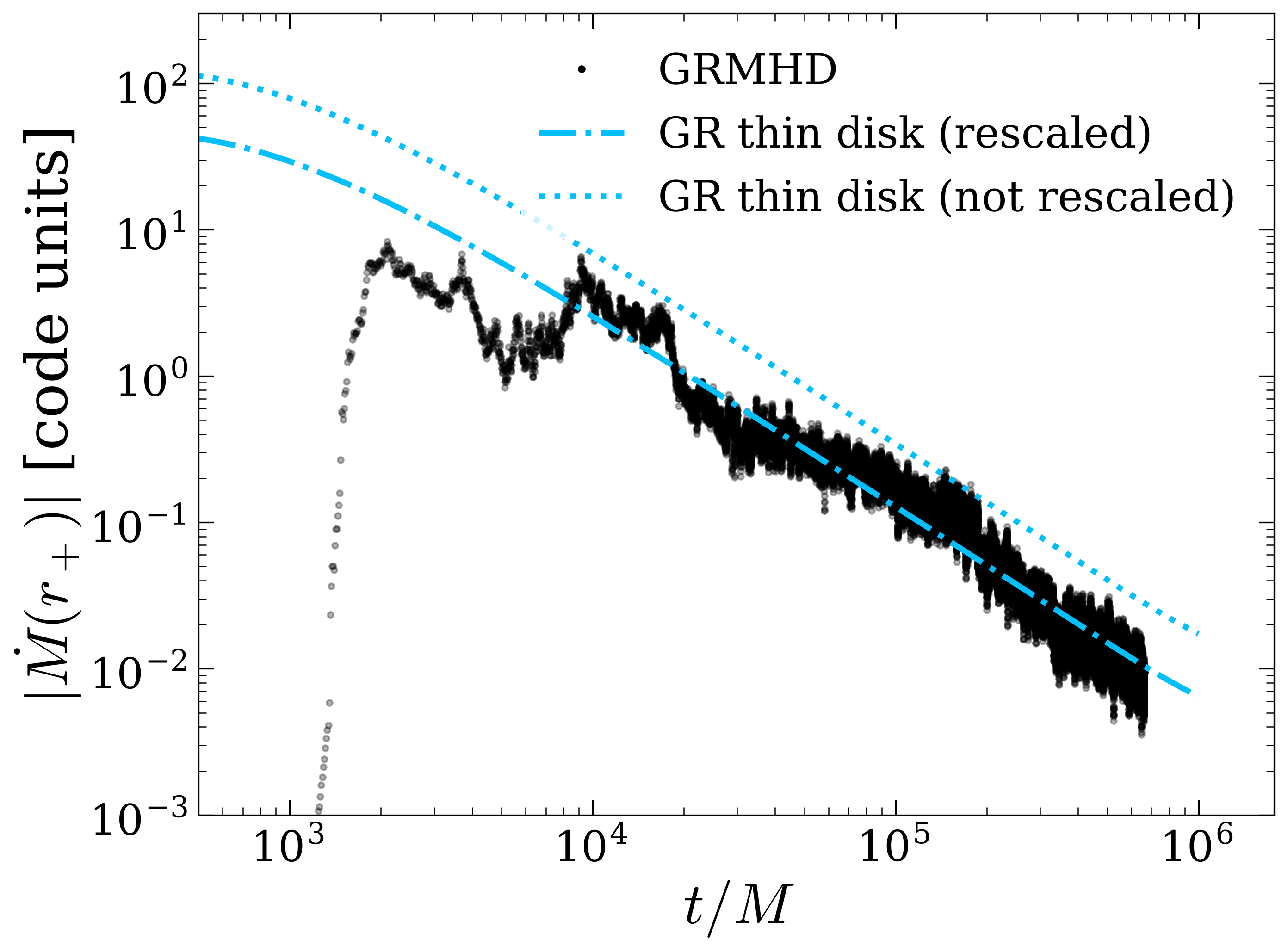}
    \includegraphics[width=0.49\linewidth]{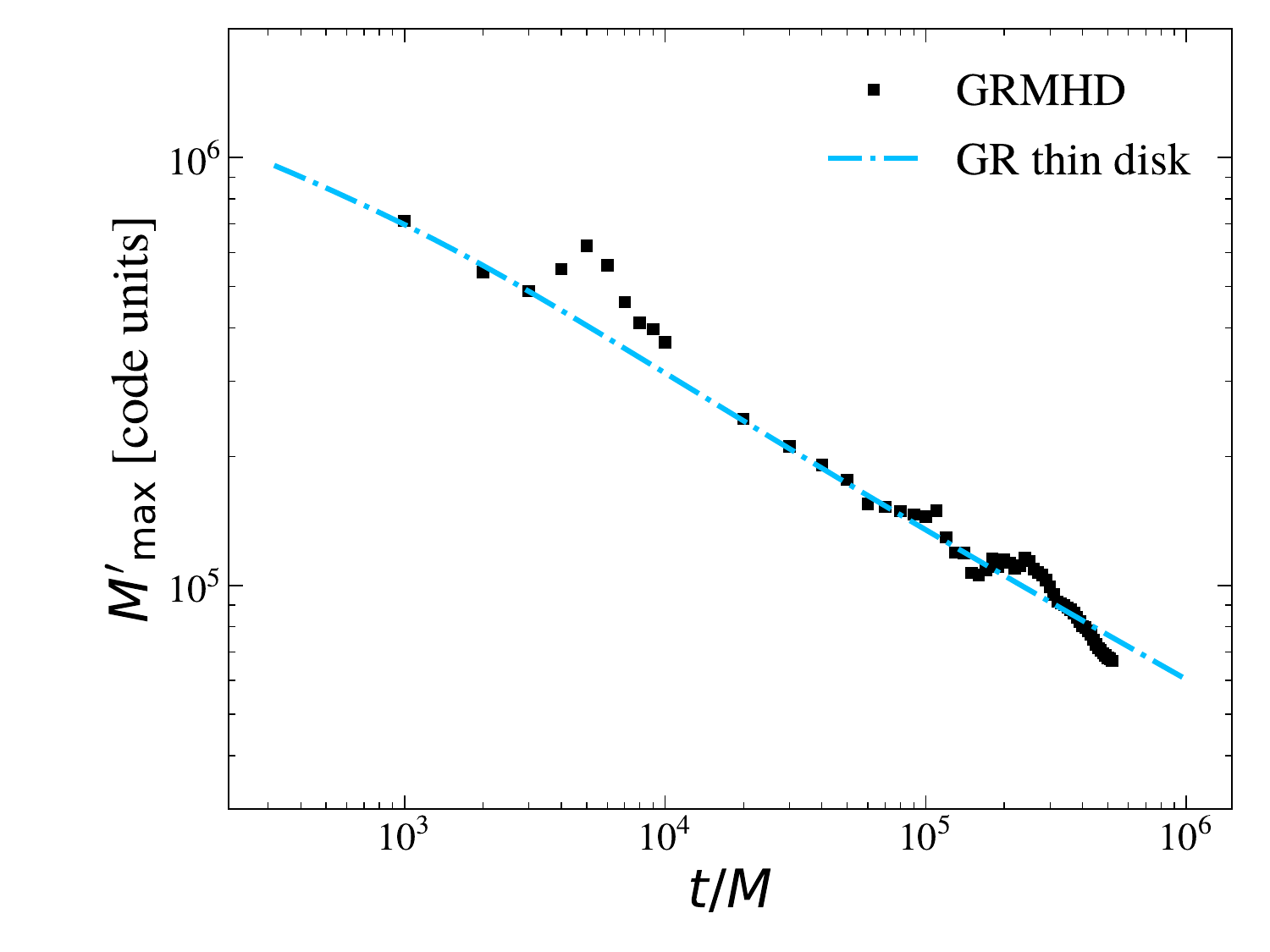}
    \includegraphics[width=0.49\linewidth]{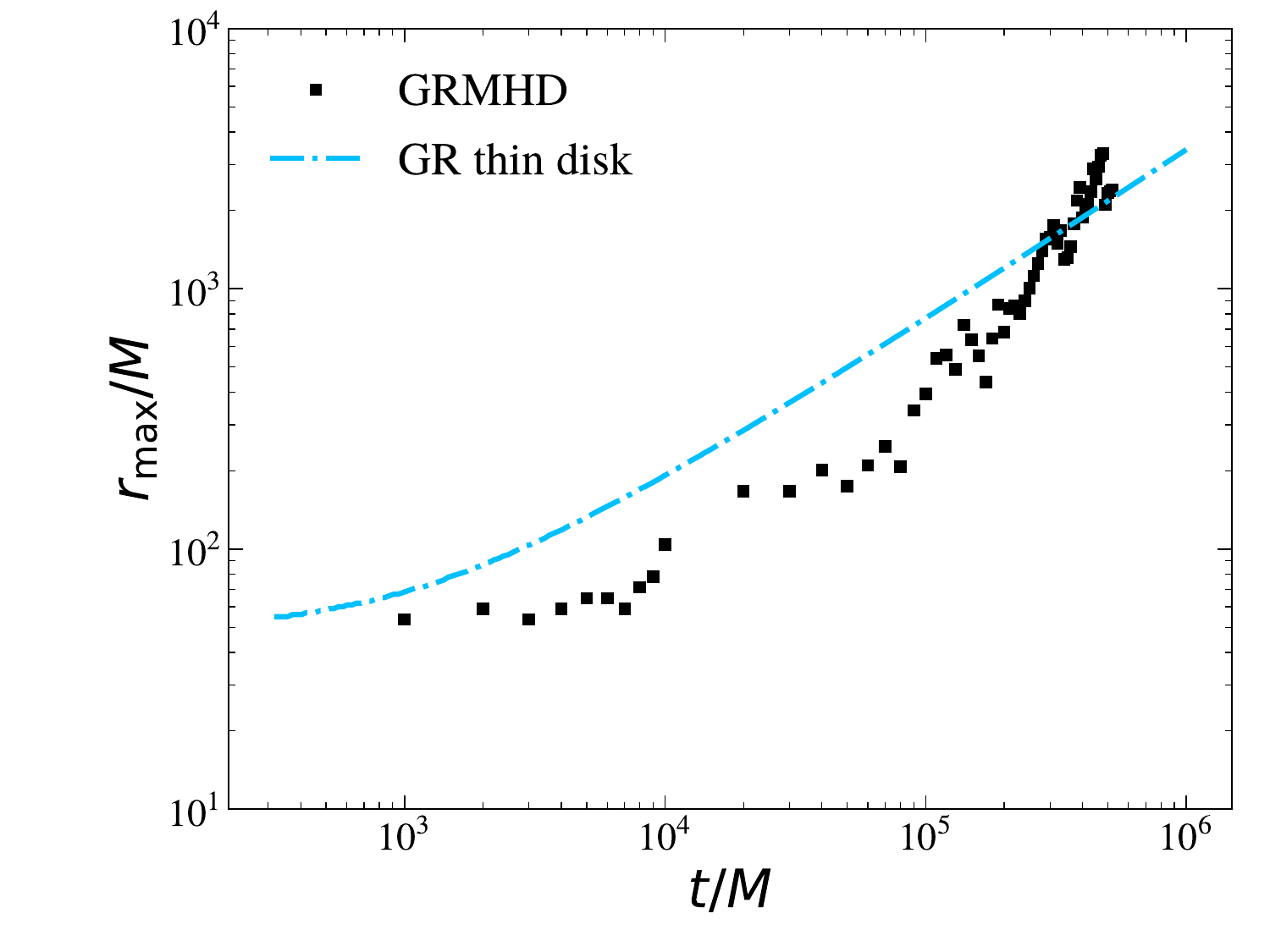}
    \includegraphics[width=0.49\linewidth]{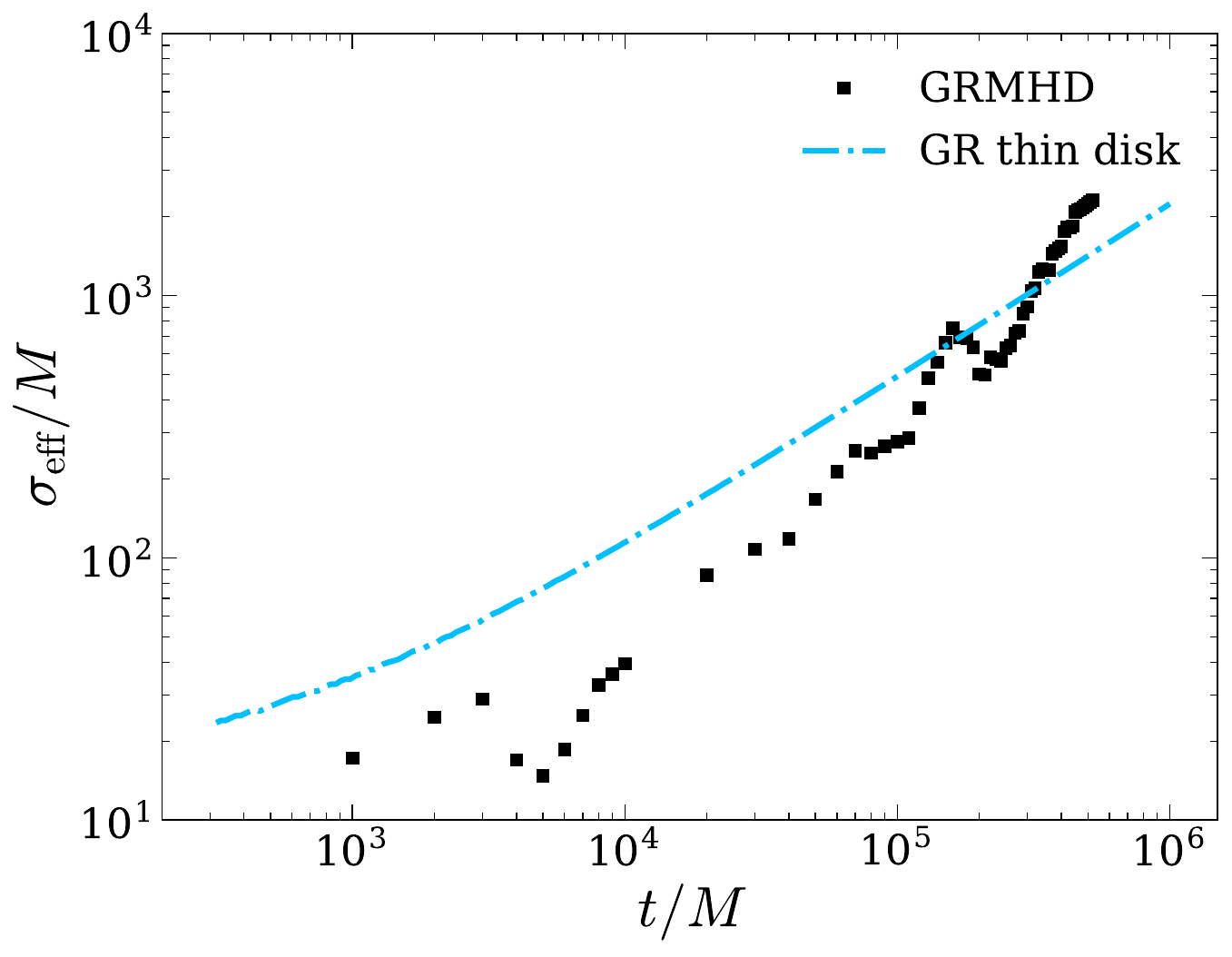}
    \caption{The four accretion flow summary statistics introduced in section \ref{sec:summarystats}, and a comparison between simulation data (black dots) and GRTDT (blue dot-dashed curves). This plot is for the traditional GRMHD run, which saves the mass accretion rate (upper left panel) at higher cadence than the three $M'$ summary statistics, which require the full simulation data to compute (as opposed to the data local to the horizon). Both $\dot M$ and $M'_{\rm max}$ are displayed in code units, while $r_{\rm max}$ and $\sigma_{\rm eff}$ are displayed in units of gravitational radii. The time axis on each plot is displayed in units of gravitational times. \revision{In the upper left panel the dotted curve shows the GRTDT horizon accretion rate with the same normalization as the mass distribution (i.e., with no free parameters), while the dot-dashed curve shows the same prediction renormalized by a factor $\sim 1/2$ to approximately match the simulation, to allow the time dependence to be compared. The difference between the two reflects the fraction of the bound mass loss which is transferred to the outflow rather than accreted (section \ref{sec:massloss}, Figure \ref{fig:horizonshare}). The effective width $\sigma_{\rm eff}$ is defined identically for both theories (half the distance between the two radii at which $M' = M'_{\rm max}/\sqrt{e}$); as neither mass distribution is Gaussian, this comparison is approximate.}  }
    \label{fig:summarystats}
\end{figure*}

\subsection{A comparison of the summary statistics}

\begin{figure*}
    \centering
    \includegraphics[width=\linewidth]{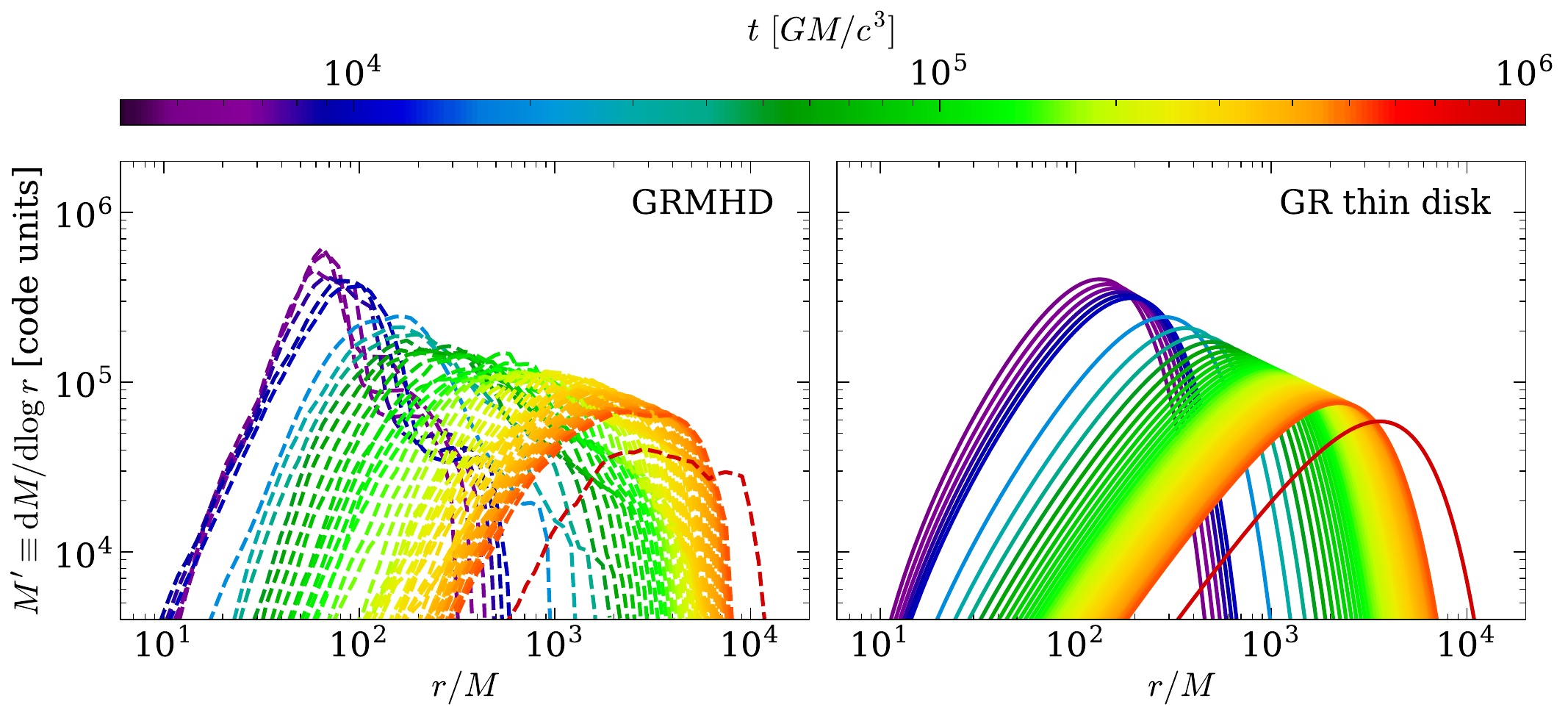}
    \caption{A visual comparison of the evolution of $M'$ for  classical GRMHD (left panel) and the equivalent evolution for GRTDT (right panel). \revision{The color of each curve denotes the time of the snapshot, as shown by the color bar shared by both panels; the snapshots span more than two orders of magnitude in time, from $t_1 \approx 5 \times 10^3 \, t_\mathrm{g}$ to $t_2 \approx 10^6 \, t_\mathrm{g}$. } There is a striking similarity in the global evolution, with clear agreement on key evolutionary features between the two techniques.   }
    \label{fig:comparison}
\end{figure*}

In Figure \ref{fig:summarystats} we show the four evolving summary statistics of the traditional GRMHD run, along with the solution of the GRTDT equations (eq. \ref{green_sup}). In the upper left panel we show the evolution of $\dot M$, which is saved at much higher cadence than the summary statistics of $M'$, which require a full data save (as opposed to a local save on horizon scales).

Looking just at the simulation data (black points), it is clear that an asymptotic power-law is a very good description for both $\dot M(t)$, and $M'_{\rm max}(t)$. The evolution of $r_{\rm max}(t)$ and $\sigma_{\rm eff}(t)$ appears well approximated by a power-law over the last $\sim$ decade in time, but otherwise shows significant curvature on this log-log plot. This can be understood by reference to the global conservation analysis (equation \ref{rmax_global}), as the specific angular momentum of the flow at the ISCO is a moderate fraction of the specific angular momentum of the initial disk (i.e., $\sqrt{r_I} \sim 6^{1/2} \sim 2.5$ is within a factor $\sim 2$ of $\sqrt{r_{\rm max, 0}} \sim 5.8$). The ratio of these two factors appears in equation (\ref{rmax_global}) as a deviation from a pure power law time dependence, which becomes less important for very large times. 

Moving to a comparison between GRMHD and GRTDT, we see that, on this summary statistic level, GRTDT does a (potentially) surprisingly good job at reproducing the numerical data. Clearly, as discussed above (and will be quantified further below), the GRTDT does a poor job at reproducing the mass accretion rate onto the black hole at early times. The mass accretion rate in GRTDT on timescales $t \lesssim 5 \times 10^3 t_\mathrm{g}$ is a factor of few larger than the peak accretion rate reached in the GRMHD run\footnote{Note that when the disk is not in a steady state the luminosity and mass accretion rate of the disk are not proportional, i.e., $L_{\rm disk} \neq \epsilon \dot M c^2$ (equality only occurs in the steady state, and is approached asymptotically at large times, with $L_{\rm disk}<\epsilon\dot M_{\rm acc}c^2$ at earlier times), and so this does not necessarily mean that the luminosity of GRTDT would be wrong by the same factor of a few at these times.}. This is why a factor $\sim 2$ more mass was required in the GRTDT initial condition to get good agreement at late times. \revision{On timescales $t \gtrsim 10^4\, t_\mathrm{g}$ the time dependence of $\dot M$ is well reproduced by GRTDT. Its normalization, however, is not: with the same normalization as is used for the mass distribution, so that no free parameters are introduced, GRTDT over-predicts the horizon accretion rate by a factor $\sim 2$ at $t \gtrsim 2\times 10^4\, t_\mathrm{g}$ (dotted curve in the upper left panel of Figure \ref{fig:summarystats}). We attribute this to the physics of mass loss, rather than to the turbulent stress parameterization. In GRTDT all of the mass lost from the disk passes through the innermost stable circular orbit, whereas in the GRMHD simulation a growing fraction of the mass lost by the bound gas is instead transferred to the unbound outflow. We quantify this in section \ref{sec:massloss}. To allow the time dependence of the two accretion rates to be compared directly, we also show the GRTDT prediction rescaled to match the simulation (dot-dashed curve).} 

The summary statistics of the mass distribution are shown in the upper right, and two lower panels.  Again, it is clear that on timescales $t \gtrsim 10^4 t_\mathrm{g}$ GRTDT is capturing the gross evolution seen in the GRMHD simulations. In particular, the amplitude of $M'_{\rm max}(t)$ is well approximated by a simple power law decay with time, as predicted in GRTDT. Both $r_{\rm max}(t)$ and $\sigma_{\rm eff}(t)$ show stochastic short-timescale variability, but also a long-timescale evolution which is in broad (factor $\lesssim 2$ level) agreement with the predictions of GRTDT. 

\subsection{A comparison of $M'(r, t)$}

\begin{figure*}
    \centering
    \includegraphics[width=0.49\linewidth]{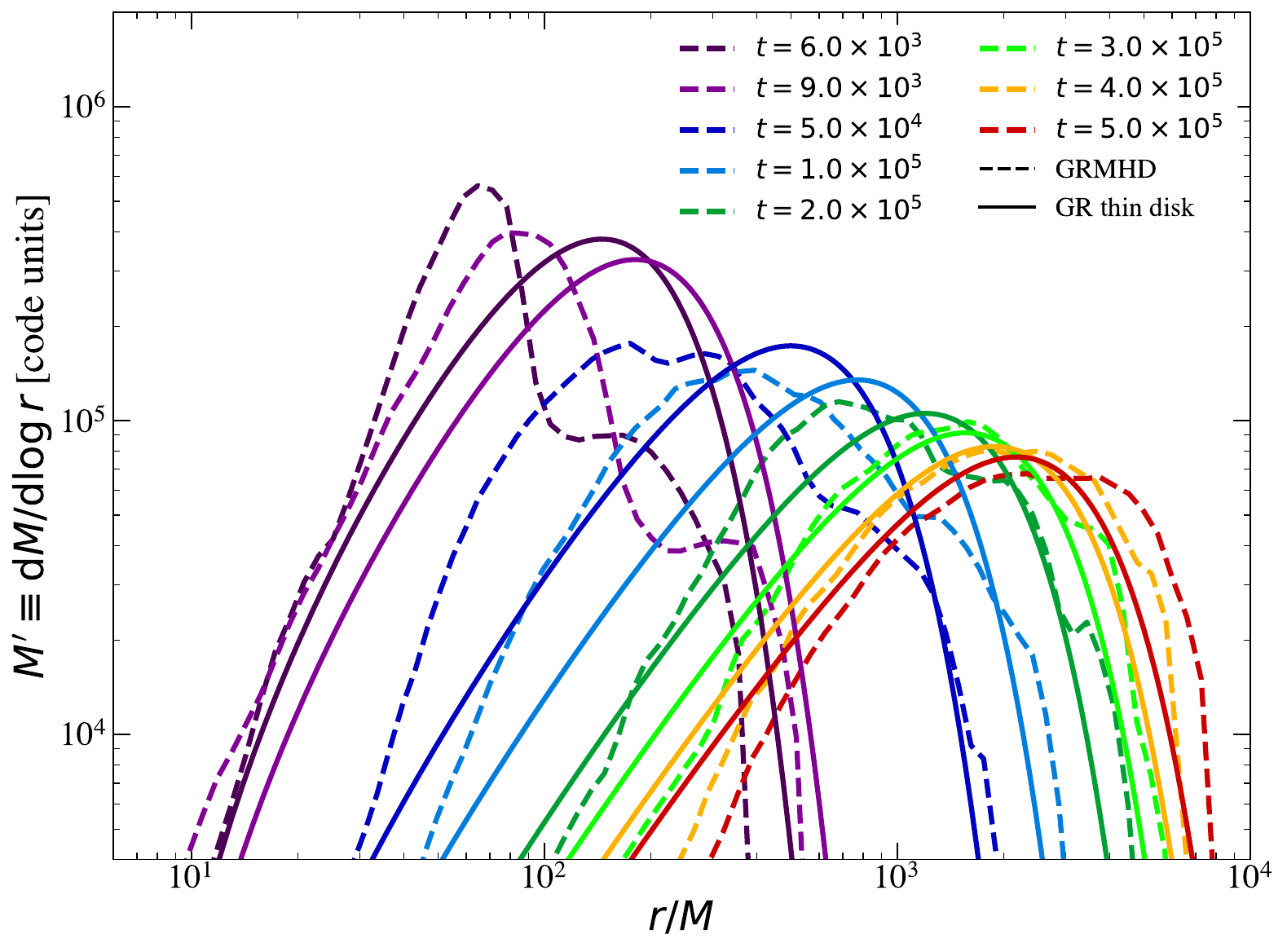}
    \includegraphics[width=0.49\linewidth]{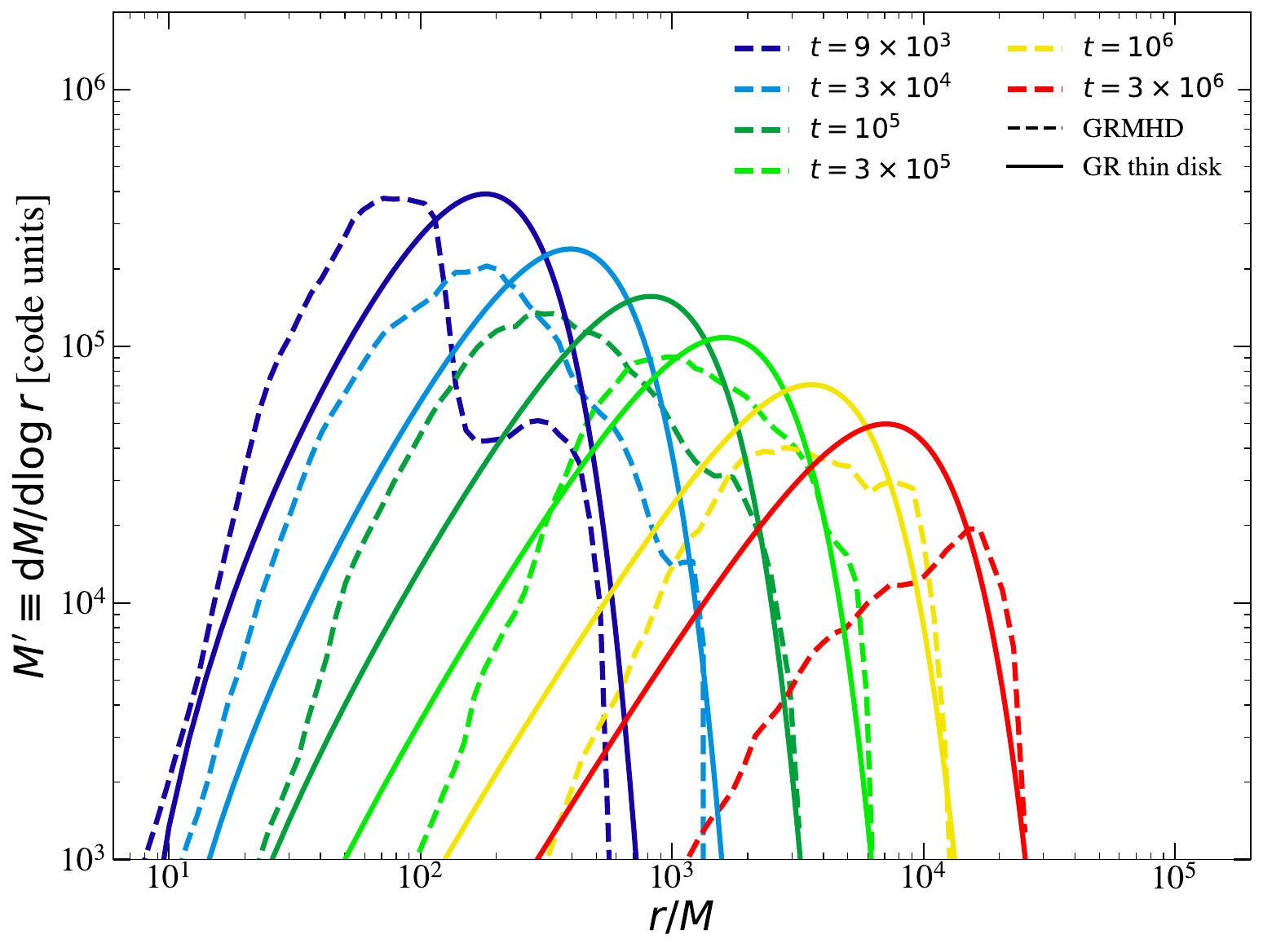}
    \caption{The evolution $M'$ from GRMHD (dashed curves), and GRTDT (solid lines), over plotted so that a more direct comparison can be performed. The left panel shows classical GRMHD, while the right panel shows the cyclic zoom run, which can be run out to even longer times. GRTDT provides a remarkably accurate description of the GRMHD evolution, despite its inherent simplifications. \revision{The evolution at the latest times is discussed in section \ref{sec:massloss}.} }
    \label{fig:overlay}
\end{figure*}

In Figure \ref{fig:comparison} we display the full evolution of the mass distribution $M'$ at a number of times in the evolution of the classical GRMHD run (\revision{left panel}), and the equivalent evolution in GRTDT (\revision{right panel}). The times here span \revision{more than two orders of magnitude, from $t_1 \approx 5\times 10^3 \,t_\mathrm{g}$ to $t_2 \approx 10^6\, t_\mathrm{g}$ (the final GRMHD snapshot being taken from the cyclic zoom run), and are indicated by the shared color bar}. 

While in any one individual snapshot the GRMHD evolution is clearly noisy (as is to be expected in a turbulent flow), there is a striking similarity in the global evolution, with clear agreement on key evolutionary features between the two techniques. The decrease in the amplitude of  the disk mass distribution, and the spreading to larger radii, are clearly in close agreement between the two theories (this of course is also seen in Figure \ref{fig:summarystats}).

A more detailed comparison can be seen in Figure \ref{fig:overlay}, where we overplot the GRTDT evolution on top of the GRMHD evolution for the classical GRMHD run (left hand panel), and the cyclic zoom run (right hand panel). The GRMHD evolution is shown by dashed curves, while the GRTDT evolution is shown by solid curves (of the same color).  It is clear from Figure \ref{fig:overlay} that GRTDT provides a remarkably accurate description of the (significantly more expensive) GRMHD simulation.

\section{Discussion}\label{sec:discussion}
In this section we discuss in detail our findings, and their implications both for theoretical modeling of accretion flows and also observations of time dependent disk systems. 
\subsection{Theoretical areas for improvement}
While GRTDT clearly provides a successful description of the gross evolution seen in GRMHD, there are certainly areas where the theory is less successful. We discuss in the following sections each specific failing, their physical origins, implications and how they may be rectified by future modeling advances. 
\subsubsection{Large density gradients}\label{sec:gradients}
The main failure of GRTDT occurred at early times (timescales shorter than $t_\star \sim {\rm few} \times 10^3 t_\mathrm{g}$), in regions of high density gradients. This can be readily understood, as GRTDT is built around the assumption that the $\overline U^t$ and $\overline U^\phi$ four velocity components are given exactly by their circular orbit values, and that the radial velocity is entirely specified by angular momentum conservation  
\begin{align}
    &\overline U^r = - {1 \over r \Sigma  \overline U_\phi '} \, {\partial \over \partial r} \left[ {r \Sigma W^r_\phi } \right] ,
\end{align}
and crucially, {\it is small compared to the rotational velocity} 
\begin{equation}
    |\overline U^r| \ll |r \overline U^\phi|,
\end{equation}
(here the factor $r$ ensures both 4-velocity components have the same dimensions). This cannot be guaranteed to be true at the earliest times. We have, assuming  the rough $W^r_\phi \sim {\rm constant}$ scaling seen in our simulation and inserting the dimensional parameters back in, that $W^r_\phi = w_{\rm code} \times \,(GM_\bullet)$ (here $w_{\rm code} \sim 0.05$ is the value measured in code units), meaning 
\begin{equation}
    |\overline U^r| \sim \sqrt{GM_\bullet\over r} \,  w_{\rm code} \left|{\partial \ln \Sigma \over \partial \ln r}\right| \ll  |r \overline U^\phi| \sim \sqrt{GM_\bullet \over r}.
\end{equation}
This result implies that thin disk theory cannot be a self consistent theory in regions of large density gradients, or specifically this theory is self-inconsistent in regions where 
\begin{equation}
     \left|{\partial \ln \Sigma \over \partial \ln r}\right| \gtrsim {1\over w_{\rm code}} ,
\end{equation}
which is clearly satisfied by the initial condition of the GRMHD simulation (see Figure \ref{fig:dens}). The initial condition of our GRMHD simulations has a density contrast between the initial torus and the background of factor $\sim 10^{10}$ which occurs over a narrow radial range $\sim {\cal O}(1-10) r_\mathrm{g}$, meaning that the predicted radial velocity is $\sim {\cal O}(1)$ times the orbital velocity. 

In regions with these large density gradients the inflow is super-Keplerian (i.e., $|\overline U^r| \gtrsim \sqrt{GM_\bullet/r}$), which is clearly inconsistent with the underlying perturbative expansion, which treats the orbital Keplerian velocity as the zeroth order dominant term, with the radial velocity assumed small in all expressions. 

This is only a problem at early times however, as  GRTDT evolution is purely diffusive, which is clearly a good approximation to the long term evolution of GRMHD (Figures \ref{fig:summarystats}, \ref{fig:comparison} and \ref{fig:overlay}), 
meaning any density gradients are relatively quickly damped, and the theory returns to a regime where it is valid. 

\subsubsection{The turbulent growth lag-time}
Another discrepancy between GRMHD and GRTDT is that GRTDT has no formal mechanism to ``wait'' for the MRI to grow over a few orbital timescales. While in an observational system this is unlikely to be a problem (as the initial condition of GRMHD is artificial, and all observed disks will already exist in a saturated MRI-turbulent state), this is an important point for direct comparison to GRMHD simulations.  In effect this leads to another route through which GRTDT starts ``too quickly'' at early times, and has too high an accretion rate when compared to GRMHD. 

We note a possible modeling resolution of this problem could be to ``switch on'' the stress in thin disk theory through a parameterization of the form 
\begin{equation}
    W^r_\phi = w(r) f(t), 
\end{equation}
where $f(t)$ incorporates some sort of lag time. This theory has Green's functions of an identical form to those used here (for suitable $w(r)$ parameterizations), but now in terms of the time coordinate 
\begin{equation}
    \lambda \equiv \int_{t_0}^t f(t') \, {\rm d}t', 
\end{equation}
rather than $t$ itself. Again, in future applications it may be of interest to take this seriously as an approach, we however do not open up this possibility for now, as we only have one set of simulations to compare to (and a large parameter space of possible $f(t)$ to explore). 
\subsubsection{Mass loss (launching of winds)}\label{sec:massloss}
In the simplified form used in this paper, GRTDT does not account for mass loss through winds/outflows launched from the disk. In the full GRMHD evolution, an $\sim {\cal O}(1)$ fraction of the initial mass was eventually launched from the disk into winds (as can be seen from the density contours and Bernoulli parameter plots; Figures \ref{fig:dens}, \ref{fig:bern} and \ref{fig:m+j}; \citealt{Guo2025}). 

On the surface level (Figures \ref{fig:summarystats}, \ref{fig:comparison}, \ref{fig:overlay}) this mass launching does not appear to lead to significant deviations in the evolution of the bound mass in GRMHD from that predicted by GRTDT. However, observation of Figure \ref{fig:m+j} shows that this is likely because\revision{, although the outflow removes a substantial fraction of the mass lost by the bound gas from $t \gtrsim 10^5\,t_\mathrm{g}$ (see below),} the time taken for \revision{it} to meaningfully \revision{alter the distribution} of the bound gas is $t \gtrsim {\rm few} \times 10^6 t_g$ (see the growth of the unbound angular momentum, Fig. \ref{fig:m+j}).  Indeed, the mass distribution $M'$ seen in the last plotted cyclic zoom run (Figure \ref{fig:overlay}) already shows a marked decrease in amplitude (at time $t = 3\times 10^6 t_g$), and the not plotted final epoch ($t = 10^7 t_g$) has effectively no remaining bound gas (see Figure \ref{fig:bern}), something which is of course not reproduced by GRTDT without outflows. 


Mass loss is clearly an element of the global GRMHD evolution which is not captured by GRTDT, and an extended theoretical framework is certainly of interest here. It would be simple to incorporate (at least formally), mass loss into the GRTDT equations, especially in the limit at which the unbound mass carries its local Keplerian angular momentum to infinity (and therefore does not impart a torque back on the disk). This limit is approximately what was observed in \cite{Guo2025}. In generality, when mass is lost from an accreting system the GRTDT  governing equation is modified to 
\begin{multline}
     {\overline U^t} {\partial \Sigma \over \partial t} - {1 \over r } {\partial \over \partial r} \left({ 1 \over \overline U_\phi ' } {\partial \over \partial r} \left[ {r \Sigma W^r_\phi} \right] \right) =\\ - F_M + {1\over r}{\partial \over \partial r}\left({r\over \overline U_\phi'} F_J\right),
\end{multline}
where the two mass loss terms have the form 
\begin{equation} 
 F_M \equiv {1\over 2\pi} \int_0^{2\pi}\int_{-H}^{+H} {\partial \over \partial z} \left( \rho U^z \right) \,  {\rm d}z \, {\rm d}\phi,
\end{equation}
(the mass carried out of the system) and 
\begin{equation}
    F_J = {1\over 2\pi} \int_0^{2\pi}\int_{-H}^{+H} \rho U^z {\partial \over \partial z} \left( U_\phi  \right) \,  {\rm d}z \, {\rm d}\phi, 
\end{equation}
(a potential torque on the system from the outflow). Performing the mass integral leaves 
\begin{equation}\label{mass}
F_M \approx 2  \rho(r, H) U^z(r, H)   ,
\end{equation}
having assumed symmetry at $z = \pm H$ and no azimuthal dependence (a reasonable assumption, and one that appears to be respected by the GRMHD simulations, see Figure \ref{fig:dens}). The torque integral is less simple to compute analytically, but the outflows seen in \cite{Guo2025} appeared to carry their Keplerian angular momentum to infinity $\partial_z U_\phi = 0 \to F_J = 0$. This suggests that most of the progress can be made by considering a parameterized $F_M$. 

Of course, in real GRMHD, $F_M$ is determined self consistently from the underlying equations of motion of the flow, but if one were to prescribe $F_M$ in much the same way as one prescribes $W^r_\phi$ (i.e., by making a simplifying scaling assumption) then accounting for this mass loss on the evolution of the bound material could be done carefully \citep[see][for exact solutions in certain limits of the Newtonian gravity analogue of this problem]{Mummery26winds2}. We postpone such an analysis to future work, to avoid overfitting $F_M$ (etc.) to results obtained from a single simulation.  \revision{Incorporating mass loss in this way would suppress the accretion rate in GRTDT, bringing it into closer agreement with the GRMHD simulation, as the following analysis indicates.}

\begin{figure*}
    \centering
    \includegraphics[width=\linewidth]{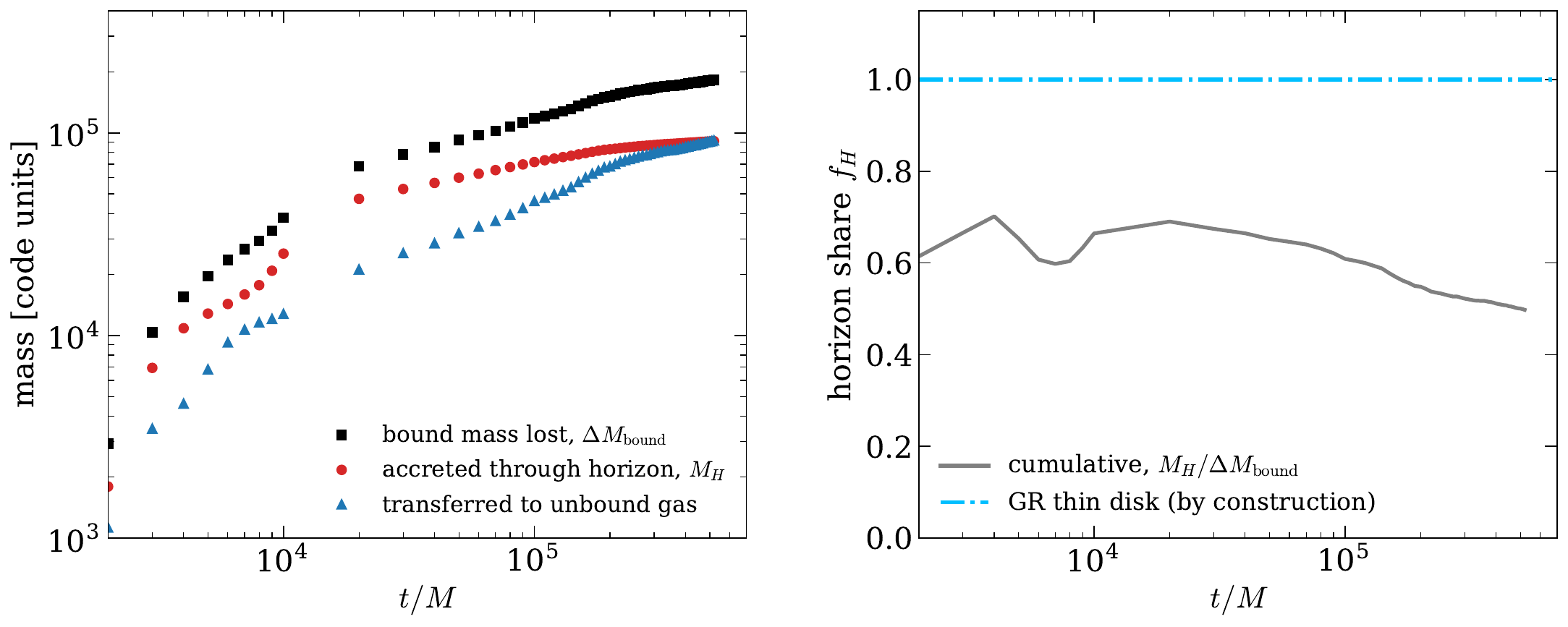}
    \caption{\revision{The fate of the mass lost by the bound gas in the standard GRMHD run. {\bf Left:} the cumulative mass lost by the bound gas, $\Delta M_{\rm bound}$ (black squares), the cumulative mass accreted through the horizon, $M_H$ (red circles), and their difference, the mass transferred to unbound gas (blue triangles). {\bf Right:} the cumulative horizon share of the bound mass loss, $M_H/\Delta M_{\rm bound}$ (grey curve). In GRTDT all of the mass lost by the disk is accreted, and so this ratio is unity by construction (blue dot-dashed line). In the simulation only half of the mass lost by $t \approx 5\times10^5\,t_\mathrm{g}$ has reached the black hole, and the instantaneous share is smaller still at late times (see text).}}
    \label{fig:horizonshare}
\end{figure*}

\begin{figure*}
    \centering
    \includegraphics[width=0.48\linewidth]{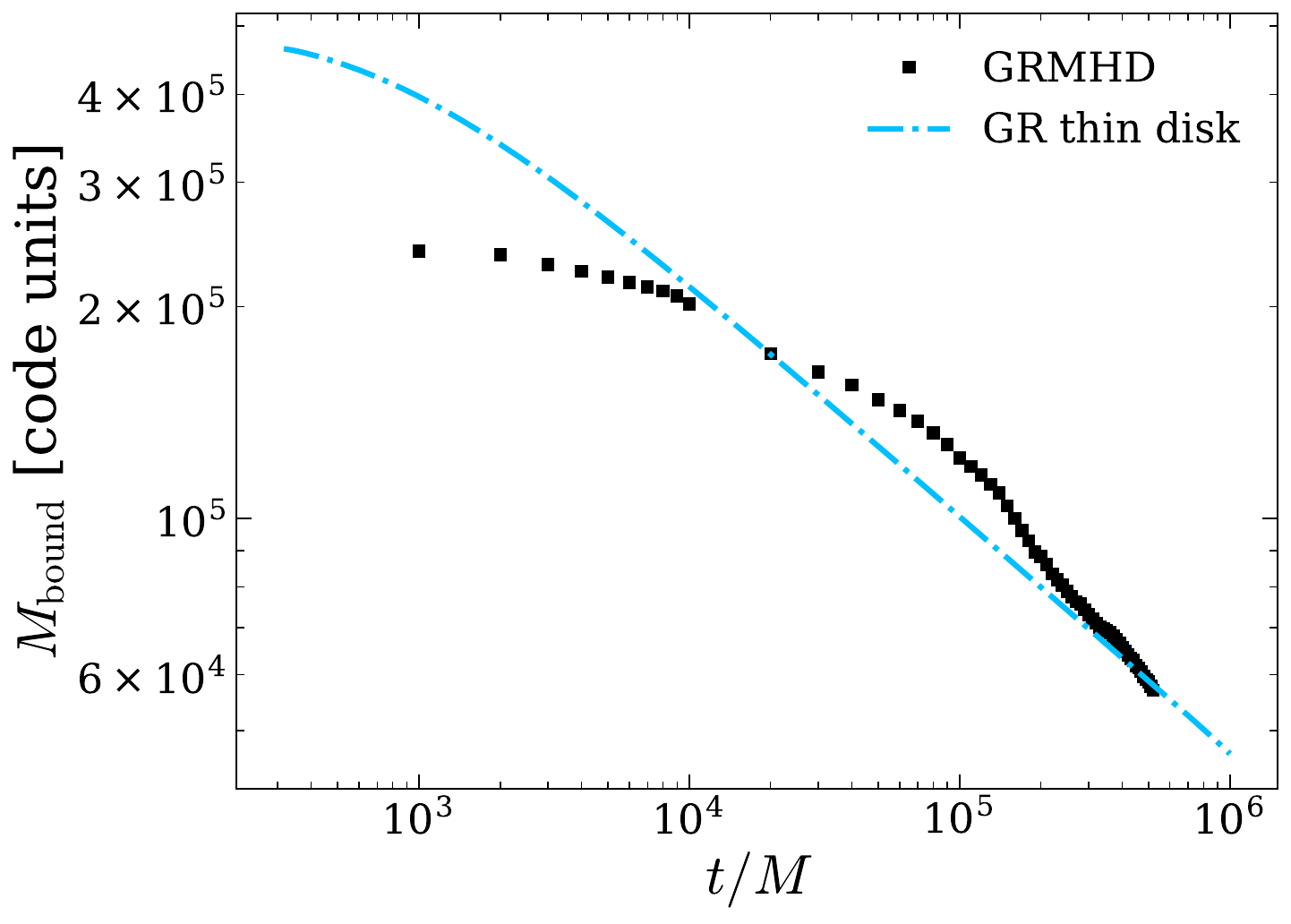}
    \includegraphics[width=0.48\linewidth]{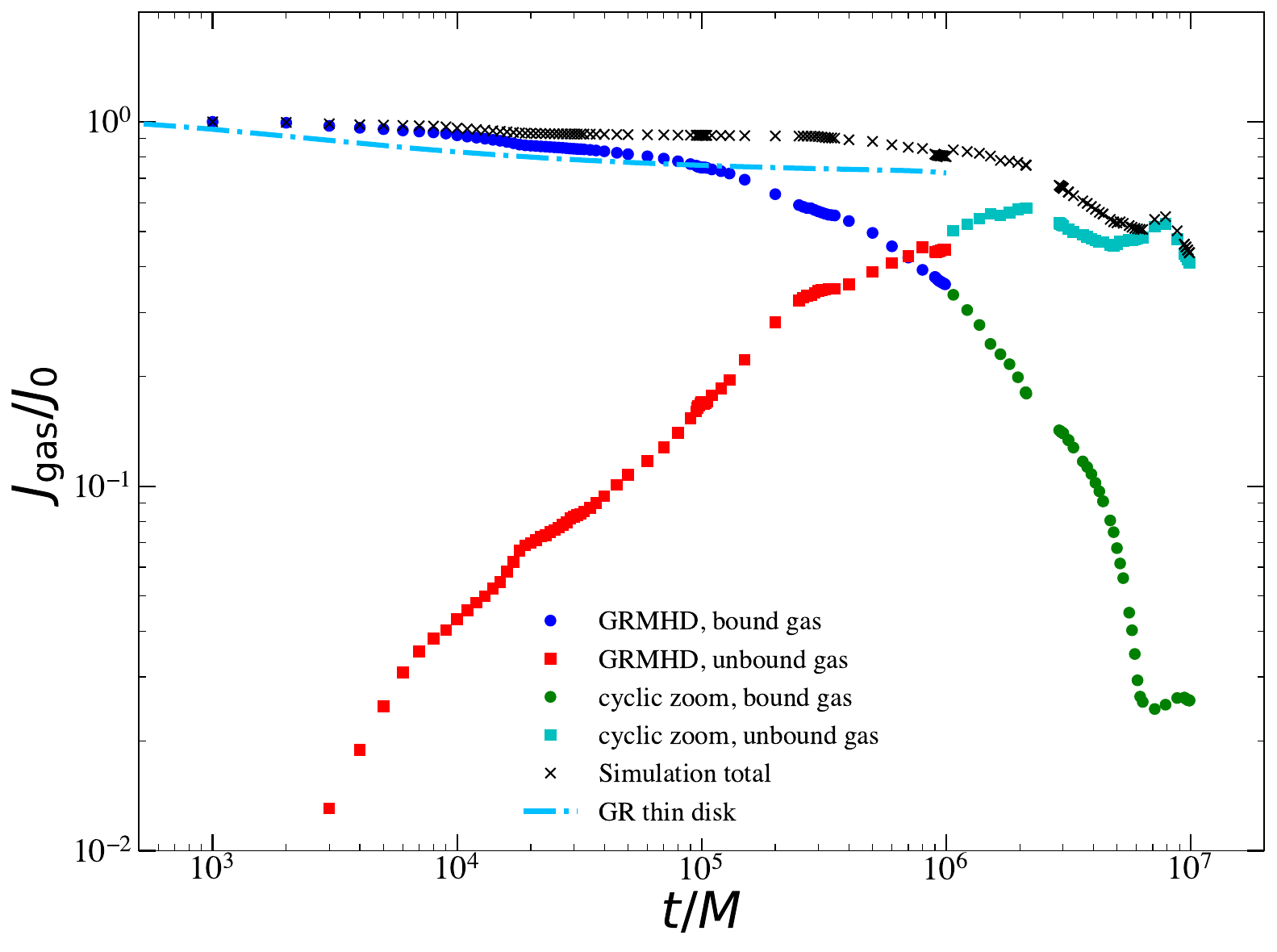}
    \caption{\revision{The evolution of the bound total mass (code units; left panel) and the angular momentum of the bound and unbound gas, normalized by the initial angular momentum of the torus (right panel). The left panel highlights the extra mass which must be included in the GRTDT initial condition for good agreement with GRMHD at late times, which compensates for the rapid accretion in GRTDT during the first $\sim 10^4\,t_\mathrm{g}$ which is not seen in GRMHD. The right panel shows the growing importance of outflows. Up to $t \sim 10^5\,t_\mathrm{g}$, the angular momentum lost by the bound gas is broadly consistent with that predicted by GRTDT, which includes only losses through the horizon (blue dot-dashed curve). At later times an increasing fraction of the angular momentum lost by the bound gas is transferred to unbound gas, and by $t \sim 10^6\,t_\mathrm{g}$ roughly half of the bound gas angular momentum loss has gone into the outflow. Beyond a few $\times 10^6\,t_\mathrm{g}$ (cyclic zoom run; green and cyan points) nearly all of the change in the angular momentum of the bound gas is due to outflows, which explains the divergence between GRTDT and GRMHD at these times.}}
    \label{fig:m+j}
\end{figure*}

\revision{The importance of this mass loss for the horizon accretion rate can be quantified directly from the simulation. In Figure \ref{fig:horizonshare} we show the cumulative mass budget of the bound gas: the total mass lost by the bound gas, $\Delta M_{\rm bound}(t) = M_{\rm bound}(0) - M_{\rm bound}(t)$, the mass accreted through the horizon, $M_H(t) = \int_0^t |\dot M(r_+)|\, {\rm d}t'$, and their difference, which is the mass transferred from bound to unbound gas (together with a small amount lost through the outer boundary of the simulation domain). In GRTDT the horizon share of the mass loss, $f_H \equiv M_H/\Delta M_{\rm bound}$, is unity by construction. In the GRMHD simulation it is not: averaged over the evolution, only half of the mass lost by the bound gas by $t \approx 5\times10^5\,t_\mathrm{g}$ has been accreted (right panel of Figure \ref{fig:horizonshare}). The instantaneous share of the mass loss rate which reaches the horizon falls with time, from $\sim 70\%$ at $t\sim 10^4\,t_\mathrm{g}$ to $\sim 40\%$ at $t \sim 10^5\,t_\mathrm{g}$ and $\sim 20\%$ by $t \sim 5\times 10^5\,t_\mathrm{g}$. Since the bound mass itself is well reproduced by GRTDT (Figure \ref{fig:m+j}), this accounts for the factor $\sim 2$ over-prediction of the horizon accretion rate by GRTDT at late times (Figure \ref{fig:summarystats}). Equivalently, at late times the outflow removes mass from the bound disk at a rate several times larger than that at which the disk accretes onto the black hole.}

\revision{Two aspects of this result are notable. First, the outflow suppresses the horizon accretion rate without substantially altering the evolution of the bound mass distribution (section \ref{sec:otherwise}): the location and amplitude of the mass distribution, and the time dependence of the accretion rate, are all reproduced by GRTDT, while the normalization of the accretion rate is not. This is consistent with an outflow which removes mass at approximately the local specific angular momentum of the disk, and so exerts little net torque on it. Second, the horizon share decreases steadily with time, as the disk spreads to larger radii. A natural interpretation is that mass is lost to the outflow over a range of radii, so that the fraction of the inflowing mass which survives to reach the black hole decreases as the radial extent of the disk grows. Time-dependent solutions of the Newtonian disk equations including such a mass-loss term exist \citep{Mummery26winds2}, and a general relativistic generalization of these solutions would allow the evolution of $f_H$ to be modeled directly; we leave this to future work.}

\subsubsection{The stress index $\mu$}
We have, as far as is possible, sought to use a stress parameterisation in this work which reflects the output of the GRMHD simulation (Fig. \ref{fig:stress}), rather than an a priori physical model (such as the $\alpha$ viscosity, which we shall discuss later). However, our choice of $W^r_\phi \approx 0.05$ (code units) is informed mainly by the right hand panel of Fig. \ref{fig:stress} (i.e., the time evolution of the total stress). Realistically, one could equally well have picked a range of numbers between $w\sim 0.01-0.1$ from inspection of Fig. \ref{fig:stress}, and our choice of the rough midpoint seen in the evolution is somewhat ad-hoc. 

Further, inspection of the left hand panel of Figure \ref{fig:stress} highlights that, within the main body of the disk (a radial range which expands with time), an equally reasonable parameterisation of the stress would be $W^r_\phi \sim r^{A}$, where $A<0$ ($A \sim - 0.3$ is a reasonable fit). 

This highlights an ambiguity in this theory, what stress parameterisation should one choose? This ambiguity persists even though we are able to measure $W^r_\phi$ in our simulations. In reality, this ambiguity is not particularly meaningful as although changing the stress index $\mu$ appears like a major change, its impact on global disk quantities is always through the mass accretion rate decay index $n$, which is a weak function of $\mu$, namely $n = (4-2\mu)/(3-2\mu)$. We show this broad lack of $\mu$ dependence in the evolving disk profiles explicitly in Appendix \ref{app:plots}. 

Perhaps more meaningful is the choice of amplitude $w$. This is because the characteristic timescale of the evolution (at a given radius) 
\begin{equation}
    t_{\rm visc} \propto {\sqrt{r^3} \over W^r_\phi(r)} \propto {1\over w} r^{3/2-\mu} \propto {t_{\rm orb}(r) \over w r^\mu},
\end{equation}
is inversely proportional to $w$. As effectively all disk profiles are only functions of the self similar time $t/t_{\rm visc}$, one's choice of $w$ impacts the evolution more notably (see Appendix \ref{app:plots} for Figures).

\subsection{\revision{Could the evolution have been otherwise?}}\label{sec:otherwise}
\revision{In the introduction we listed a number of physically plausible ways in which the evolution of our simulated torus could have departed from that of a thin disk, while remaining entirely consistent with the conservation of mass and angular momentum. Having presented the comparison between GRMHD and GRTDT, we now return to each of these possibilities in turn, and ask what its signature would have been and what the simulation shows. We wish to stress that none of these outcomes was excluded a priori by the conservative nature of the underlying equations; each has precedent in the literature in some physical regime, and the extent to which each is (or is not) realized in our simulation is the main physical content of the comparison.}

\revision{{\it Non-local removal of angular momentum.} The most significant alternative to diffusive evolution is one in which the angular momentum lost by accreting gas is carried away by an outflow, or by large scale magnetic torques, rather than being transported outwards through the disk. Such an evolution is perfectly conservative, but is qualitatively different from that of a thin disk: in the limit in which a magnetized wind removes the angular momentum, the disk accretes without spreading, and the characteristic radius of the mass distribution remains fixed or even decreases with time \citep[e.g.,][]{BlandfordPayne82, BaiStone13, Tabone22}. Our simulations launch a substantial outflow, which unbinds an ${\cal O}(1)$ fraction of the disk mass over the evolution, and so this was a realistic possibility. The signature would have been a stalling of the growth of $r_{\rm max}$ and $\sigma_{\rm eff}$. Instead, $r_{\rm max}$ grows from $\approx 50\,r_\mathrm{g}$ to $\sim 10^4\,r_\mathrm{g}$ by $t \sim 3\times 10^6\,t_\mathrm{g}$, with $\sigma_{\rm eff}$ growing similarly, and over the period in which the comparison can be made this growth proceeds at the rate predicted by GRTDT, to within a factor $\sim 2$ (Figures \ref{fig:summarystats} and \ref{fig:overlay}). This is not because the outflow carries negligible angular momentum. By $t\sim 10^6\,t_\mathrm{g}$ roughly half of the angular momentum lost by the bound gas has been transferred to unbound gas rather than through the horizon (Figure \ref{fig:m+j}). Rather, it appears that the outflow removes angular momentum along with mass at approximately the local specific angular momentum of the disk, as was found by \citet{Guo2025}, and so exerts little net torque on the remaining bound gas. In this limit an outflow reduces the normalization of the disk without changing the character of its evolution. We note also that the turbulent stress which we measure is positive at all radii and times (Figure \ref{fig:stress}), and so angular momentum is transported outwards through the bound gas, as in a thin disk.}

\revision{{\it Non-local transport by large-scale magnetic fields.} A closely related, and in some ways more serious, concern is specific to the magnetically arrested state. In a MAD the magnetic flux threading the black hole is dynamically important, and the field is capable of transporting mass and angular momentum in ways that are not captured by any local stress. Magnetic flux is periodically ejected from the horizon region in eruptions which propagate outwards through the disk; ordered large-scale field can exert torques at one radius which are set by conditions at another; and magnetic pressure and tension forces can be comparable to gravity, so that the gas need not follow the near-Keplerian orbits assumed by GRTDT, nor have a radial velocity set only by angular momentum transport. Any of these would violate the locality assumption on which GRTDT rests, a possibility which we flagged in section \ref{sec:assumptions}. Three features of our simulation indicate that these effects do not control the evolution of the bulk of the disk mass. First, the magnetic field is dynamically dominant only close to the black hole: the Alfv\'en speed exceeds the Keplerian speed only within $r \sim 10\,r_\mathrm{g}$, falling to $v_A^2/v_K^2 \sim 0.03$--$0.1$ at $r \gtrsim 100\,r_\mathrm{g}$ (Figure \ref{fig:stress_mil}), while at all times $t \gtrsim 10^4\,t_\mathrm{g}$ essentially all of the disk mass resides at these larger radii. Second, the stress which we measure locally, rather than one fitted to the evolution, reproduces the rate at which the disk evolves and spreads; were angular momentum transported predominantly by non-local channels, a locally measured stress would under-predict this rate. Third, the horizon accretion rate follows a smooth power law in time, with fluctuations about this trend which are log-normally distributed and which obey a linear rms--flux relation (section \ref{sec:variability}). This is the expected signature of local, multiplicative fluctuations, rather than of a mass flux dominated by discrete eruption events. This is consistent with the MAD simulation of \citet{Chatterjee22}, also around a non-spinning black hole, in which flux eruptions drive a strong vertical flux of angular momentum into disk winds, but the average outward transport of angular momentum through the disk remains primarily radial. We cannot, however, exclude a role for this physics in two respects. At the earliest times, $t \lesssim 10^4\,t_\mathrm{g}$, the disk is compact and overlaps the magnetically dominated region, and this may contribute to the early-time discrepancy between GRMHD and GRTDT which we attribute principally to large density gradients and the growth of the turbulence (section \ref{sec:gradients}). And flux eruptions are known to eject mass, so may contribute to the transfer of bound mass to the outflow which suppresses the horizon accretion rate (section \ref{sec:massloss}).}

\revision{{\it A stress with an intrinsic scale, or no simple description at all.} The utility of GRTDT rests on the turbulent stress being describable by a simple parameterization. In principle the stress in a turbulent, magnetized flow could have an arbitrarily complicated dependence on position and time, depend on the history of the flow, or be dominated by intermittent events, in which case it could still be measured but no simple closure would capture the evolution. Instead, a single constant stress amplitude, measured from the simulation rather than fitted to its evolution, reproduces the evolution of both $\dot M$ and $M'$ to within a factor $\sim 2$ over more than two decades in time, and the measured stress is close to constant in radius and only weakly dependent on time (Figure \ref{fig:stress}). A particular case is a stress carrying an intrinsic scale. GRTDT with a scale-free stress predicts that the accretion rate, and the amplitude, location and width of the mass distribution, all evolve as power laws in time at late times, with a single index $n$. If the stress carried an intrinsic scale, this self-similarity would be broken, and the accretion rate would show a break or a curvature as the disk mass moved through that scale. A radial structure of this kind is present in our simulations: the stress rises sharply within $r \lesssim 10\,r_\mathrm{g}$ (Figure \ref{fig:stress}), where the radial magnetic field is strongest (Figure \ref{fig:stress_mil}). But this region contains a negligible fraction of the disk mass once $t \gtrsim 10^4\,t_\mathrm{g}$, and it leaves no visible imprint on the late time evolution, which follows a power law in time over more than a decade and a half in time (Figure \ref{fig:summarystats}).}

\revision{{\it Decay of the turbulence as the disk spreads.} In our simulations the initial magnetic field is confined to the initial torus, which extends only to $r \approx 100\,r_\mathrm{g}$. By $t \sim 3\times10^6\,t_\mathrm{g}$ the peak of the bound mass distribution lies at $r \sim 10^4\,r_\mathrm{g}$, and the outer edge of the disk at $r \approx 2.5\times 10^4\,r_\mathrm{g}$, more than two orders of magnitude beyond any initially magnetized region. As the disk spreads, the magnetic flux it carries is distributed over an ever larger area, and it was not obvious a priori that turbulence capable of transporting angular momentum would be sustained at the radii into which the disk spreads \citep[cf.][]{Salvesen16}. Had it not been, the stress would have decayed with time, and the outward spreading of the disk would have slowed and eventually stalled. We find instead that the stress at large radii remains a fixed fraction of the local Keplerian scale, with the magnetic and thermal energy densities saturating at approximately fixed fractions of $v_K^2$ at $t = 10^6\,t_\mathrm{g}$ (Figure \ref{fig:stress_mil}), and that the bound-mass-averaged stress, after an initial modest decline (by a factor of a few up to $t\sim 3\times 10^5\,t_\mathrm{g}$), remains approximately constant to $t \sim 4\times 10^6\,t_\mathrm{g}$ while the disk spreads by a further order of magnitude in radius (Figure \ref{fig:stress}).}

\revision{{\it Reshaping of the mass distribution by outflows.} Even an outflow which exerts no net torque on the disk could alter the shape of the bound mass distribution, if it removed mass preferentially from some radii. The signature of such preferential mass loss would be a systematic difference between the shapes of the GRMHD and GRTDT mass distributions, rather than simply their normalizations. Up to $t\sim 10^6\,t_\mathrm{g}$ we see no such systematic difference (Figure \ref{fig:overlay}). The GRMHD profiles are noisier and somewhat flatter-topped than the smooth GRTDT profiles, as expected in a turbulent flow, but the radial extent and the locations of the inner and outer edges of the mass distribution are well reproduced. The outflows do, however, matter for the total disk mass, and they substantially reduce the rate at which mass reaches the black hole: by late times only $\sim 20\%$ of the mass lost by the bound gas is accreted, the remainder being transferred to the outflow (Figure \ref{fig:horizonshare}), and as a result GRTDT over-predicts the horizon accretion rate by a factor $\sim 2$ while reproducing its time dependence. At later times, $t \gtrsim 3\times 10^6\,t_\mathrm{g}$, the outflows appear to dominate the evolution of the remaining bound gas, and the agreement with GRTDT breaks down (section \ref{sec:massloss}).}

\revision{{\it Thick disk kinematics and vertical structure.} Finally, GRTDT assumes that the gas follows circular, Keplerian orbits, and describes the disk only through height-integrated quantities. Neither assumption is obviously appropriate for a flow with $H/r \sim 1$, in which pressure support makes the rotation sub-Keplerian, and in which the vertical structure of the flow (including any meridional circulation) is absent by construction from the height-averaged theory. We do not attempt to characterize these effects directly here. We note, however, that the evolution of $r_{\rm max}$ derived in section \ref{sec:analysis} assumes that the specific angular momentum of the gas is Keplerian, and that this evolution is reproduced by the simulation. Whatever the detailed vertical and rotational structure of the flow, its net effect on the radial redistribution of the disk mass therefore appears to be small, at least at the level of accuracy of our comparison. As discussed in section \ref{sec:successes}, this is consistent with the fact that the corrections which appear at ${\cal O}(H/r)^2$ enter the global angular momentum budget only weakly.}

\revision{Taken together, the picture which emerges is that the alternatives to diffusive evolution which were most plausible a priori are either not realized in our simulation (a stress with an intrinsic scale or no simple description, a decaying turbulence, or dynamically important large-scale fields in the bulk of the disk), or are realized in a way which does not alter the character of the evolution of the bound gas (an outflow which removes angular momentum at the local specific angular momentum of the disk), but which does reduce the horizon accretion rate by a factor $\sim 2$ relative to GRTDT. GRTDT fails where its explicit assumptions fail: at the earliest times, where large density gradients drive radial velocities comparable to the orbital velocity (section \ref{sec:gradients}), and at the latest times, where outflows dominate the angular momentum budget. We stress that these conclusions are drawn from a single simulation, of a single magnetic field geometry, around a non-spinning black hole. In particular, it seems entirely possible that a different initial field geometry, or a spinning black hole capable of launching a jet, would produce outflows which exert a net torque on the disk, and therefore a qualitatively different evolution. A suite of long-duration simulations exploring these parameters would provide a valuable test of the generality of our results.}

\subsection{Successes of analytical theories}\label{sec:successes}
While it is clear that GRTDT has its limitations, particularly at the earliest times in the evolution of an accretion flow, it is clear from the results presented in this paper that GRTDT provides a remarkably good description of the long-time evolution of \revision{this} black hole accretion flow as described by GRMHD. 

In section \ref{sec:analysis} we showed that GRTDT predicts that the four summary statistics of an evolving accretion flow should all follow power law time dependence at large times, with only a single power law index $n$ of interest in the problem.  The physics of this result was that a local and scale-free turbulent transport of angular momentum (i.e., tensor component $W^r_\phi \propto r^\mu \rho^\eta$) always results in a power law decay of the mass accretion rate $\dot M \sim t^{-n}$, from which one can show that the global conservation laws of mass and angular momentum imply power law decays of the amplitude of the mass distribution and power law growth of the size of the disk. A more detailed theory of GRTDT, which explicitly solves the evolution equation for the density, shows that the width of the mass distribution also grows with a power law in time, simply a result of the purely diffusive (in angular momentum space) nature of GRTDT. 

The same four summary statistics are clearly power law functions of time in GRMHD (Figures \ref{fig:long_time_grmhd}, \ref{fig:summarystats}), and a detailed comparison shows that these power law behaviors are well approximated by GRTDT (Figure \ref{fig:summarystats}). The evolution of the full disk mass distribution $M'(r, t)$ shows that even in the details, GRTDT and GRMHD produce remarkably consistent evolving profiles. 

It is worth keeping in mind that the disk in our GRMHD simulation is not, in a traditional sense, particularly ``thin''. The initial aspect ratio of the flow is not small $H/r \sim {\cal O}(1)$ and, as the simulation does not contain radiation, remains broadly this thick throughout the evolution. The magnetic field in the inner regions of this flow was also dynamically strong (MAD), again not obviously a limit in which a simple thin disk theory might be expected to perform well.  The fact that a thin disk theory reproduces the broader evolutionary features of this full GRMHD evolution in some detail therefore implies that the corrections which appear at order ${\cal O}(H/r)^2$, such as pressure forces and the explicit vertical dependence to the variables themselves, do not appear to modify the long-term evolution of \revision{this} flow in a particularly important manner. 

\revision{Part of this success can be understood physically. Global conservation laws strongly constrain the bulk of the disk evolution: once the accretion rate is known to decay as a power law, mass and angular momentum conservation fix the evolution of the amplitude and location of the mass distribution (section \ref{sec:analysis}). Local effects such as pressure gradients and inner-disk magnetic field strengths enter the global angular momentum budget only weakly, through the modification of the angular momentum of each element of disk material from $\rho U^tU_\phi$ to $\rho U^tU_\phi(1+(c_S^2+v_A^2)/c^2)$. But conservation alone does not determine the decay index $n$, nor the evolution of the width of the mass distribution, nor indeed whether the evolution is diffusive at all (section \ref{sec:otherwise}). These are set by the nature of the angular momentum transport, and it is the empirical finding that this transport is local and outwards which allows GRTDT to reproduce the evolution in detail.}

The final piece of physics which makes GRTDT successful is that the turbulent stress extracted from our full GRMHD simulations is a relatively weak function of both radius and time (and therefore also density; see Figure \ref{fig:stress}). In principle $W^r_\phi$ (in simulations) could have shown an arbitrarily complicated spatio-temporal profile, preventing thin disk theory from being a practically useful simplification.  


\subsection{Implications for the $\alpha$-disk formalism }\label{sec:alpha}
The success of GRTDT in reproducing \revision{a full GRMHD simulation} over many decades in radius,  time and the amplitude of the evolving disk density, suggests that the general spirit of the $\alpha$-disk formalism (i.e., the existence of some local and scale free stress, which is well defined and drives the evolution) is successful. Of course, this is a different question from whether any particular {\it choice} of the $\alpha$-prescription successfully reproduces the evolution. We stress at the outset of this section a point to which we will return later, which is that the following $\alpha$-models all assume the presence of radiative losses, something not a part of our GRMHD simulations. 

Indeed, assuming the explicit \cite{SS73} $\alpha$ disk model with the common assumption that the gas pressure dominates (i.e., the gas pressure sets the sound speed, and that $W^r_\phi = \alpha r c_S^2$), it can be shown that \citep[e.g.,][]{Cannizzo90, Pringle91} 
\begin{equation}
    W^r_\phi \propto \Sigma^{2/3}\, r^{1/2} . 
\end{equation}
This does not appear to correspond well to the stress measured from the simulation (Figure \ref{fig:stress}). Note that while the peak  amplitude of $M'$ only drops by $\sim 1$ order of magnitude as the disk spreads to larger radii,  this actually corresponds to a  $\sim 5$ orders of magnitude drop in $\Sigma$ (as the disk also got a factor $\sim 100$ larger). A scaling of $W^r_\phi \sim r^{1/2} \Sigma^{2/3}$ would then predict a drastic (factor $\sim 10^2-10^3$) drop in the stress, which is not observed (Figure \ref{fig:stress}).  

Similar results hold for any generic ``traditional'' $\alpha$-model scaling, which as a rule all depend on the surface density to some non-zero power.  For example, another seemingly natural choice, a radiation pressure dominated $\alpha$-model, results in the unfortunate scaling $W^r_\phi \propto \Sigma^{-2}$, which leads to a diffusion equation for the disk density with a negative diffusion coefficient, an evolution which is inherently unstable \citep[it leads to clumping, not spreading, of the disk density][]{LightmanEardley74} and therefore clearly does not describe the evolution seen in our GRMHD simulations which is well described by a diffusive theory. 

What is likely going on here is that the turbulent transport of angular momentum is entirely magnetically dominated, and it is therefore the Alfv\'en velocity $v_A$ which sets the characteristic velocity scale of the turbulence. An example of an attempt to incorporate explicit magnetic fields into the $\alpha$-disk formalism goes back to \cite{Begelman07} and uses an Alfv\'en speed set equal to the geometric mean of the gas-pressure sound speed and Keplerian rotation speed, leading to
\begin{equation}
    W^r_\phi \propto \Sigma^{2/7}\, r^{3/14}, 
\end{equation}
which, with its weaker surface density dependence (at fixed radius), is somewhat better in keeping with our results than a pure gas pressure (it only predicts a factor $\sim 10$ drop in the stress at the location of disk mass maximum). However, if we examine the stress at a fixed radius $r\sim 100r_g$ we can also disfavor this description. The stress at $r\sim 100r_g$ shows almost no evolution over the temporal range $t=3\times 10^5-10^7t_g$ when the density of the disk drops by many orders of magnitude ($r\sim 100r_g$ goes from the location of the maximum $M'$ to well within the spreading disk during this time). 

It is important to emphasize that the above $\alpha$ stress models all assume that the saturated state of the disk turbulence is in some meaningful sense limited by a disk scale height set with radiative losses included. The simulations of \cite{Guo2025} have no such losses, and so it is not surprising that we do not find agreement between the stress needed to reproduce \cite{Guo2025} and these thin-disk stress prescriptions.   Testing the latter against simulations will require 
long timescale simulations with radiation.

\begin{figure*}
    \centering
    \includegraphics[width=0.49\linewidth]{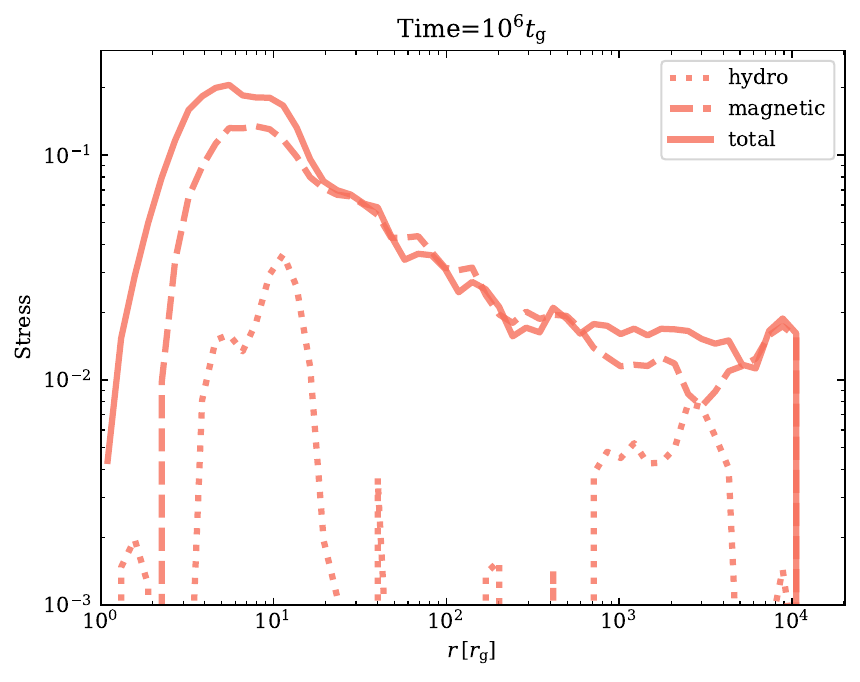}
    \includegraphics[width=0.49\linewidth]{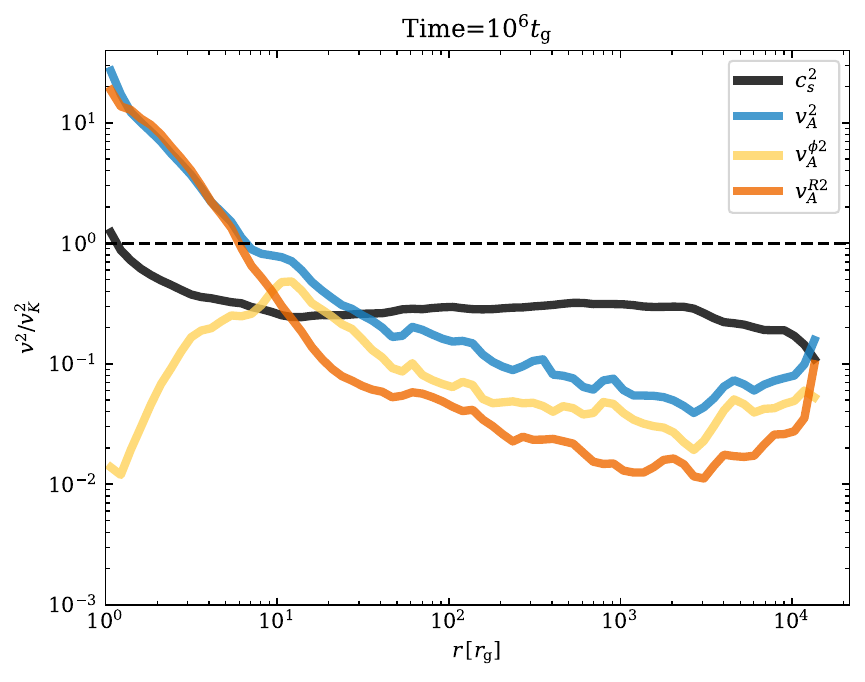}
    \caption{Properties of the turbulent stress (left) and various normalized velocities (right) in the disk at a time $t = 10^6 t_g$. On the left we demonstrate that despite the thermal pressure dominating in the disk, the stress is dominated at all radii by magnetic stresses. On the right we highlight the relevant velocity scales in the problem. In our MAD flow the radial magnetic field ramps up near to the horizon, while as a result of the zero spin of the hole the azimuthal component drops markedly. At larger scales $r\gtrsim 100r_g$ the magnetic and thermal velocities reach \revision{approximately} fixed fractions of the Keplerian velocity, indicating a\revision{n approximately} fixed disk aspect ratio.   }
    \label{fig:stress_mil}
\end{figure*}

In fact, it is somewhat natural in a non-radiative magnetically dominated flow for $W^r_\phi$ to be only weakly radially dependent at large radii, as we now argue. Let us assume that the magnetic fields dominate the turbulent stress, and that a mixture of magnetic and gas pressure determine vertical hydrostatic balance. Move to a Newtonian gravity for convenience, and note that vertical hydrostatic balance becomes (where $\beta_{\rm sat}$ is the saturated value of the usual plasma beta parameter) 
\begin{equation}
    - {\partial \over \partial z} \left({B^2 \over 8\pi} (1 + \beta_{\rm sat})\right) - \rho {\partial \over \partial z} \Phi \approx 0,
\end{equation}
or for $\Phi = -GM_\bullet/r$ (spherical $r$), 
\begin{equation}
    {\partial \over \partial z} \left({B^2 \over 8\pi} (1 + \beta_{\rm sat})\right) \approx - {GM_\bullet \rho z \over r^3} .
\end{equation}
We then ``integrate'' this expression by taking $z \sim (\partial/\partial z)^{-1} \sim H$, and $r \approx R$ (cylindrical radius), to find 
\begin{multline}
    v_A^2 = {B^2 \over 4\pi \rho } \approx \left({H\over R}\right)^2 {GM_\bullet \over R} {1 \over  (1 + \beta_{\rm sat})} \\ \approx \left({H\over R}\right)^2  {v_K^2\over  (1 + \beta_{\rm sat})},
\end{multline}
where $v_K = \sqrt{GM_\bullet/R}$ is the Keplerian rotational velocity. Then, if the stress is magnetically dominated, we have 
\begin{equation}
    W^{R\phi} = - \left\langle {B^R B^\phi \over 4\pi \rho}\right\rangle \equiv {\cal P}^{R\phi} \, v_A^2, 
\end{equation}
where we define the projector 
\begin{equation}
    {\cal P}^{R\phi} \equiv - { \left\langle B^R B^\phi \right\rangle  \over B^2} .
\end{equation}
Then combining these results we have
\begin{align}
    W^r_\phi &=  r W^{R\phi}, \nonumber\\ 
    &\approx {\cal P}^{R\phi}\left({v_A\over v_K}\right)^2 G M_\bullet, \\
    &\approx {\cal P}^{R\phi}\left({H\over R}\right)^2 {G M_\bullet \over  (1 + \beta_{\rm sat})}. 
\end{align}
We have written this final stress parameterisation out twice; the middle expression always holds whenever magnetic fields dominate the stress, while the final aspect ratio expression only holds when the magnetic field also plays an important role in vertical hydrostatic balance. One therefore needs to estimate $v_A$ and ${\cal P}^{R\phi}$ to make further progress. 

We envisage two possible saturation arguments. Far from the black hole, in a flow without radiative losses, there is likely no natural scale for the aspect ratio $H/R$ to ``aim for'', and it is plausible that the saturated magnetic energy density will be just set by an initial internal energy density/scale height associated with the initial condition, i.e.,  $H/R \approx \theta_0 \equiv (H/R)_{t=0}$. In which case the saturated magnetic field would satisfy
\begin{equation}
    {v_A^2 \over v_K^2} \approx {\theta_0^2 \over  (1 + \beta_{\rm sat})}, \quad {\rm or} \quad B^2 \approx {4\pi \rho \theta_0^2 v_K^2 \over  (1 + \beta_{\rm sat})}.
\end{equation}
The question then becomes about the spatio-temporal structure of ${\cal P}^{R\phi}$. Close to a spinning black hole one might expect a ramping up of ${\cal P}^{R\phi}$ owing to field lines being dragged with the horizon, but far from a black hole it is perhaps more plausible that ${\cal P}^{R\phi}$ is  a weak function of disk radius, as there seems no obvious scale over which the tangled $R\phi$ field correlations might change. Then, with no radial dependence to ${\cal P}^{R\phi}$ and $\theta_0$ fixed, $W^r_\phi \sim {\rm constant}$ emerges. 

Indeed, an equipartition guess of ${\cal P}^{R\phi}\sim 1/3$, along with an initial aspect ratio of $\theta_0 = (H/R)_{t=0} \sim 1/2$ and a saturation of $\beta_{\rm sat} \sim 3$ (like that seen in our simulation), would then give a value \revision{$W^r_\phi \sim 0.02$}, certainly in keeping with what we observe\revision{ (Figure \ref{fig:stress})}. An obvious test of this saturation argument would be to run another long time GRMHD simulation with a much thinner initial condition (such as a \citealt{Chakrabarti85} initial condition) and extract the saturated magnetic stress from that simulation. 

Alternatively, it is plausible that $v_A$ at all radii is set by the inner disk/near horizon saturation of the field, which then falls off with some self similar radial profile to large distances (i.e., the magnetic field at all disk radii is controlled by near horizon physics). This framework still facilitates a natural future numerical experiment, as one could compare the natural evolutionary timescales of MAD and SANE disks, which are known to saturate their radial field at different levels through the normalized horizon magnetic flux parameter (here an integral over ${\rm d}^2 S$ is an integral over the upper black hole horizon hemisphere) 
\begin{equation}
    \phi_{\rm BH} \equiv {{\iint} B^r\, {\rm d^2}S \over \left({\iint \rho U^r\, {\rm d}^2S}\right)^{1/2}} \sim {v_A^R \over c} .
\end{equation}
As the \cite{Znajek77} regularity condition requires that in a degenerate magnetosphere the ratio $B^\phi/B^r$ must take a specific value at the horizon\footnote{Somewhat unfortunately, the only time this ratio approaches zero is for a Schwarzschild black hole, the precise geometry we have here. Future simulations over a range of spins are certainly warranted and interesting. }, one might estimate in GRMHD that $v_A^\phi \sim v_A^r$ on near horizon scales, such that the amplitude of the stress at all radial scales
\begin{equation}
    W^r_\phi \sim \phi_{\rm BH}^2.
\end{equation}
If this saturation argument is correct, then one would expect two distinct evolutionary timescales for MAD and SANE disks, as 
\begin{equation}
    t_{\rm evol} \sim 1/w \sim 1/\phi_{\rm BH}^2, 
\end{equation}
and $\phi_{\rm BH}$ is known to be bi-modal. Some circumstantial evidence in favor of this argument is provided by \cite{Wong25}, who found systematically faster inflow for MAD simulations compared to SANE simulations, although further numerical analysis is required. 

Observing the properties of the turbulent stress and velocity scales seen in our simulation at large times (Figure \ref{fig:stress_mil}), we see that there are elements of both saturation arguments present in our solution. At large radial scales both the gas pressure (or sound speed $c_s^2/v_K^2$) and magnetic pressure (or Alfv\'en speed $v_A^2/v_K^2$) reach \revision{approximately} fixed fractions of $v_K^2$, indicating a\revision{n approximately} constant aspect ratio of order 1, much like that associated with the initial condition. However, closer to the horizon (within \revision{$r \lesssim 20\,r_g$} at this late time) we see a dramatic transition away from these constant ratios, with the radial component of the field ramping up. There is, at this late stage, barely any mass located on these near horizon scales, with most of the matter in the constant ratio region. This likely explains why a constant $w$ describes well the evolution of the bulk of the matter despite this near-horizon physics, but a future suite of simulations testing different magnetic field geometries will provide the final say. 

Note that in a classical $\alpha$-viscosity formalism one has $W^r_\phi = \alpha (H/R)^2 GM_\bullet$, meaning that while our derivation is motivated differently, all that has been modified in practical terms is the change from a parameter named $\alpha$ to one named ${\cal P}^{R\phi} = - \left\langle B^R B^\phi / B^2 \right\rangle$, with a different physical interpretation.


\begin{figure*}
    \centering
    \includegraphics[width=0.49\linewidth]{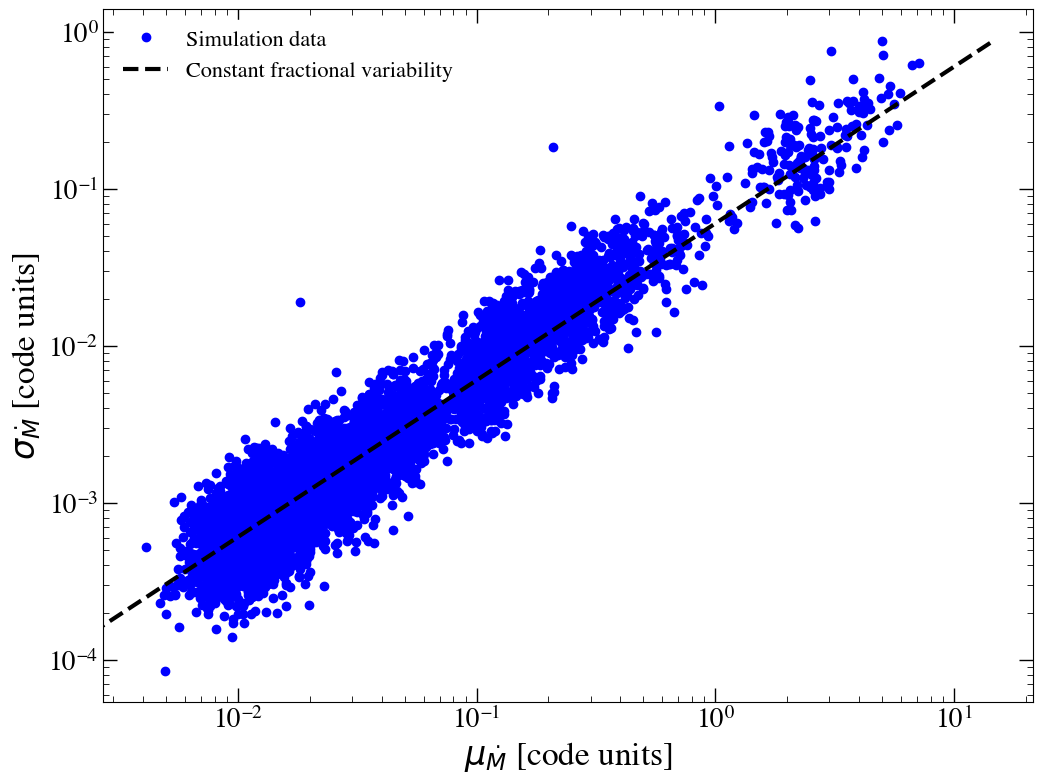}
    \includegraphics[width=0.48\linewidth]{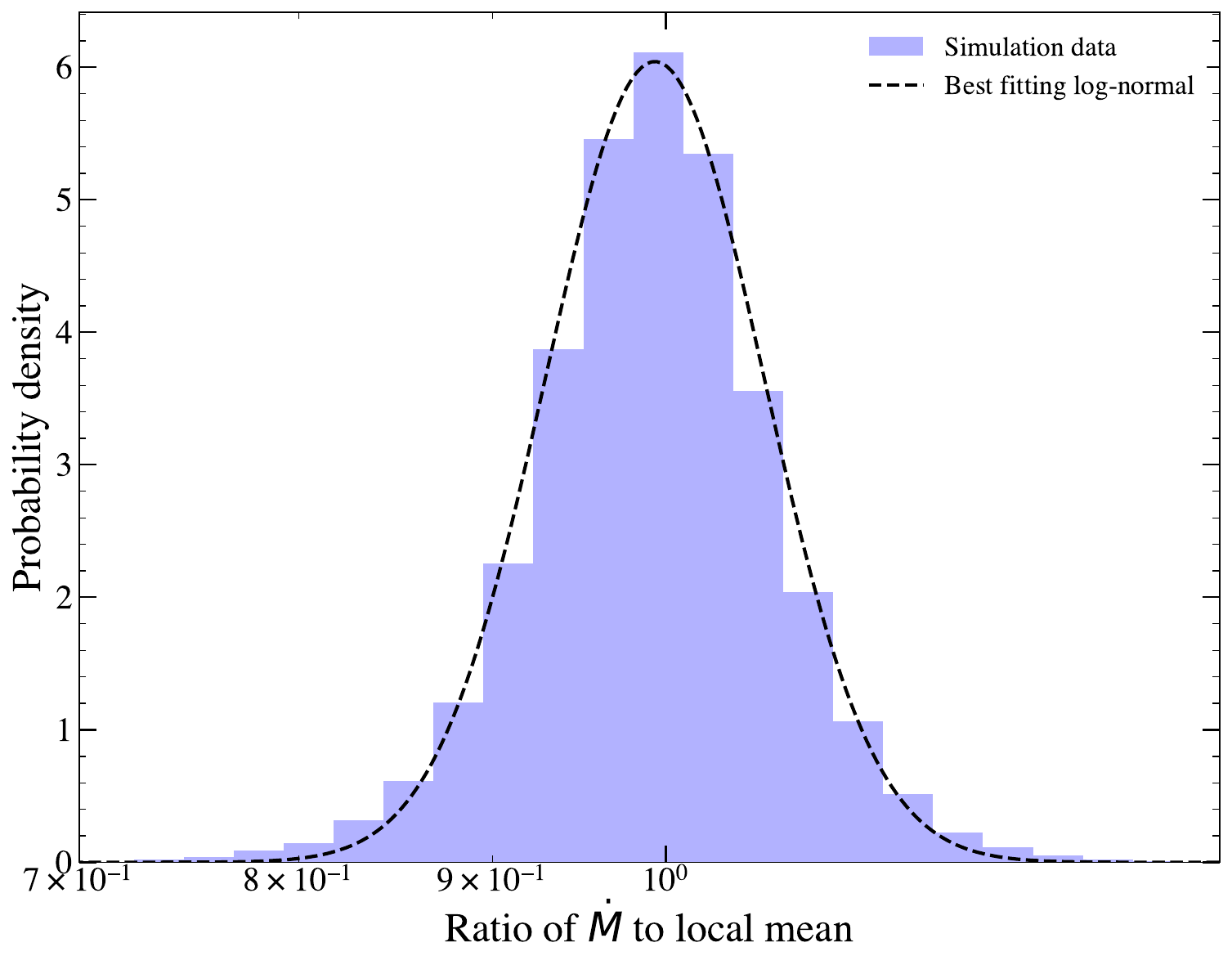}
    \caption{The properties of the variability of the horizon mass accretion rate in the GRMHD simulation, and simple predictions motivated by GRTDT. In the left hand panel we show the ``RMS-flux'' relationship (see text) of the simulation data (blue points), calculated by binning the mass accretion rate in the simulation (e.g., Figure \ref{fig:summarystats}) into bins with 10 points in each bin\revision{ (bins of duration $100\,t_\mathrm{g}$)}. The black dashed line is a simple constant fractional variability line $\sigma_{\dot M} = 0.06 \mu_{\dot M}$, which clearly provides an adequate description of the data over 3 orders of magnitude in $\dot M$. The right hand panel shows the distribution of the ratio of $\dot M$ to the local temporal mean of $\dot M$ in the simulation (on a logarithmic horizontal axis; blue histogram), and a log-normal distribution (black dashed curve). This log-normal distribution provides an excellent description of the observed distribution.    }
    \label{fig:variability}
\end{figure*}

\subsection{Short timescale variability}\label{sec:variability}

A by-product of the simulations presented in this work is the first opportunity to study the short-timescale variability of the horizon accretion rate of a realistic GRMHD simulation over many decades in time. This is an important intersection between observational and theoretical studies of the accretion process, as stochastic variability is ubiquitously observed from accreting black hole systems. This variability can be as rapid as the local orbital timescale of the very inner disk, or orders of magnitude longer than any physical process which is involved in the direct production of the disk's luminosity.  Broadly similar gross disk variability properties are also observed across different systems which span a huge array of length scales: from the compact disks in Galactic X-ray binaries \citep[e.g.,][]{Uttley01}, all the way up to the colossal disks in  active galactic nuclei \citep[e.g.,][]{Vaughan2011}. 



Connections between observed variability and theoretical models of the accretion process have thus far been limited to simple thin disk theory.  Indeed, GRTDT can be utilized to study short timescale variability  within a framework known as the ``theory of propagating fluctuations'' \citep{Lyubarskii97, Mushtukov2018, Mummery23b}. One of the predictions of which, with good observational support \citep[e.g.,][]{Uttley05},  is that the distribution of $\dot M$ about its evolving mean should be {\it log-normal}. This simple analytical prediction of thin disk theory has been confirmed with stochastic viscous-hydro simulations (within the thin disk theory framework) in both one \citep{Turner21} and two spatial dimensions \citep{Turner23}. This log-normal prediction has two simple implications which we can test with our long time GRMHD data. 

The first is that the accretion rate onto the black hole should follow a ``linear RMS-flux relationship'' (\revision{this is a property of the log-normal distribution, although not unique to it: any fluctuations whose amplitude scales with the mean produce such a relation, and so it is the shape of the distribution of $R$ which tests log-normality specifically}). What this means is that if the time series of $\dot M$ is binned into a series of temporal bins $\Delta t_j$, then the mean of $\dot M$ in that bin (denoted $\mu_{\dot M}$) should be linearly proportional to the standard deviation of $\dot M$ in that bin (denoted $\sigma_{\dot M}$). The second prediction is that if one computes the ratio of the values of the accretion rates within each bin to the mean of the accretion rates in that bin (i.e., $R_j = \{\dot M_j\}/\langle \{ \dot M_j\}\rangle$ for a list $\{\dot M_j\}$ of accretion rates), then $R$ should be log-normally distributed. 

Both of these predictions are recovered in our long timescale GRMHD run, as can be seen in Figure \ref{fig:variability}. On the left hand panel we show the ``RMS-flux'' relationship of the simulation data (blue points), calculated by binning the mass accretion rate in the simulation (e.g., Figure \ref{fig:summarystats}) into bins with 10 points in each bin\revision{ (i.e., bins of duration $100\,t_\mathrm{g}$)}. The black dashed line is the simple constant fractional variability line $\sigma_{\dot M} = 0.06 \mu_{\dot M}$, which clearly provides an adequate description of the data over 3 orders of magnitude in $\dot M$. The right hand panel shows the distribution of $R$ in the simulation (on a logarithmic horizontal axis; blue histogram), and a log-normal distribution (black dashed curve). This log-normal distribution provides an excellent description of the observed distribution.  \revision{These bins are much shorter than the timescale of the secular evolution of $\dot M$ (Figure \ref{fig:summarystats}), which therefore contributes negligibly to the variability within a bin.} 

\revision{The variability of the horizon accretion rate is also strongly time-correlated. To quantify this, we first remove the secular decline of the accretion rate by fitting $\ln|\dot M| = a + b\ln t$ over $2\times10^4 < t < 6.6\times10^5\,t_\mathrm{g}$, and take the residual $\delta(t) = \ln|\dot M| - (a + b\ln t)$, which is the fractional fluctuation of $\dot M$ about its power-law trend. The horizon accretion rate is recorded every $10\,t_\mathrm{g}$, and we estimate the power spectrum $P(f)$ of $\delta(t)$ with Welch's method, using segments of length $\approx 4\times10^4\,t_\mathrm{g}$ with a Hann window. The resulting spectrum is averaged in logarithmically spaced frequency bins, and fitted with a smoothly broken power law
\begin{equation}
P(f) = \frac{A\,(f/f_b)^{-\alpha_1}}{1 + (f/f_b)^{\alpha_2 - \alpha_1}} ,
\end{equation}
which tends to $P \propto f^{-\alpha_1}$ for $f \ll f_b$ and $P \propto f^{-\alpha_2}$ for $f \gg f_b$. The fit is performed in $\ln P$ and restricted to $f < 2\times10^{-2}\,t_\mathrm{g}^{-1}$, above which the spectrum flattens as it approaches the Nyquist frequency of the output cadence.}

\begin{figure}
    \centering
    \includegraphics[width=\linewidth]{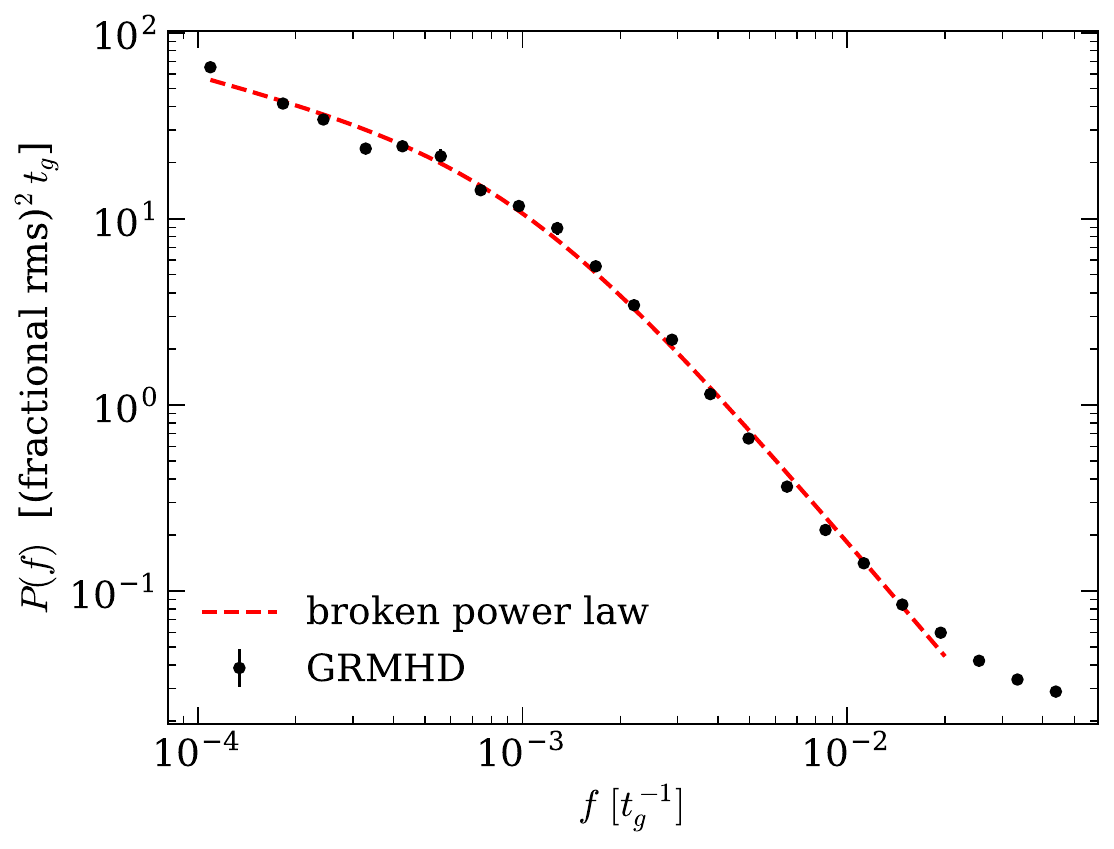}
    \caption{\revision{The power spectrum of the fractional fluctuations of the horizon accretion rate about its secular decline, for the standard GRMHD run over $2\times10^4 < t < 6.6\times10^5\,t_\mathrm{g}$ (black points), and a broken power-law fit (red dashed curve; see text). The flattening at the highest frequencies, $f \gtrsim 2\times10^{-2}\,t_\mathrm{g}^{-1}$, approaches the Nyquist frequency of the $10\,t_\mathrm{g}$ output cadence and is excluded from the fit.}}
    \label{fig:psd}
\end{figure}

\revision{In Figure \ref{fig:psd} we show the power spectrum of the fractional fluctuations of $\dot M$ about its secular power-law decline, computed from the standard run over $2\times10^4 < t < 6.6\times10^5\,t_\mathrm{g}$. The spectrum is well described by a broken power law, $P(f) \propto f^{-0.45\pm0.13}$ below a break frequency $f_b \approx (1.0 \pm 0.25)\times10^{-3}\,t_\mathrm{g}^{-1}$, steepening to $P(f) \propto f^{-2.06\pm0.09}$ above it. This form is robust to the method used to remove the secular trend and to the length of the segments over which the spectrum is estimated, which change $f_b^{-1}$ only within the range $\approx 900$--$1400\,t_\mathrm{g}$. Such band-limited, red-noise spectra, flat at low frequencies and steepening to $P \propto f^{-2}$ above a break, are characteristic of the observed variability of accreting black holes \citep[e.g.,][]{Uttley01, Vaughan2011}, and are a natural expectation of the propagating fluctuations paradigm \citep{Lyubarskii97, IngramDone2011,Mushtukov2018,Mummery23b}, in which the break reflects the characteristic timescale of the innermost regions of the flow; relativistic propagating-fluctuation models produce power spectra of broadly this form \citep[e.g., figure 16 of][]{Mummery23b}.}

\revision{Two properties of this break are notable. First, it does not evolve with time: fitting the spectrum in shorter windows gives break timescales of $\sim 10^3$--$2\times10^3\,t_\mathrm{g}$ with no systematic trend, while over the same period the peak of the disk mass distribution moves outwards by a factor $\sim 20$. The break is therefore set by the inner flow, within a few tens of $r_\mathrm{g}$, rather than by the bulk of the disk. Second, it is not straightforward to associate the break with a single radius. Its timescale is comparable to the inflow timescale of the inner $\sim 10$--$20\,r_\mathrm{g}$ for the stress amplitude measured in our simulation, to orbital periods at $\sim 30\,r_\mathrm{g}$, and to the extent of the region in which the magnetic field is dynamically dominant (Figure \ref{fig:stress_mil}), in which magnetic flux eruptions may also contribute; the order-unity factors involved in relating a break frequency to a radius do not allow us to distinguish between these. The break timescale is also comparable to orbital timescales at the radius of the initial torus, but it is unchanged while the density there falls by more than two orders of magnitude, and so does not appear to reflect a memory of the initial mass distribution. Whether it depends on the initial magnetic flux, which is set by the initial condition, could be tested with simulations of different initial tori.}

The results presented here therefore strengthen the simple ``propagating fluctuations'' paradigm, and suggest that the inclusion of log-normally distributed fluctuations around a smoothly evolving GRTDT mean evolution should capture much of the stochastic nature of accreting systems. 

\newpage
\subsection{Observational implications}
The results of this work have important implications for observations of intrinsically time-dependent black hole accretion flows. The most obvious system which is described by such a time dependent accretion flow are tidal disruption events (TDEs). TDEs occur when an unfortunate star is perturbed onto a near-radial orbit about the supermassive black hole, and upon coming sufficiently close the self gravity of the star $a_\star \sim GM_\star / R_\star^2$ is overcome by the tidal gravity of the black hole  $a_T \sim GM_\bullet R_\star/r^3$, i.e., within a radius $r_T \sim R_\star (M_\bullet/M_\star)^{1/3}$. 

The initial evolution of the tidal debris of the star is complicated, but there is a large amount of observational evidence that the debris then settles down into an evolving accretion flow \citep[e.g.][]{vanVelzen19, MumBalb20a, Guolo24, Mummery_et_al_2024, Guolo25, Guolo25c}. At times beyond $\sim 1$ year post disruption, all of the observed luminosity from such systems is well described (from optical--X-ray frequencies) as being sourced by an evolving disk \citep{Guolo25c}. 

It is important to note that the observational timescales of these systems $\Delta t_{\rm obs} \sim 1\, {\rm month}-10$ years, correspond, for a typical black hole mass of a TDE candidate of $M_\bullet \approx 4 \times 10^6 M_\odot$, to gravitational times of $\Delta t_{\rm obs} \sim 10^5-10^7\, t_\mathrm{g}$, \revision{precisely} the timescales probed by our simulations. The (early) timescales over which GRTDT deviates from GRMHD ($\sim {\rm few} \times 10^3 t_\mathrm{g}$) correspond for a typical TDE black hole to a relatively short period $(\sim$ days). 

The good agreement between GRMHD and GRTDT at late times is an important result, as GRTDT has been used as the basis for a wide range of data modeling techniques of TDE systems \citep[e.g.,][]{MumBalb20a, mummery2024fitted, Nicholl24, Guolo25, Goodwin25, Chakraborty25}. One particularly important element of TDE evolution is a long-lived light curve ``plateau'' (near time independent phase) seen at times $\gtrsim 1$ year in the optical-UV bands \citep{vanVelzen19, MumBalb20a}. This results (in the GRTDT framework) from a combination of disk cooling due to mass loss through the inner edge (which would decrease the optical-UV flux) and disk spreading to larger radii (increasing the emitting area, and therefore the flux at optical-UV frequencies). These two effects cancel out in GRTDT, leading to a near time-independent evolutionary phase \citep{Mummery_et_al_2024}. 

\revision{Our results also bear on the physical parameters inferred from such fits. In GRTDT all of the mass lost by the disk is accreted, so that a disk mass inferred from the observed evolution of the inner-disk accretion rate (and hence luminosity) implicitly assumes that the horizon share of the mass loss is unity. In our simulation, by contrast, only $\sim 20$--$70\%$ of the mass lost by the bound disk reaches the black hole, with the remainder unbound into an outflow, while the time dependence of the accretion rate is nonetheless reproduced by GRTDT. The horizon share is likely to depend on the thickness of the disk, its magnetic field geometry and the spin of the black hole, and radiatively efficient disks, such as those expected in TDEs, are thinner than the disk simulated here and may launch much weaker outflows. We therefore do not suggest that the factor found here applies to observed systems. Rather, our result highlights that if outflows are present, a disk mass inferred from a GRTDT fit to an accretion-rate-driven light curve is a lower limit on the mass processed through the disk, and that the shape of the light curve alone cannot distinguish between these cases.} 

While our GRMHD simulations do not contain radiation physics and so we cannot determine ``light curves'' from first principles, the key evolutionary phases which drive this plateau phase in TDE systems are all present in our long timescale simulations. The density (and therefore temperature) and accretion rate of the inner disk regions drop with time, and the bulk of the disk mass spreads to progressively larger radii.  \revision{Within the GRTDT framework, the disk in our simulations would therefore be expected to produce a long-lived plateau phase at optical/UV frequencies, as is observed in TDE light curves.}

\section{Conclusions}\label{sec:conclusions}
In this paper we have used cutting edge, long-timescale GRMHD simulations to examine the long time evolution of black hole accretion flows (out to an unprecedented $\sim 10^6-10^7$ gravitational times). In particular we have shown how the gross features of this evolution can be understood with reference to global conservation laws which can be formulated from the equations of GRMHD. Moving to an approximate GRTDT framework, the evolution of this system can be reproduced on an even finer level of detail. 

This work has a range of important implications for both observational and theoretical work.  Firstly, we have highlighted several features of GRTDT which can be explored/improved in future work. We have shown that the GRMHD evolution over the first $\sim$ ``viscous'' timescale (a misnomer of course) is poorly reproduced by GRTDT, a result of self inconsistencies of this theory in regions of large density gradients. Further, our GRMHD simulations launch winds which ultimately remove $\sim {\cal O}(1)$  of the disk mass in an unbound outflow\revision{. These winds do not substantially alter the evolution of the bound mass distribution, but they suppress the horizon accretion rate by a factor $\sim 2$ relative to GRTDT}. We have further shown that the turbulent stress $W^r_\phi$ of the GRMHD flow is only a weak function of both radius and time.

Incorporating both of these physical effects should be the subject of future studies, and it will be particularly interesting to see \revision{whether a theory incorporating mass loss to outflows (section \ref{sec:massloss}) can reproduce both the early-time evolution and the normalization of the horizon accretion rate}. 

Another key result of this work is that we have \revision{recovered} the old observational \citep{Uttley05} and theoretical \citep{Lyubarskii97, Turner21} prediction that the variability in the mass accretion rate onto the black hole should be log-normally distributed about its (evolving) mean. \revision{We have also shown that this variability has a band-limited red-noise power spectrum, flat below a break at $f_b \approx 10^{-3}\,t_\mathrm{g}^{-1}$ and steepening to $P \propto f^{-2}$ above it, with a break frequency which does not evolve as the disk spreads, and which is therefore set by the inner flow.} This key numerical result opens up the possibility of extending GRTDT with log-normal  perturbations driven at each radius, which may offer insight into the observed variability structure of accreting systems (in a way which is much more easily calculable than full GRMHD simulations). 

Our results are strongly supportive of the GRTDT framework, and its utility in describing the bulk features of evolving black hole accretion flows. \revision{The fact that over nearly three decades in time, during which the accretion rate falls by three orders of magnitude and the disk spreads by more than two orders of magnitude in radius, a simple analytical theory can reproduce so much of the evolution of a relativistic, magnetically arrested fluid which is not thin, is a non-trivial result, and one which was not guaranteed by the conservation laws which both theories share.} \revision{If these results hold more generally,} GRTDT will have \revision{clear} utility in describing observational systems where a full exploration of GRMHD parameters would be prohibitively expensive. This includes soft X-ray flares from Galactic X-ray binary systems, and multi-wavelength light curves of tidal disruption events. 

Observational implications of our work include a general support for a spreading-and-cooling disk interpretation of optical/UV light curve ``plateaus'' in tidal disruption events. 

\section*{Acknowledgements}
A.M. acknowledges support from the Ambrose Monell Foundation, the W.M. Keck Foundation and the John N. Bahcall Fellowship Fund at the Institute for Advanced Study. A.M. is grateful for various conversations with George Wong throughout the completion of this work. 
We acknowledge the EuroHPC Joint Undertaking for awarding this project access to the EuroHPC supercomputer LUMI, hosted by CSC (Finland) and the LUMI consortium through a EuroHPC Regular Access call.
The authors are pleased to acknowledge that the work reported on in this paper was substantially performed using the Princeton Research Computing resources at Princeton University, which is consortium of groups led by the Princeton Institute for Computational Science and Engineering (PICSciE) and Office of Information Technology's Research Computing.

\bibliography{andy, minghao}{}
\bibliographystyle{aasjournal}

\appendix

\section{Global conservation law analysis}\label{sec:global_app}
\subsection{Constructing a global conservation law}
To construct a global conservation law in relativity one must be somewhat careful, as in general the integral of a covariant quantity is not itself a covariant quantity. By working with sufficient factors of $|g|$ however this can be accounted for. If we have an (otherwise completely general) 4-vector $V^\mu$ and a metric $g_{\mu\nu}$ in coordinate basis $x^\mu$ then the ``covariant derivative'' is given by 
\begin{equation}
    \nabla_\mu V^\mu = {1\over \sqrt{|g|}} {\partial \over \partial x^\mu} \left(\sqrt{|g|} V^\mu \right) ,
\end{equation}
where $|g|$ is the (absolute value of the) determinant of the metric. This means that 
\begin{equation}
    \iiiint_{\cal V}  (\nabla_\mu V^\mu ) \, \sqrt{|g|} \, {\rm d}^4 x = \iiiint_{\cal V}  {\partial \over \partial x^\mu} \left(\sqrt{|g|} V^\mu \right) \, {\rm d}^4 x  = {\underset {\quad\quad\,\,\, \partial {\cal V}} \oiiint} \sqrt{|g|} V^\mu n_\mu \, {\rm d}^3 x, 
\end{equation}
where the final step is a simple application of Gauss' law (as integrals and partial derivatives are inverse operations, even in a curved space). This final result is covariant, as both $\sqrt{|g|} {\rm d}^4 x$ and $\nabla_\mu V^\mu$ are covariant quantities (in the first integral). In these expressions ${\cal V}$ is a general 4-dimensional spacetime volume, bounded by a 3-dimensional spacetime surface $\partial {\cal V}$ with normal $n_\mu$. In other words a single volume integral of the covariant divergence can be transformed into a (sum of) surface integrals. 

\subsection{Mass conservation}
We can apply this result directly to the local expression of mass conservation in the fluid 
\begin{equation}
    \nabla_\mu (\rho U^\mu) = 0,
\end{equation}
where $\rho$ is the rest mass density and $U^\mu$ the fluid 4-velocity. Integrating over the volume defined by ${\cal V} = [0, t] \times [r_+, \infty] \times [0, \pi] \times [0, 2\pi]$ (i.e., spherical in space from the horizon to infinity and linear in time as measured at infinity), we find
\begin{multline}
    \iiiint_{\cal V} \sqrt{|g|} \nabla_\mu (\rho U^\mu) \, {\rm d}^4 x = \int_{r_+}^\infty \int_0^{\pi} \int_0^{2\pi} \left[ \left.(\rho U^t)\right|_{t=t} - \left.(\rho U^t)\right|_{t=0} \right] \sqrt{|g|} \, {\rm d}\phi \, {\rm d}\theta \, {\rm d}r \\ - \int_{0}^t \int_0^{\pi} \int_0^{2\pi} \left.\left(\sqrt{|g|} \rho U^r\right)\right|_{r=r_+}  \,  {\rm d}\phi \, {\rm d}\theta \, {\rm d} t' = 0 ,
\end{multline}
where we have assumed that $\sqrt{|g|} \rho$ vanishes as $r \to \infty$ (i.e., the total mass is finite). This has a clear physical interpretation. Defining the time-dependent disk mass
\begin{equation}
    M_d(t) \equiv \int_{r_+}^\infty \int_0^{\pi} \int_0^{2\pi} \left.\left(\sqrt{|g|} \rho U^t\right)\right|_{t=t}\, {\rm d}\phi \, {\rm d}\theta \, {\rm d}r ,
\end{equation}
and the accretion rate
\begin{equation}
    \dot M \equiv \int_0^{\pi} \int_0^{2\pi} \left.\left(\sqrt{|g|} \rho U^r\right)\right|_{r=r_+, \, t=t}  \,  {\rm d}\phi \, {\rm d}\theta ,
\end{equation}
we find 
\begin{equation}
    M_d(t) - M_d(0) = \int_0^t \dot M(t') \, {\rm d}t'. 
\end{equation}
\subsection{Angular momentum conservation}
A very similar expression can be derived for angular momentum conservation. The Kerr metric is independent of the coordinate $\phi$, meaning that there is a Killing vector $k^\mu = \delta^\mu_\phi$ which by definition satisfies $\nabla_\mu k_\nu + \nabla_\nu k_\mu = 0$ (meaning that $\nabla_\mu k_\nu$ is an antisymmetric tensor). The 4-vector 
\begin{equation}
    J^\mu \equiv k_\nu T^{\mu \nu} ,
\end{equation}
where $T^{\mu\nu}$ is the stress energy tensor, satisfies 
\begin{equation}
    \nabla_\mu J^\mu = T^{\mu\nu} \nabla_\mu k_\nu + k_\nu \nabla_\mu T^{\mu\nu} = 0,
\end{equation}
where the first term vanishes on symmetry grounds and the second by the conservation of energy momentum. Thus 
\begin{multline}
    \iiiint_{\cal V} \sqrt{|g|} \nabla_\mu J^\mu \, {\rm d}^4 x = \int_{r_+}^\infty \int_0^{\pi} \int_0^{2\pi} \left[ \left.(J^t)\right|_{t=t} - \left.(J^t)\right|_{t=0} \right] \sqrt{|g|} \, {\rm d}\phi \, {\rm d}\theta \, {\rm d}r  \\ - \int_{0}^t \int_0^{\pi} \int_0^{2\pi}  \left.\left(\sqrt{|g|}J^r\right)\right|_{r=r_+} \,  {\rm d}\phi \, {\rm d}\theta \, {\rm d} t' = 0 .
\end{multline}
Where we have again assumed that $\sqrt{|g|}J^r\to 0$ sufficiently quickly so that the integral at infinity vanishes (i.e., the total angular momentum of the system is finite). Using the stress energy tensor of ideal GRMHD we have that
\begin{equation}
    J^\mu = \left(\rho + {P+e + b^2/4\pi\over c^2} \right) U^\mu U_\phi + \left(P + {b^2 \over 8\pi}\right) \delta^{\mu}_{\phi} - {b^\mu b_\phi \over 4\pi } .
\end{equation}
This means we can define the disk's angular momentum
\begin{equation}
    J_d(t) \equiv \int_{r_+}^\infty \int_0^{\pi} \int_0^{2\pi} \sqrt{|g|} \left. \left[\rho U^t U_\phi \left(1 + {c_s^2 + v_A^2 \over c^2} \right) - {b^tb_\phi\over 4\pi} \right] \right|_{t=t}  \, {\rm d}\phi \, {\rm d}\theta \, {\rm d}r ,
\end{equation}
and the angular momentum accretion rate (by analogy with mass accretion rate)
\begin{equation}
    \dot J \equiv \int_0^{\pi} \int_0^{2\pi}\left. \left(\sqrt{|g|} \left[\rho U^r U_\phi \left(1 + {c_s^2 + v_A^2 \over c^2} \right) - {b^rb_\phi\over 4\pi} \right] \right) \right|_{r=r_+} \, {\rm d}\phi \, {\rm d}\theta ,
\end{equation}
so that 
\begin{equation}
    J_d(t) - J_d(0) = \int_0^t \dot J(t') \, {\rm d}t'. 
\end{equation}
In these expressions we have defined $c_s^2 \equiv (P+e)/\rho$ and $v_A^2 \equiv {b^2 / 4\pi\rho}$. Finally, note that the integral of the transport term $b^rb_\phi$ over the spherical  surface at the horizon must vanish, as a result of causality.

\section{Green's function solutions}\label{app:sol}
While exact solutions to the Newtonian thin disk equations have been known for a long time \citep[e.g.,][]{LBP74, Pringle81}, they are  inappropriate for a detailed comparison to a GRMHD simulation, as they have unfortunate properties in the limit $r \to 0$, which clearly are not relevant for a black hole system. One could simply curtail these solutions at a finite radius (to mimic the ISCO, or event horizon, etc.), but this does not take into account the modifications to the dynamics of the flow induced by a relativistic space time. 

Fortunately, accurate asymptotic solutions of the full GRTDT equations are known, provided one makes the simplifying assumption that the evolution equation is linear (or equivalently neglect a density dependence in $W^r_\phi$; i.e., takes $\eta = 0$). A Green's function solution of the GRTDT density equation was derived by \cite{Mummery23a}, and has the functional form  
\begin{multline}\label{green_s}
G_\Sigma(x, \tau |x_0) =  {M_d(0) \over 2\pi r_I^2 c_0}  \sqrt{x^{-\beta} f_\beta(x) \exp\left(-{1 \over x} \right) \left(1 - {2\over x}\right)^{5/2 - 3/4\beta} } \\
 {x^{-3/4 - \mu} \over \tau} \exp\left({-f_\beta(x)^2 - f_\beta(x_0)^2 \over  4\tau} \right) I_{1\over 4\beta} \left({ f_\beta(x) f_\beta(x_0) \over  2\tau}\right),
\end{multline}
where 
\begin{equation}
f_\beta(x) = {x^\beta \over 2 \beta} \sqrt{1 - {2\over x}}\left[1  - {x^{ - 1} \over { (\beta - 1)}}\, {}_2F_1\left(1, {3\over 2}-\beta; 2-\beta; {2\over x}\right) \right] + {2^{\beta - 2} \over \beta (\beta - 1)}\sqrt{\pi} {\Gamma(2-\beta)  \over  \Gamma({3/ 2} - \beta)}. 
\end{equation}
and 
\begin{equation}
    \beta \equiv {3 - 2\mu \over 4}, \quad x \equiv {2 r \over r_I}, \quad x_0 \equiv {2 r_0 \over r_I} , \quad \tau \equiv {t \over t_\star}, \quad t_\star \equiv {\sqrt{GM_\bullet r_I^3} \over w (1-r_I/r_0)}, 
\end{equation}
and 
\begin{equation}
c_0 = x_0^{(1 + 14\mu)/8} \left(1 - {2\over x_0}\right)^{3/4 - 3/8\beta} \exp\left(-{1\over 2x_0}\right) \sqrt{1\over f_\beta(x_0)} ,
\end{equation}
and $M_d(0)$ is the total mass of the disk at $t = 0$. 
A reader familiar with the Newtonian disk equations will recognize the second line of equation \ref{green_s} as having the same broad functional form as the \cite{LBP74} solutions. In these equations $I_p(z)$ is the modified Bessel function of order $p$, ${}_2F_1(a, b;c;z)$ is the Hypergeometric function, and $\Gamma(z)$ is the usual gamma function. Note that the black hole spin only enters this solution implicitly through the ISCO radius $r_I$. 

From this solution we can derive all of the scaling arguments which we intuited from a global conservation analysis, only now with an explicit expression for $n$. Firstly, the time dependence of the accretion rate through the innermost stable circular orbit is found by taking the $x \to 2$ limit of the spatial gradient of this Green's function, which as $f_\beta(x\to2) \to 0$ means using the small argument expansion of the Bessel function $I_p(z)\sim z^p$, and therefore (see \citealt{Mummery23a} for the full functional form of the accretion rate)
\begin{equation}
    \dot M(t) \propto {1 \over t^{1+1/4\beta}} \exp\left(-{ f_\beta(x_0)^2 \over  4\tau} \right) ,
\end{equation}
or 
\begin{equation}
    \dot M(t) \propto t^{-n}, \quad n \equiv {4-2\mu \over 3-2\mu}, \quad t\to \infty.
\end{equation}
The time dependence of the location of the disk outer edge can be similarly found from the point at which the exponential decay begins to cut off the disk density, i.e., 
\begin{equation}
    f_\beta(x_{\rm out}) \sim t^{1/2} \to r_{\rm out} \sim t^{1/2\beta} \sim t^{2/(3-2\mu)} \sim t^{2n-2} .
\end{equation}
While the amplitude of $M'$ can be found from the scaling of $x^2 G_\Sigma(x, t)$ at this outer scale (i.e., at the point where $f_\beta(x) \sim t^{1/2}$), where again we make use of the small argument Bessel function expansion, to find 
\begin{equation}
    M'_{\rm max} \sim {x_{\rm out}^{5/4 -\mu} \over t^{1+1/8\beta}} \sim t^{-1/(3-2\mu)} \sim t^{1-n} .
\end{equation}
What we learn from this solution, which cannot be inferred from global conservation equations, is that the equivalent width of the mass distribution inherits the same time dependence as the outer disk radius, namely 
\begin{equation}
    \sigma_{\rm eff}(t) \sim t^{2/(3-2\mu)} \sim t^{2n-2} .
\end{equation}
This confirms the intuition of the global analysis, that there is only one key microphysical parameter in the problem, namely the index $n$. 

\section{Dependence of long-term profiles on the parameterisation of $W^r_\phi$}\label{app:plots}
In this Appendix we highlight some of the sensitivity of our solutions to the stress parameterization $W^r_\phi$. We begin with the precise choice of the value of $w$ (assuming $\mu=0$ throughout). 

\begin{figure*}
    \centering
    \includegraphics[width=0.49\linewidth]{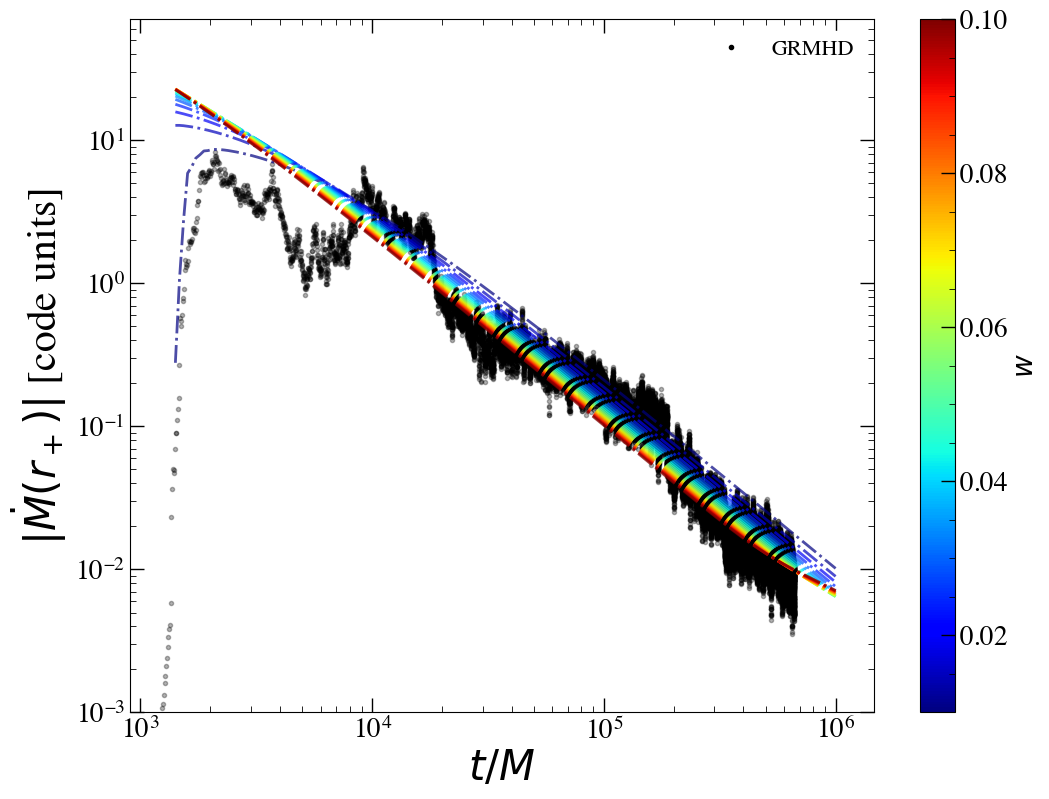}
    \includegraphics[width=0.49\linewidth]{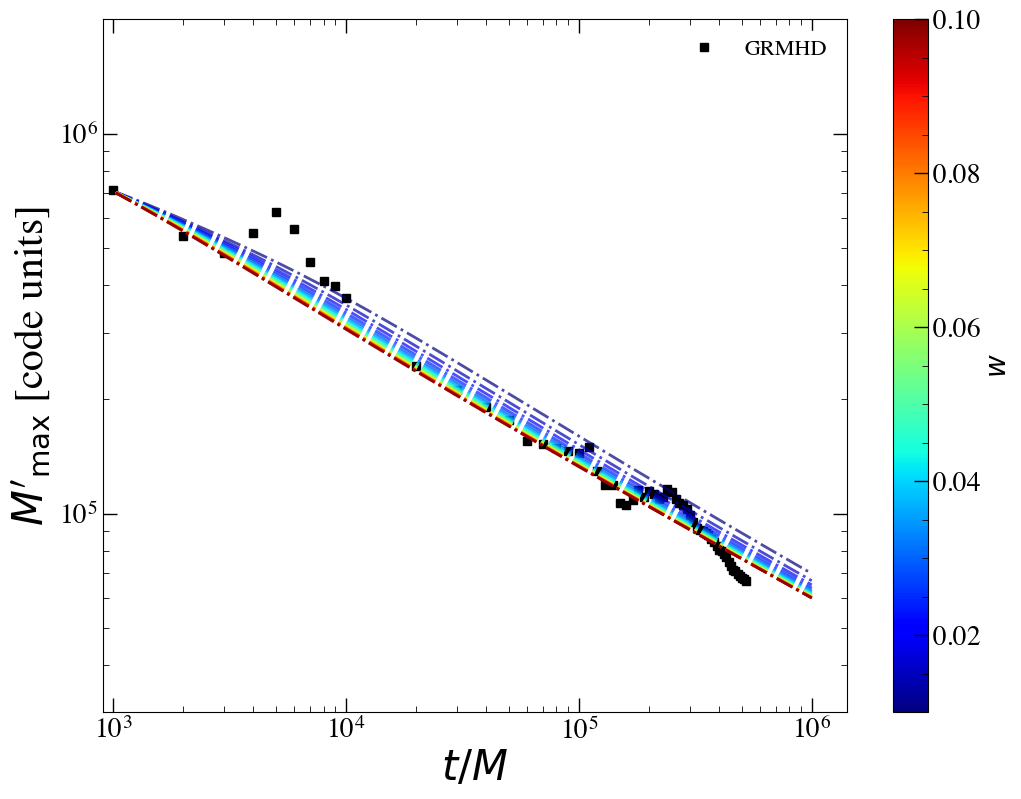}
    \includegraphics[width=0.49\linewidth]{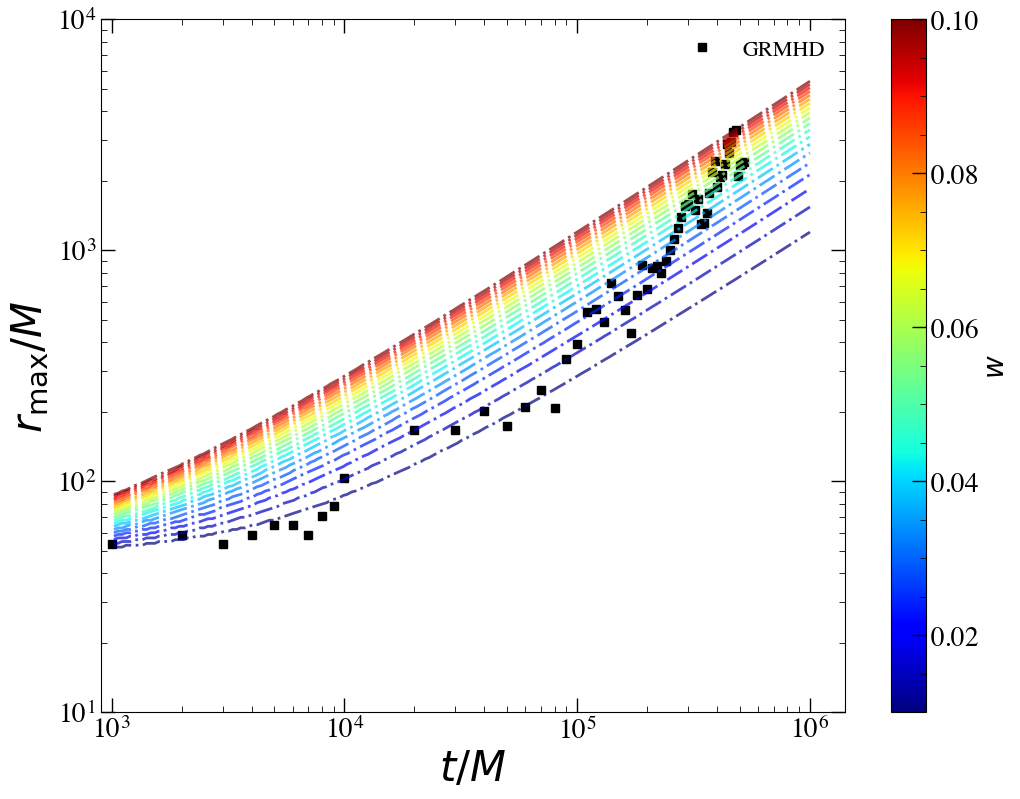}
    \includegraphics[width=0.49\linewidth]{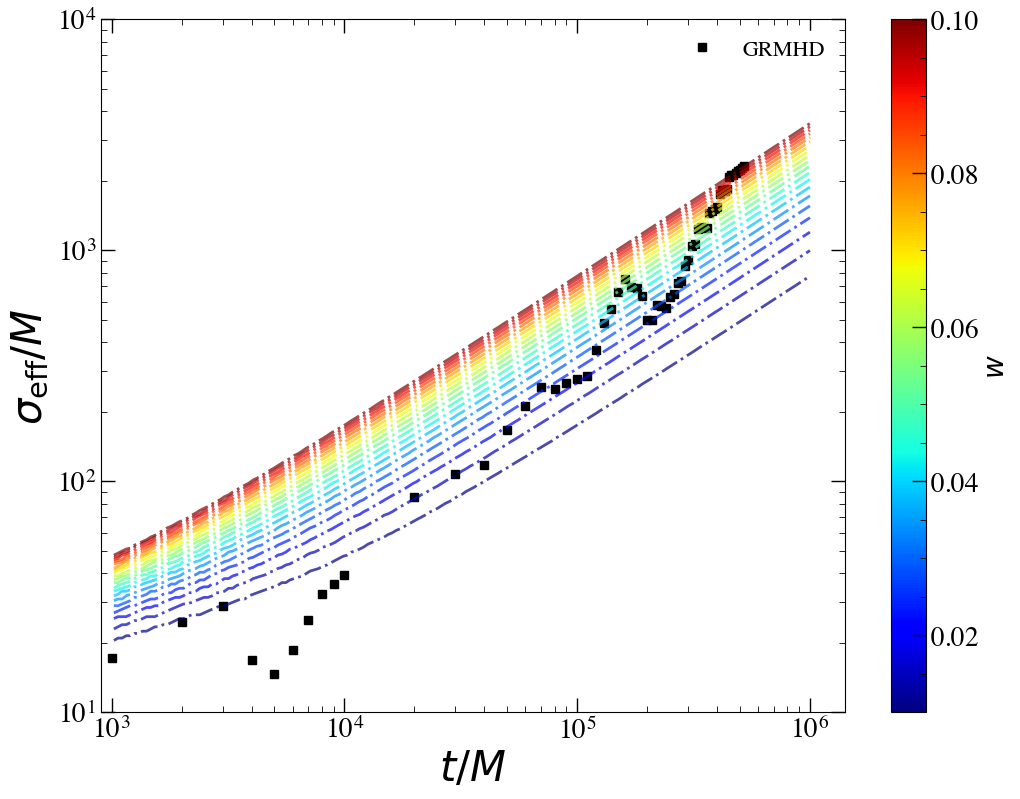}
    \caption{The impact of varying a constant turbulent stress amplitude $w$ on the evolving summary statistics. GRMHD values are plotted by black points, while the different profiles are denoted by line color (see color bar). Larger values of $w$ cause faster evolution (spreading, particularly notable in $r_{\rm max}$ and $\sigma_{\rm eff}$), a simple result of the evolutionary time of the systems scaling as $t_{\rm evol}\propto 1/w$. The peak accretion rate is strongly dependent on $w$, as it broadly scales as $\dot M_{\rm pk} \sim M_{\rm disk, 0}/t_{\rm evol}\sim w$, but at late times the difference is minimal. \revision{The GRTDT accretion rates are renormalized by a common factor $\sim 1/2$, as in Figure \ref{fig:summarystats}.}   }
    \label{fig:C1}
\end{figure*}

In Figure \ref{fig:C1} we demonstrate the impact of varying $w$ on the four summary statistics. We normalize each solution to have the same value of $M'_{\rm max}$ at $t = 10^3 t_g$. Larger values of $w$ cause faster evolution (spreading, particularly notable in $r_{\rm max}$ and $\sigma_{\rm eff}$), a simple result of the evolutionary time of the systems scaling as $t_{\rm evol}\propto 1/w$. The peak accretion rate is strongly dependent on $w$, as it broadly scales as $\dot M_{\rm pk} \sim M_{\rm disk, 0}/t_{\rm evol}\sim w$, but at late times the difference is minimal.  This minimal dependence at late times can be seen from the fact that $\dot M(t\gg t_{\rm evol})\sim \dot M_{\rm pk} (t/t_{\rm evol})^{-n} \sim M_{\rm disk, 0} t_{\rm evol}^{n-1}\sim w^{1-n}$, a weakly decaying function of $w$. 

We note, briefly, that a transition from a lower stress at early times $w_{\rm early} \sim 10^{-2}$ to a larger stress at later times $w_{\rm late} \sim 10^{-1}$ through a smooth interpolation function $f(t)$ (how to do this is discussed in the text) would likely perform best, but until further analysis on the evolution of $W^r_\phi$ across different simulations this likely represents overfitting. 

\begin{figure*}
    \centering
    \includegraphics[width=0.49\linewidth]{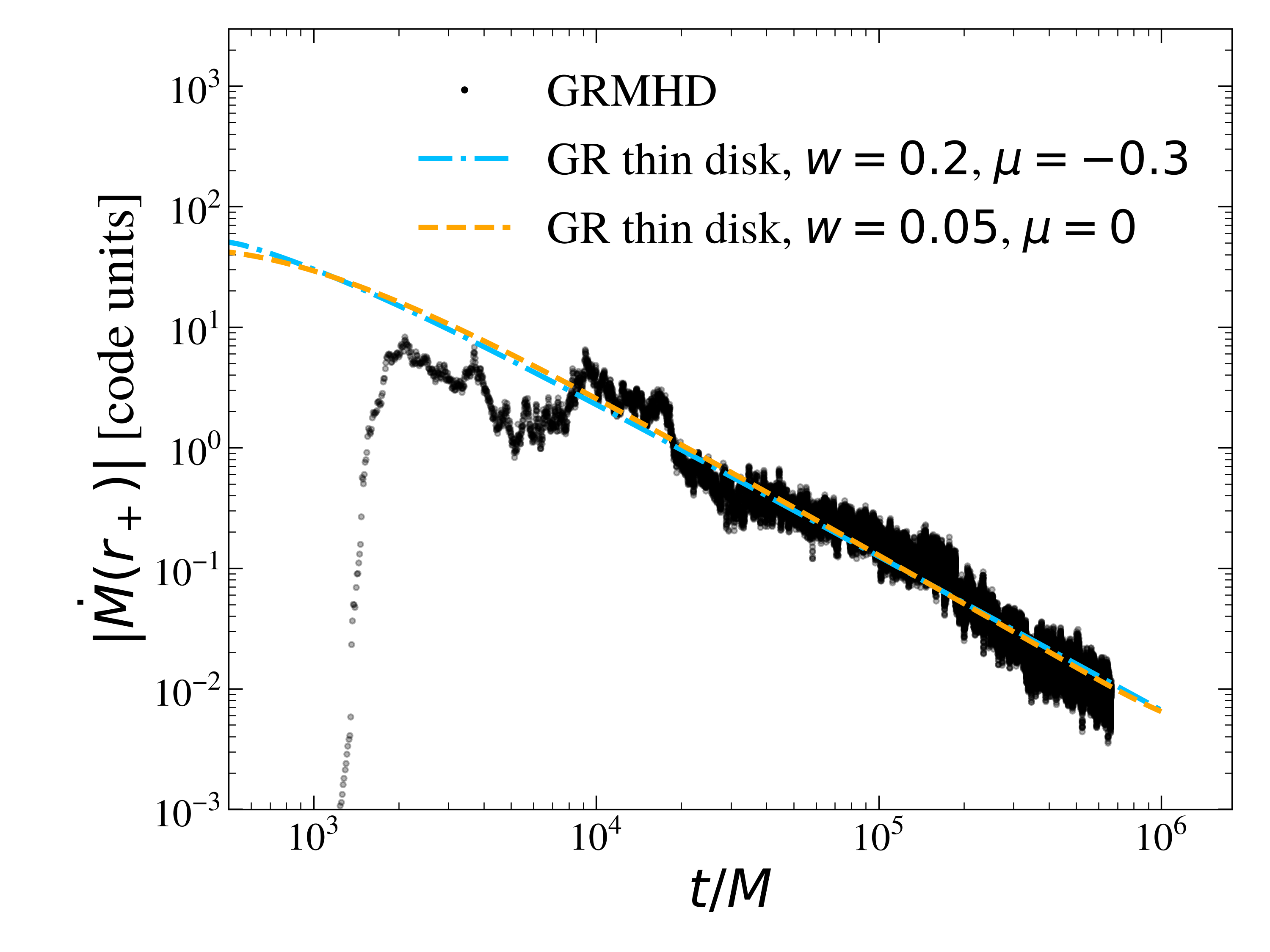}
    \includegraphics[width=0.49\linewidth]{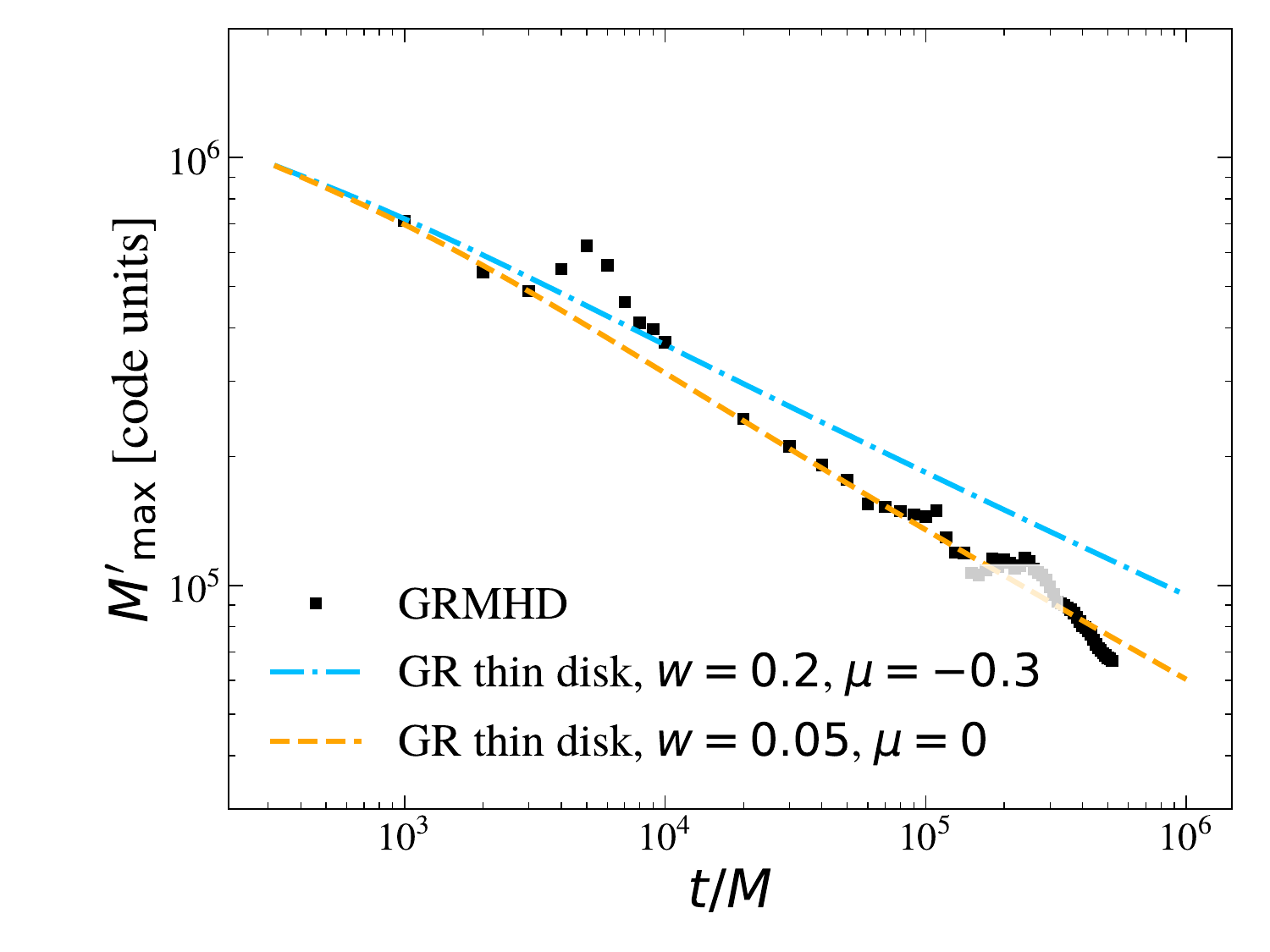}
    \includegraphics[width=0.49\linewidth]{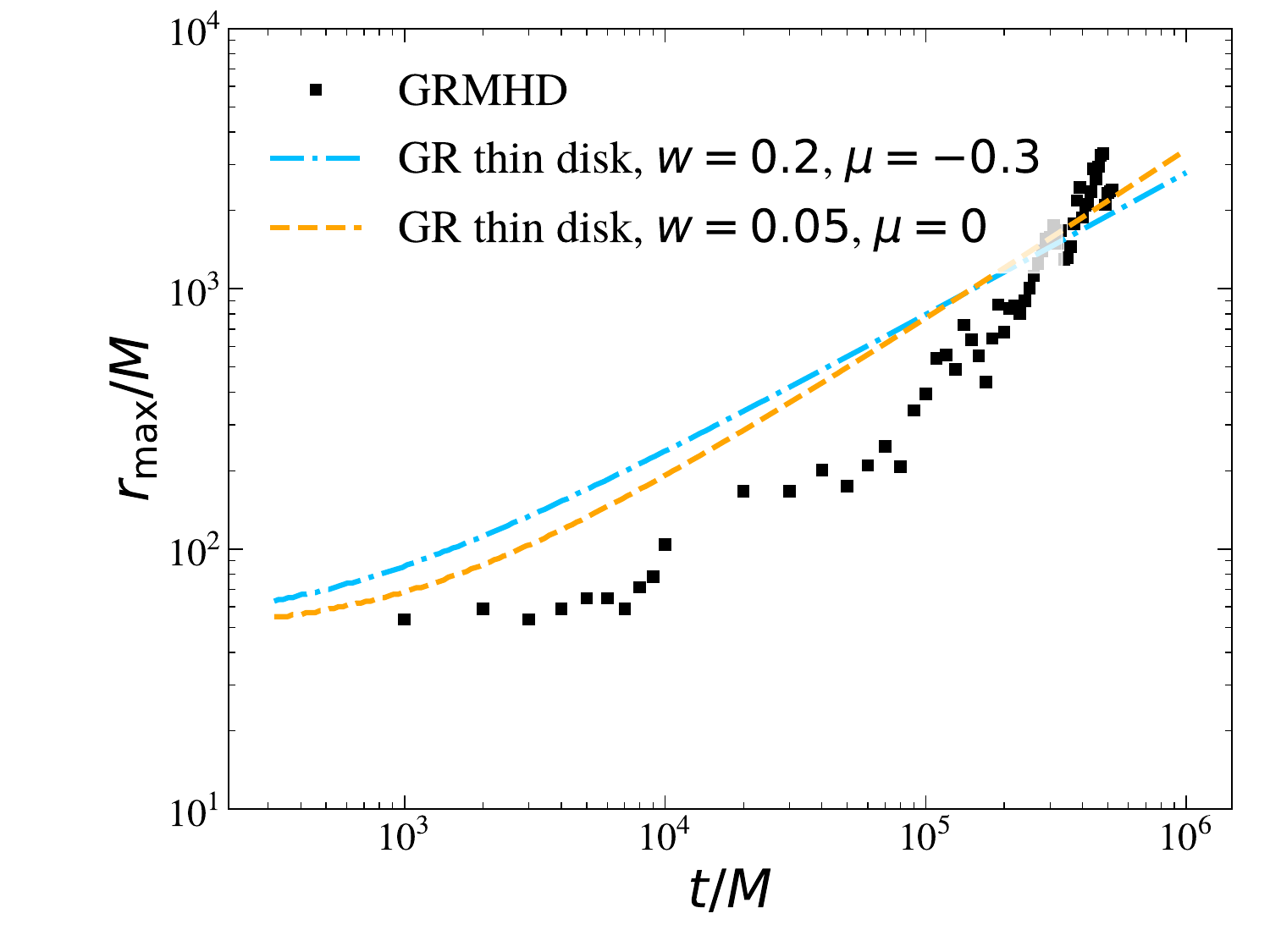}
    \includegraphics[width=0.49\linewidth]{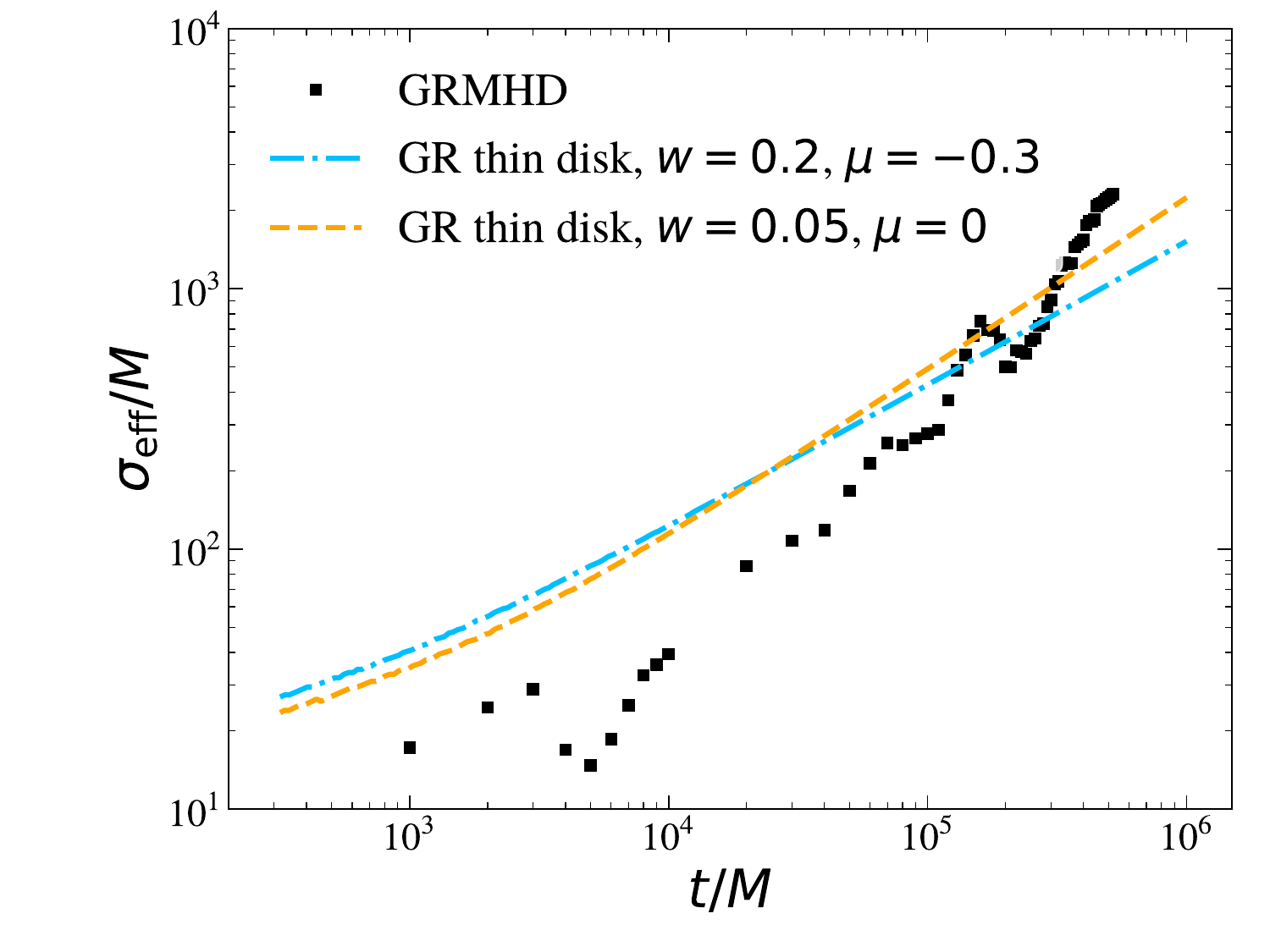}
    \caption{The evolution of the summary statistics for two plausible parameterizations of the turbulent stress seen in our simulation, $W^r_\phi = 0.05$ (orange), and $W^r_\phi = 0.2 r^{-0.3}$ (blue). The summary statistics are broadly insensitive (on these log plots) to such a change. More detailed comparison is required to distinguish the two. \revision{The GRTDT accretion rates are renormalized by a common factor $\sim 1/2$, as in Figure \ref{fig:summarystats}.}  }
    \label{fig:C2}
\end{figure*}

In Figure \ref{fig:C2} we highlight the (broad lack of) dependence of the summary statistics on the precise stress parameterisation. We choose as a different example $W^r_\phi = 0.2 r^{-0.3}$ (code units), which is a plausible description of the radial dependence seen in our simulation (Fig. \ref{fig:stress}). The summary statistics are broadly insensitive (on these log plots) to such a change, although of course there are factor $\sim 2$ differences. 

\begin{figure}
    \centering
    \includegraphics[width=0.5\linewidth]{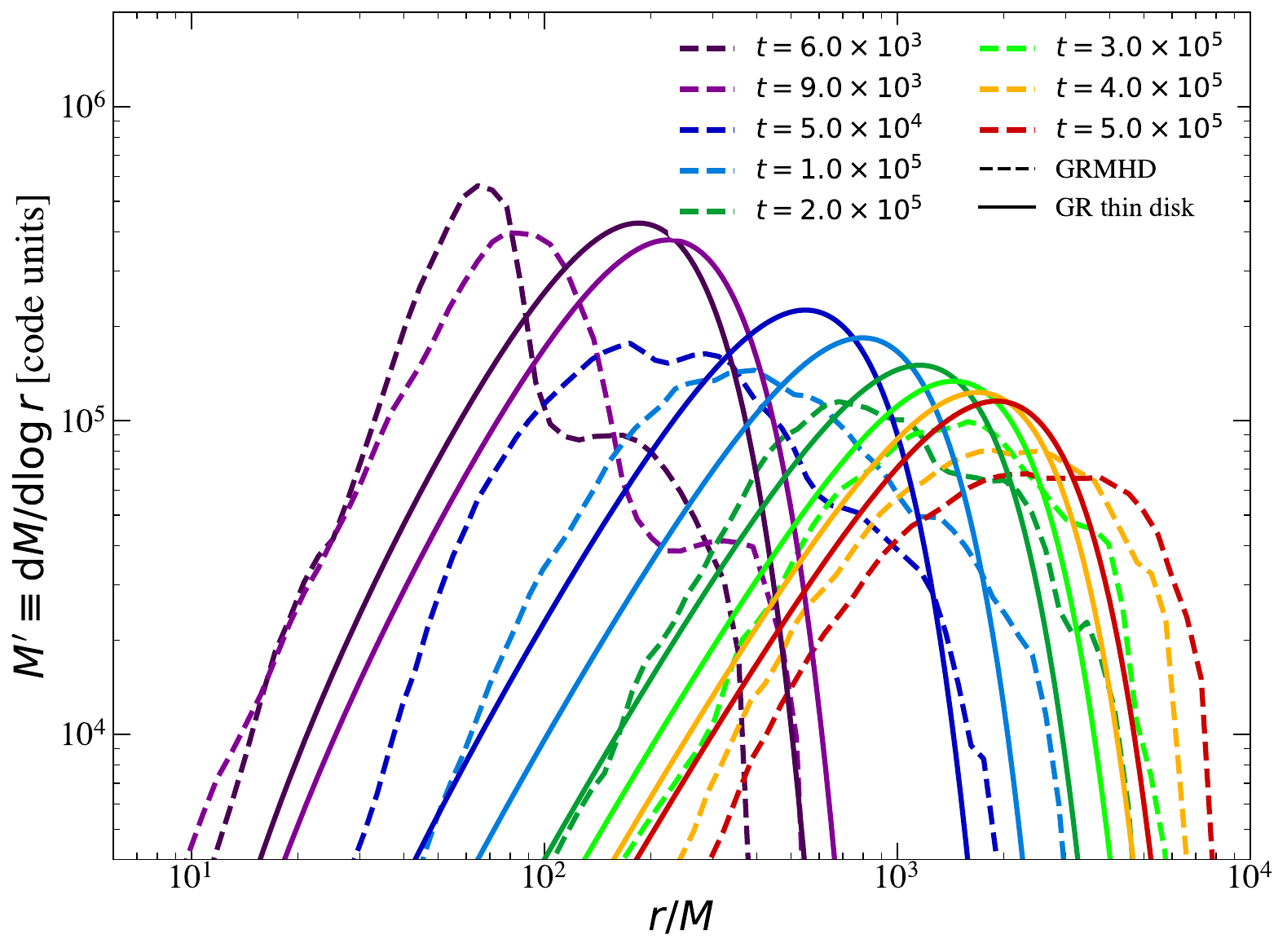}
    \caption{A detailed comparison between the GRMHD mass distribution and that of GRTDT for $W^r_\phi = 0.2 r^{-0.3}$. We see that the evolution of this disk is a less good approximation of the GRMHD simulation. This is likely because in the region of the disk with the bulk of the material the stress really is better approximated by a constant, with the apparent global rise at small radii a relativistic effect which only impacts low density (near horizon) regions. }
    \label{fig:C3}
\end{figure}

A more detailed comparison of $M'$ however shows that this is indeed a worse description of the system. This is shown in Figure \ref{fig:C3}. We see that the global evolution of the disk mass distribution is less well described by a radially-dependent stress. This is likely because in the region of the disk with the bulk of the material (large radii) the stress really is better approximated by a constant, with the apparent global rise at small radii a relativistic effect which only impacts low density (near horizon) regions. In this sense, the near horizon physics appears to have less effect on the evolution than the large scale saturation. This should be tested with future numerical simulations.

\section{\revision{Comparison of the standard and cyclic zoom simulations}}\label{app:cyclic}
\revision{The cyclic zoom method, and a series of tests of it against standard GRMHD simulations, are presented in \citet{Guo2025}. Those tests include hydrodynamic and magnetized Bondi accretion, magnetically arrested tori around black holes of a range of spins evolved to $10^5\,t_\mathrm{g}$, and the long-term evolution of the torus configuration used in the present paper, compared with a standard simulation over $10^6\,t_\mathrm{g}$ \citep[][their section 3.5]{Guo2025}. To make the present paper self-contained, we here compare directly the two simulations analyzed in this work, over the period in which both are available.}

\begin{figure*}
    \centering
    \includegraphics[width=\linewidth]{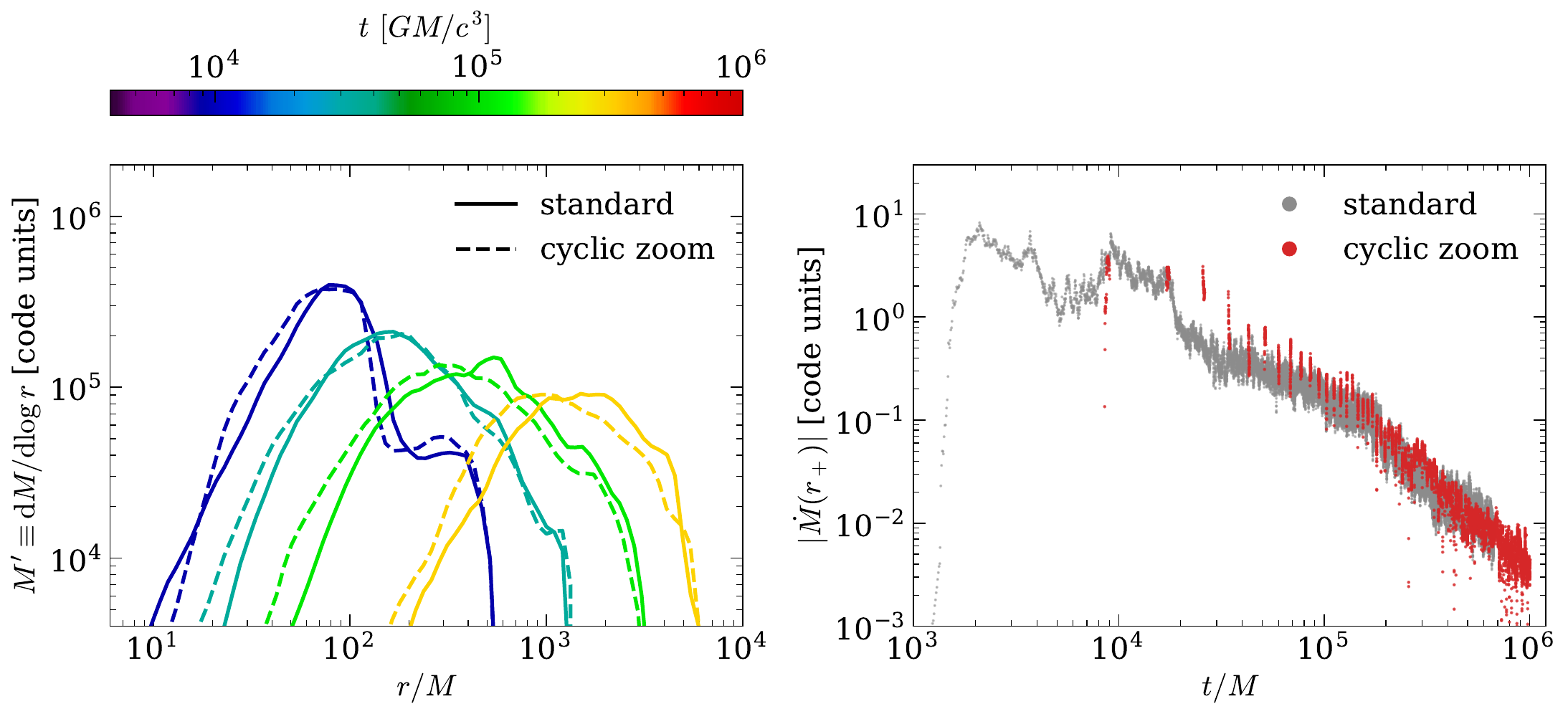}
    \caption{\revision{A direct comparison of the standard GRMHD and cyclic zoom simulations over the period in which both are available. {\bf Left:} the bound mass distribution $M'$ at four matched times, for the standard run (solid) and the cyclic zoom run (dashed); the color of each curve denotes its time, on the same scale as Figure \ref{fig:comparison}. {\bf Right:} the horizon accretion rate in the standard run (grey) and the cyclic zoom run (red; available only during the finest-level phase of each zoom cycle). }
    \label{fig:cyclic_compare}}
\end{figure*}

\revision{In Figure \ref{fig:cyclic_compare} we show the bound mass distribution $M'$ from the two simulations at four matched times between $t \approx 9\times10^3$ and $3\times10^5\,t_\mathrm{g}$ (left panel), and the horizon accretion rate from both simulations on the same axes (right panel). In the cyclic zoom run the horizon is resolved only during the finest-level phase of each cycle, and so the accretion rate is only available during these phases.}

\revision{The bound mass distributions of the two runs agree well. At the four matched times, the peak amplitude $M'_{\rm max}$ agrees to within $10\%$, the total bound mass to within $7\%$, and the effective width $\sigma_{\rm eff}$ to within $\sim 20\%$. The location of the peak, $r_{\rm max}$, differs by up to a factor $\sim 2$ at $t\sim 10^5\,t_\mathrm{g}$. This reflects the fact that the mass distribution is broad and flat-topped at these times, so that the location of its maximum is a noisy statistic in either run (see also Figure \ref{fig:summarystats}), rather than any systematic difference in where the mass resides: the inner and outer edges of the distributions agree closely.}

\revision{The horizon accretion rates agree to within a factor $\sim 1.5$ for $t \gtrsim 3\times 10^4\,t_\mathrm{g}$, comparable to the amplitude of the stochastic variability in either run. The cyclic zoom accretion rate is systematically somewhat higher (by a median factor $\sim 1.2$--$1.5$) over $3\times10^4 \lesssim t \lesssim 3\times10^5\,t_\mathrm{g}$, with the two runs converging at later times. We have checked that the accretion rate measured in the cyclic zoom run shows no systematic transient following each re-refinement of the horizon region: the median accretion rate during the first quarter of each finest-level phase is, on average, within $5\%$ of that during the last quarter.}

\revision{Over the period in which a comparison is possible, the cyclic zoom run reproduces the evolution of the bound mass distribution of the standard run to an accuracy well within the level of agreement between either run and GRTDT, and it reproduces the horizon accretion rate to within the level of its stochastic variability. We do not use the cyclic zoom run to study accretion rate variability, which is not expected to be captured accurately by the method because the horizon is resolved only intermittently \citep{Guo2025}. Beyond $10^6\,t_\mathrm{g}$ no standard run is available, and the reliability of the method cannot be tested directly. We note three considerations relevant to this regime. First, the method is not conservative to machine precision within the de-refined mask region, although \citet{Guo2025} find that the resulting differences in the total mass are $\lesssim 1\%$ over $10^6\,t_\mathrm{g}$. Second, \citet{Guo2025} found that for a non-spinning black hole the cyclic zoom method overestimates the electromagnetic feedback power by a factor $\sim 2$, which could plausibly enhance the unbinding of gas over long times. Third, the number of zoom levels was increased at $t = 10^6$ and $3\times10^6\,t_\mathrm{g}$ to follow the expanding disk. The marked decline in the bound mass at $t \gtrsim 3\times10^6\,t_\mathrm{g}$, and the transition to an almost entirely unbound flow by $t\sim 10^7\,t_\mathrm{g}$ (Figures \ref{fig:dens}, \ref{fig:bern} and \ref{fig:m+j}), most likely reflects the growing dominance of outflows at late times (Figure \ref{fig:m+j}), but given these considerations we cannot exclude a contribution from the limitations of the method in this regime. As in the main text, we therefore restrict quantitative comparisons with GRTDT to times before this transition.}

\end{CJK*}
\end{document}